\documentclass[twocolumn,usenatbib]{mnras}

\usepackage{newtxtext,newtxmath}

\usepackage[T1]{fontenc}

\DeclareRobustCommand{\VAN}[3]{#2}
\let\VANthebibliography\thebibliography
\def\thebibliography{\DeclareRobustCommand{\VAN}[3]{##3}\VANthebibliography}

\usepackage{graphicx}	
\usepackage{amsmath}	
\usepackage{enumitem}
\usepackage{arydshln}
\usepackage[figuresleft]{rotating}
\usepackage{caption}

\newcommand{\fsps}{$\mathrm{fs~s^{-1}}$}
\newcommand{\msun}{\ifmmode\mbox{M}_{\odot}\else$\mbox{M}_{\odot}$\fi}
\newcommand{\Vsun}{\ifmmode\mbox{V}_{\odot}\else$\mbox{V}_{\odot}$\fi}

\newcommand{\maspy}{$\rm mas~yr^{-1}$}
\newcommand{\kmps}{$\rm km~s^{-1}$}

\newcommand{\psrpi}{\ensuremath{\mathrm{PSR}\pi}}
\newcommand{\mspsrpi}{\ensuremath{\mathrm{MSPSR}\pi}}

\newcommand{\rcs}{$\chi_{\nu}^{2}$}

\newcommand{\psrb}{PSR~J0030$+$0451}
\newcommand{\psrc}{PSR~J0610$-$2100}
\newcommand{\psrd}{PSR~J0621$+$1002}
\newcommand{\psrea}{PSR~J1012$+$5307}
\newcommand{\psreb}{PSR~J1024$-$0719}
\newcommand{\psrfa}{PSR~J1518$+$4904}
\newcommand{\psrfb}{PSR~J1537$+$1155}
\newcommand{\psrga}{PSR~J1640$+$2224}
\newcommand{\psrgb}{PSR~J1643$-$1224}
\newcommand{\psrha}{PSR~J1721$-$2457
}
\newcommand{\psrhb}{PSR~J1730$-$2304}
\newcommand{\psri}{PSR~J1738$+$0333}
\newcommand{\psro}{PSR~J1824$-$2452A}
\newcommand{\psrka}{PSR~J1853$+$1303}
\newcommand{\psrl}{PSR~J1910$+$1256}

\newcommand{\psrmb}{PSR~J1918$-$0642}
\newcommand{\psrkb}{PSR~J1939$+$2134}

\newcommand{\Psrb}{J0030$+$0451}
\newcommand{\Psrc}{J0610$-$2100}
\newcommand{\Psrd}{J0621$+$1002}
\newcommand{\Psrea}{J1012$+$5307}
\newcommand{\Psreb}{J1024$-$0719}
\newcommand{\Psrfa}{J1518$+$4904}
\newcommand{\Psrfb}{J1537$+$1155}
\newcommand{\Psrga}{J1640$+$2224}
\newcommand{\Psrgb}{J1643$-$1224}
\newcommand{\Psrha}{J1721$-$2457
}
\newcommand{\Psrhb}{J1730$-$2304}
\newcommand{\Psri}{J1738$+$0333}
\newcommand{\Psro}{J1824$-$2452A}
\newcommand{\Psrka}{J1853$+$1303}
\newcommand{\Psrl}{J1910$+$1256}
\newcommand{\Psrma}{J1911$-$1114}
\newcommand{\Psrmb}{J1918$-$0642}
\newcommand{\Psrkb}{J1939$+$2134}
\newcommand{\Psrna}{J0437$-$4715}
\newcommand{\Psrnb}{J1713$+$0747}

\newcommand{\multilinecomment}[1]{}

\title[The \mspsrpi\ results and implications]{The \mspsrpi\ catalogue: VLBA astrometry of 18 millisecond pulsars}

\author[H.~Ding et al.]{
H. Ding,$^{1,2}$\thanks{E-mail: hdingastro@hotmail.com}
A. T. Deller,$^{1,2}$\thanks{E-mail: adeller@astro.swin.edu.au}
B. W. Stappers,$^{3}$
T. J. W. Lazio,$^4$
D. Kaplan,$^{5}$
S. Chatterjee,$^6$
W. Brisken,$^7$
\newauthor
J. Cordes,$^6$
P. C. C. Freire,$^8$
E. Fonseca,$^{9,10}$
I. Stairs,$^{11}$
L. Guillemot,$^{12,13}$
A. Lyne,$^{3}$
I. Cognard,$^{12,13}$
\newauthor
D. J. Reardon,$^{1,2}$
G. Theureau$^{12,13,14}$
\\
$^{1}$Centre for Astrophysics and Supercomputing, Swinburne University of Technology, John St, Hawthorn, VIC 3122, Australia\\
$^{2}$ARC Centre of Excellence for Gravitational Wave Discovery (OzGrav), Australia\\
$^{3}$Jodrell Bank Centre for Astrophysics, Department of Physics and Astronomy, The University of Manchester, Manchester M13 9PL, UK\\
$^4$Jet Propulsion Laboratory, California Institute of Technology, 4800 Oak Grove Blvd, Pasadena, CA  91109 USA\\
$^{5}$Center for Gravitation, Cosmology and Astrophysics, Department of Physics, University of Wisconsin-Milwaukee, Milwaukee, WI 53201\\
$^{6}$Cornell Center for Astrophysics and Planetary Science, Cornell University, Ithaca, NY 14853, USA\\
$^7$National Radio Astronomy Observatory, P.O. Box O, Socorro NM 87801, USA\\
$^8$Max-Planck-Institut für Radioastronomie, auf dem Hügel 69, Bonn D-53121\\
$^9$Department of Physics and Astronomy, West Virginia University, Morgantown, WV 26506-6315, USA\\
$^{10}$Center for Gravitational Waves and Cosmology, Chestnut Ridge Research Building, Morgantown, WV 26505, USA\\
$^{11}$Dept. of Physics and Astronomy, Unviersity of British Columbia, 6224 Agricultural Road, Vancouver, BC V6T 1Z1\\
$^{12}$Observatoire Radioastronomique de Nan{\c c}ay, Observatoire de Paris, Universit\'e PSL, Universit\'e d’Orl\'eans, CNRS, 18330 Nan{\c c}ay, France\\
$^{13}$Laboratoire de Physique et Chimie de l'Environnement et de l'Espace, Universit\'e d’Orl\'eans/CNRS, 45071 Orl\'eans Cedex 02, France\\
$^{14}$LUTH, Observatoire de Paris, Universit\'e PSL, Universit\'e Paris Cit\'e, CNRS, 92195 Meudon, France\\
}

\date{Accepted XXX. Received YYY; in original form ZZZ}

\pubyear{2015}

\begin{document}
\label{firstpage}
\pagerange{\pageref{firstpage}--\pageref{lastpage}}
\maketitle

\begin{abstract}
With unparalleled rotational stability, millisecond pulsars (MSPs)
serve as ideal laboratories for numerous astrophysical studies, many of which require precise knowledge of the distance and/or velocity of the MSP.
Here, we present the astrometric results for 18 MSPs of the ``\mspsrpi" project focusing exclusively on astrometry of MSPs, which includes the re-analysis of 3 previously published sources.
On top of a standardized data reduction protocol, more complex strategies (i.e., normal and inverse-referenced 1D interpolation) were employed where possible to further improve astrometric precision.
We derived astrometric parameters 
using {\tt sterne}, a new Bayesian astrometry inference package that allows the incorporation of prior information based on pulsar timing where applicable.  We measured significant ($>3\,\sigma$) parallax-based distances for 15 MSPs, including $0.81\pm0.02$\,kpc for \psrfa\ --- the most significant model-independent distance ever measured for a double neutron star system. 
For each MSP with a well-constrained distance, we estimated its transverse space velocity and radial acceleration. Among the estimated radial accelerations, the updated ones of \psrea\ and \psri\ impose new constraints on dipole gravitational radiation and the time derivative of Newton's gravitational constant. Additionally, significant angular broadening was detected for \psrgb, which offers an independent check of the postulated association between the HII region Sh~2-27 and the main scattering screen of \psrgb.
Finally, the upper limit of the death line of $\gamma$-ray-emitting pulsars is refined with the new radial acceleration of the hitherto least energetic $\gamma$-ray pulsar \psrhb. 

\end{abstract}

\begin{keywords}
radio continuum: stars -- stars: kinematics and dynamics -- gravitation -- gamma-rays: stars -- pulsars: individual: \psrb, \psrc, \psrd, \psreb, \psrfb, \psrka, \psrl, \psrmb, \psrkb\ 
\end{keywords}



\section{Introduction}
\label{sec:intro}


\subsection{Millisecond pulsars: a key for probing theories of gravity and detecting the gravitational-wave background}
\label{subsec:gravity_test_intro}

Pulsars are an observational manifestation of neutron stars (NSs) that emit non-thermal electromagnetic radiation while spinning  \citep{Hewish69,Gold68,Pacini68}. 
Over 3000 radio pulsars have been discovered to date throughout the Galaxy and the nearest members of the Local Group \citep{Manchester05}.
Due to the large moment of inertia of pulsars, the pulses we receive on Earth from a pulsar exhibit highly stable periodicity. 
By measuring a train of pulse time-of-arrivals (ToAs) of a pulsar and comparing it against the model prediction, a long list of model parameters can be inferred \citep[e.g.][]{Detweiler79,Helfand80}.
This procedure to determine ToA-changing parameters is known as pulsar timing, hereafter referred to as timing.

In the pulsar family, recycled pulsars (commonly refereed to as millisecond pulsars, or MSPs), have the shortest rotational periods. 
They are believed to have been spun-up through the accretion from their donor stars during a previous evolutionary phase as a low-mass X-ray binary (LMXB) \citep{Alpar82}. 
As the duration of the recycling phase (and hence the degree to which the pulsar is spun-up) can vary depending on the nature of the binary, there is no clear spin period threshold that separates MSPs from canonical pulsars. In this paper, we define MSPs as pulsars with spin periods of $\lesssim 40$\,ms and magnetic fields $\lesssim10^{10}$~G. This range encompasses most partially recycled pulsars with NS companions, such as \psrfb\ (also known as PSR~B1534$+$12) and \psrfa.
Compared to non-recycled pulsars, ToAs from MSPs can be measured to higher precision due to both the narrower pulse profiles and larger number of pulses.  Additionally, MSPs exhibit more stable rotation \citep[e.g.][]{Hobbs10}; both factors promise a lower level of random timing noise. 
Consequently, MSPs outperform non-recycled pulsars in the achievable precision for
probing theories underlying ToA-changing astrophysical effects. In particular, MSPs provide the hitherto most precise tests for gravitational theories \citep[e.g.][]{Kramer21a,Zhu19,Freire12}.
Einstein's theory of general relativity (GR) is the simplest form among a group of possible candidate post-Newtonian gravitational theories. 
The discovery of highly relativistic double neutron star (DNS) systems \citep[e.g.][]{Hulse75,Wolszczan91,Burgay03,Lazarus16,Stovall18,Cameron18}
and their continued timing have resulted in many high-precision tests of GR and other gravity theories (\citealp{Fonseca14,Weisberg16,Ferdman20}, and especially \citealp{Kramer21a}). The precise timing, optical spectroscopy and VLBI observations of pulsar-white-dwarf (WD) systems have, in addition, achieved tight constraints on several classes of alternative theories of gravity \citep{Deller08,Lazaridis09,Freire12,Antoniadis13,Ding20,Guo21,Zhao22}.

Gravitational Waves (GWs) are changes in the curvature of spacetime (generated by accelerating masses), which propagate at the speed of light. Individual GW events in the Hz---kHz range have been detected directly with GW observatories (e.g. \citealp{Abbott16}; see the third Gravitational-Wave Transient Catalog\footnote{\url{https://www.ligo.org/science/Publication-O3aFinalCatalog/}}), and indirectly using the orbital decay of pulsar binaries \citep[e.g.][]{Taylor82,Weisberg16,Kramer21a,Ding21a}. 
Collectively, a gravitational wave background (GWB), formed with primordial GWs and GWs generated by later astrophysical events \citep{Carr80}, is widely predicted, but has not yet been confirmed by any observational means. In the range of $10^{-9}\,\mathrm{Hz}-0.1$\,Hz, supermassive black hole binaries are postulated to be the primary sources of the GWB \citep{Sesana08}. 
In this nano-hertz regime, the most stringent constraints on the GWB are provided by pulsar timing \citep{Detweiler79}. 

To enhance the sensitivity for the GWB hunt with pulsar timing, and to distinguish GWB-induced ToA signature from other sources of common timing ``noise'' (e.g., Solar-system planetary ephemeris error, clock error and interstellar medium, \citealp{Tiburzi16}), a pulsar timing array (PTA), composed of MSPs scattered across the sky (see \citealp{Roebber19} for spatial distribution requirement), is necessary \citep{Foster90}. 
After 2 decades of efforts, no GWB has yet been detected by a PTA, though common steep-spectrum timing noise (in which GWB signature should reside) has already been confirmed by several radio PTA consortia \citep{Arzoumanian20,Goncharov21,Chen21,Antoniadis22}. At $\gamma$-rays, a competitive GWB amplitude upper limit was recently achieved using the Fermi Large Area Telescope with 12.5 years of data \citep{Fermi-LAT-Collaboration22}.


\subsection{Very long baseline astrometry of millisecond pulsars}
\label{subsec:MSP_VLBI_astrometry}

In timing analysis, 
astrometric information for an MSP (reference position, proper motion, and annual geometric parallax) can form part of the global ensemble of parameters determined from ToAs. 
However, the astrometric signatures can be small compared to the ToA precision and/or covariant with other parameters in the model, especially for new MSPs that are timed for less than a couple of years \citep{Madison13}.
Continuing to add newly discovered MSPs into PTAs is considered the best pathway to rapidly improve the PTA sensitivity \citep{Siemens13}, and is particularly important for PTAs based around newly commissioned high-sensitivity radio telescopes \citep[e.g.][]{Bailes20}.
Therefore, applying priors to the astrometric parameters can be highly beneficial for the timing of individual MSPs (especially the new ones) and for enhancing PTA sensitivities \citep{Madison13}.

Typically, the best approach to independently determine precise astrometric parameters for MSPs is the use of phase-referencing \citep[e.g.][]{Lestrade90,Beasley95} very long baseline interferometry (VLBI) observations, which can achieve sub-mas positional precision (relative to a reference source position) for MSPs in a single observation. By measuring the sky position of a Galactic MSP a number of times and modeling the position evolution,
VLBI astrometry can obtain astrometric parameters for the MSP.
Compared to pulsar timing, VLBI astrometry normally takes much shorter time 
to reach a given astrometric precision \citep[e.g.][]{Brisken02,Chatterjee09,Deller19}. 

One of the limiting factors on searching for the GWB with PTAs is the uncertainties on the Solar-system planetary ephemerides (SSEs) \citep{Vallisneri20}, which are utilized to convert geocentric ToAs to ones measured in the (Solar-system) barycentric frame (i.e., the reference frame with respect to the barycentre of the Solar system). 
Various space-mission-driven SSEs have been released mainly by two SSE providers --- the NASA Jet Propulsion Laboratory \citep[e.g.][]{Park21} and the IMCCE \citep[e.g.][]{Fienga20a}.
In pulsar timing analysis, adopting different SSEs may lead to discrepant timing parameters \citep[e.g.][]{Wang17}. On the other hand, VLBI astrometry measures offsets with respect to a source whose position is measured in a quasi-inertial (reference) frame defined using remote quasars \citep[e.g.][]{Charlot20}. Although VLBI astrometry also relies on SSEs to derive annual parallax, it is robust against SSE uncertainties. In other words, for VLBI astrometry, using different SSEs in parameter inference would not lead to a noticeable difference in the inferred parameters. Therefore, VLBI astrometry of MSPs can serve as an objective standard to be used to discriminate between various SSEs. Specifically, if an SSE is inaccurate, the barycentric frame based on the SSE would display rotation with respect to the quasar-based frame. This frame rotation can be potentially detectable by comparing VLBI positions of multiple MSPs against their timing positions \citep{Chatterjee09,Wang17}. 
By eliminating inaccurate SSEs, VLBI astrometry of MSPs can suppress the SSE uncertainties, and hence enhance the PTA sensitivities.

Besides the GWB-related motivations, interferometer-based astrometric parameters (especially distances to MSPs) have been adopted to sharpen the tests of gravitational theories for individual MSPs \citep[e.g.][]{Deller09,Deller18,Guo21,Ding21a}.
Such tests are normally made by comparing the model-predicted and observed post-Keplerian (PK) parameters that quantify excessive gravitational effects beyond a Newtonian description of the orbital motion.
Among the PK parameters is the orbital decay $\dot{P}_\mathrm{b}$ (or the time derivative of orbital period). The intrinsic cause of $\dot{P}_\mathrm{b}$ in double neutron star systems is dominated by the emission of gravitational waves, which can be predicted using the binary constituent masses and orbital parameters \citep[e.g.][]{Lazaridis09,Weisberg16}.
To test this model prediction, however, requires any extrinsic orbital decay $\dot{P}_\mathrm{b}^\mathrm{ext}$ due to relative acceleration between the pulsar and the observer to be removed from the observed $\dot{P}_\mathrm{b}$. 
Such extrinsic terms depend crucially on the proper motion and the distance of the pulsar, however these (especially the distance) can be difficult to estimate from pulsar timing. Precise VLBI determination of proper motions and distances can yield precise estimates of these extrinsic terms and therefore play an important role in orbital-decay tests of gravitational theories.
Likewise, Gaia astrometry on nearby pulsar-WD systems can potentially serve the same scientific goal, though the method is only applicable to a small number of pulsar-WD systems where the WDs are sufficiently bright for the Gaia space observatory (see Section~\ref{subsec:Gaia_results}). 


Last but not least, pulsar astrometry is crucial for understanding the Galactic free-electron distribution, or the Galactic free-electron number density $n_\mathrm{e}(\vec{x})$ as a function of position. An $n_\mathrm{e}(\vec{x})$ model is normally established by using pulsars with well determined distances as benchmarks. As the pulsations from a pulsar allow precise measurement of its dispersion measure (DM), the average $n_\mathrm{e}$ between the pulsar and the Earth can be estimated given the pulsar distance. 
Accordingly, a large group of such benchmark pulsars across the sky would enable the establishment of an $n_\mathrm{e}(\vec{x})$ model. 
In a relevant research field, extragalactic fast radio bursts (FRBs) have been used to probe intergalactic medium distribution on a cosmological scale \citep[e.g.][]{Macquart20,Mannings21}, which, however, demands the removal of the DMs of both the Galaxy and the FRB host galaxy. The Galactic DM cannot be determined without a reliable $n_\mathrm{e}(\vec{x})$ model, which, again, calls for precise astrometry of pulsars across the Galaxy.

\subsection{The \mspsrpi\ project}
\label{subsec:mspsrpi}

Using the Very Long Baseline Array (VLBA), the \psrpi\ project tripled the sample of pulsars with precisely measured astrometric parameters \citep{Deller19}, but included just three MSPs. 
The successor project, \mspsrpi, is a similarly designed VLBA astrometric program targeting exclusively MSPs. 
Compared to canonical pulsars, MSPs are generally fainter. To identify MSPs feasible for VLBA astrometry, a pilot program was conducted, which found 31 suitable MSPs. 
Given observational time constraints, we selected 18 MSPs as the targets of the \mspsrpi\ project, focusing primarily on sources observed by pulsar timing arrays. The 18 MSPs are listed in Table~\ref{tab:MSPs} along with their spin periods $P_\mathrm{s}$ and orbital periods $P_\mathrm{b}$ (if available) that have been obtained from the ATNF Pulsar Catalogue\footnote{\label{footnote:PSRCAT}\url{https://www.atnf.csiro.au/research/pulsar/psrcat/}} \citep{Manchester05}. 
The astrometric results for 3 sources (\psrea, \psrfb, \psrga) involved in the project have already been published \citep{Vigeland18,Ding20,Ding21a}. 
In this paper, we present the astrometric results of the remaining 15 MSPs studied in the \mspsrpi\ project. We also re-derived the results for the 3 published MSPs, in order to ensure consistent and systematic astrometric studies. 

\begin{table*}
\raggedright
\caption{List of pulsars and phase calibrators}
\label{tab:MSPs}
\resizebox{\textwidth}{!}{
\begin{tabular}{lcccccllllr}
\hline
\hline
PSR & $P_\mathrm{s}$ & $P_\mathrm{b}$ & Gating & Project Codes & Primary phase calibrator & $\Delta_\mathrm{PC-IBC}$ $\,^{f}$& Secondary phase calibrator & IBC $\,^{a}$ & $\Delta_\mathrm{psr-IBC}$ $\,^{b}$ & $S_\mathrm{unres}^\mathrm{IBC}$ $\,^{*}$\\
 & (ms) & (d) & gain & & & (deg) & & code & (arcmin) & (mJy)\\
\hline
\Psrb\ & 4.87 & --- & 1.75 & BD179B, BD192B & ICRF~J002945.8$+$055440 & 0.97 & FIRST~J003054.6$+$045908 & 00027 & 10.1 & 20.6\\
\Psrc\ & 3.86 & 0.29 & 1.85 & BD179C, BD192C & --- & --- & NVSS~J061002$-$211538$\,^{c}$ & 00238 & 15.4 & 112.8 \\
\Psrd\ & 28.85 & 8.3 & 2.15 & BD179D, BD192D & ICRF~J061909.9$+$073641 & 2.84 & NVSS~J062153$+$102206 & 00303 & $21.0 \! \Rightarrow \! 2.9$ $\,^{d}$ & 18.1\\
\Psrea\ $^{**}$ & 5.26 & 0.60 & 1.43 & BD179E, BD192E & ICRF~J095837.8$+$503957  & 3.38 & NVSS~J101307$+$531233 & 00462 & 7.51 & 20.8\\
\Psreb\ & 5.16 & --- & 1.69 & BD179E, BD192E & ICRF~J102838.7$-$084438 & 1.59 & FIRST~J102526.3$-$072216 & 00529 & 12.2 & 11.7\\
\Psrfa\ & 40.93 & 8.6 & 2.27 & BD179F, BD192F & ICRF~J150644.1$+$493355 & 1.78 & NVSS~J151733$+$491626 & 00691 & $13.8 \! \Rightarrow \! 6.5$ $\,^g$ & 35.6 \\
\Psrfb\ $^{**}$ & 37.90 & 0.42 & 1.62 & BD179F, BD192F, BD229 & ICRF~J154049.4$+$144745 & 3.18 & FIRST~J153746.2$+$114215 & 00840 & 16.3 & 19.2\\
\Psrga\ $^{**}$ & 3.16 & 175 & 1.90 & BD179G, BD192F & ICRF~J164125.2$+$225704 & 0.79 & NVSS~J164018$+$221203 & 00920 & 12.1 & 98.0 \\
\Psrgb\ & 4.62 & 147 & 1.56 & BD179G, BD192G & ICRF~J163845.2$-$141550 & 2.49 & NVSS~J164515$-$122013   & 01120 & 24.3 & 6.0\\
\Psrha\ & 3.50 & --- & 1.78 & BD179H, BD192H & ICRF~J172658.9$-$225801 & 2.47 & NVSS~J172129$-$250538 & 01188 & 10.1 & 7.4\\
\Psrhb\ & 8.12 & --- & 1.86 & BD179H, BD192H & ICRF~J172658.9$-$225801 & 0.76 & NVSS~J172932$-$232722 & 01220 & 25.5 & 37.9 \\
\Psri\ & 5.85 & 0.35 & 1.98 & BD179I, BD192I & ICRF~J174037.1$+$031147 & 0.66 & NVSS~J173823$+$033305 & 01313 & 7.5 & 6.5 \\
\Psro\ & 3.05 & --- & 1.53 & BD179O, BD192O & ICRF~J182057.8$-$252812 & 0.61 & NVSS~J182301$-$250438 & 01433 & $23.8 \! \Rightarrow \! 3.7$ $\,^{d}$ & 3.3 \\
\Psrka\ & 4.09 & 116 & 1.64 & BD179K, BD192K & ICRF~J185250.5+142639 & 1.51 & NVSS~J185456$+$130110 & 01535 & 14.6 & 5.9 \\
\Psrl\ & 4.98 & 58.5 & 2.65 & BD179L, BD192L & ICRF~J191158.2$+$161146 & 3.16 & NVSS~J190957$+$130434 & 01769 & 8.7 & 4.9\\
\Psrma\ & 3.63 & 2.7 & 1.67 & BD179M, BD192M & ICRF~J190528.5$-$115332 & 1.76 & NVSS~J191233$-$113327 & 01816 & 21.9 & 23.4\\
\Psrmb\ & 7.65 & 10.9 & 2.12 & BD179M, BD192M & ICRF~J191207.1$-$080421 & 2.14 & NVSS~J191731$-$062435 & 01846 & 26.1 & 50.7\\
\Psrkb\ & 1.56 & --- & 1.35 & BD179K, BD192K & ICRF~J193510.4$+$203154  & 1.88 & NVSS~J194104$+$214913$\,^{e}$ & 01647 & $24.5 \! \Rightarrow \! 4.1$ & 10.6\\
 & & & & &   & 1.94 & NVSS~J194106$+$215304$\,^{e}$ & 01648 & $27.4 \! \Rightarrow \! 2.7$ & 1.8\\

\hline
\multicolumn{11}{l}{$\bullet$ The image models for the primary and secondary calibrators listed here are publicly available\textsuperscript{\ref{footnote:calibrator_models}}.}\\
\multicolumn{11}{l}{$\bullet$ $P_\mathrm{s}$ and $P_\mathrm{b}$ stand for spin period and orbital period, respectively.}\\
\multicolumn{11}{l}{$^*$ Unresolved flux intensity of the secondary phase calibrator at 1.55\,GHz.}\\
\multicolumn{11}{l}{$^{**}$ Published in \citet{Ding20,Ding21a,Vigeland18}.}\\
\multicolumn{11}{l}{$^a$ Secondary phase calibrators are named IBC\textit{XXXXX} in the BD179 and BD192 observing files, where ``\textit{XXXXX}'' represents a unique 5-digit IBC code.}\\
\multicolumn{11}{l}{$^b$ Angular separation between target and secondary calibrator.}\\
\multicolumn{11}{l}{$^c$ NVSS~J061002$-$211538, close to the pulsar on the sky, is bright enough to serve as primary phase calibrator.}\\
\multicolumn{11}{l}{$^d$ As 1D interpolation is applied, the pulsar-to-virtual-calibrator separation is also provided after ``$\Rightarrow$'' (see Section~\ref{subsec:dualphscal}).}\\
\multicolumn{11}{l}{$^e$ Here, inverse phase referencing is adopted, where the ``secondary phase calibrators'' are ultimately the targets (see Section~\ref{subsec:sophisticated_data_reduction}).}\\
\multicolumn{11}{l}{$^f$ Angular separation between primary and secondary calibrator.}\\
\multicolumn{11}{l}{$^g$ The NVSS~J151815$+$491105, a 4.5-mJy-bright source 6\farcm5 away from the pulsar, is used as the final reference source (see Section~\ref{sec:data_reduction}).}\\

\end{tabular}
}
\end{table*}

Along with the release of the catalogue results, this paper covers several scientific and technical perspectives. 
First, this paper explores novel data reduction strategies such as inverse-referenced 1D phase interpolation (see Section~\ref{subsec:sophisticated_data_reduction}). Second, a new Bayesian astrometry inference package is presented (see Section~\ref{sec:parameter_inference}).
Third, with new parallax-based distances and proper motions, we discriminate between the two prevailing $n_\mathrm{e}(\vec{x})$ models (see Section~\ref{subsubsec:DM_distances}), and investigate the kinematics of MSPs in Section~\ref{subsec:v_t}. 
Fourth, with new parallax-based distances of two MSPs, we re-visit the constraints on alternative theories of gravity (see Section~\ref{sec:orbital_decay_tests}).
Finally, discussions on individual pulsars are given in Section~\ref{sec:individual_pulsars}, which includes a refined ``death line'' upper limit of $\gamma$-ray pulsars (see Section~\ref{subsec:J1730}).
The study of SSE-dependent frame rotation, which depends on an accurate estimation of the reference points of our calibrator sources in the quasi-inertial VLBI frame, requires additional multi-frequency observations and will be presented in a follow-up paper.

Throughout this paper, we abide by the following norms unless otherwise stated. {\bfseries 1)} The uncertainties are provided at 68\% confidence level. {\bfseries 2)} Any mention of flux density refers to unresolved flux density $S_\mathrm{unres}$ in our observing configuration (e.g., a 10-mJy source means $S_\mathrm{unres}=10$\,mJy). {\bfseries 3)}
All bootstrap and Bayesian results adopt the 50th, 16th and 84th percentile of the marginalized (and sorted) value chain as, respectively, the estimate and its 1-$\sigma$ error lower and upper bound.
{\bfseries 4)} Where an error of an estimate is required for a specific calculation but an asymmetric error is reported for the estimate, the mean of upper and lower errors is adopted for the calculation.
{\bfseries 5)} VLBI positional uncertainties will be broken down into the uncertainty of the offset from a chosen calibrator reference point, and the uncertainty in the location of that chosen reference point. This paper focuses on the relative offsets which are relevant for the measurement of proper motion and parallax, and the uncertainty in the location of the reference source is presented separately.



\section{Observations and Correlation}
\label{sec:observations}

As is mentioned in Section~\ref{subsec:MSP_VLBI_astrometry}, to achieve high-precision pulsar astrometry requires the implementation of a VLBI phase referencing technique. There are, however, a variety of such techniques, including the normal phase referencing, relayed phase referencing, inverse phase referencing and interpolation. These techniques are described and discussed in Chapter~2 of \citealp{Ding22a}.
Generally, a given phase referencing approach and hence observational setup maps directly to a corresponding data reduction procedure, though occasionally other data reduction opportunities could arise by chance (see Section~\ref{sec:data_reduction}).

The \mspsrpi\ project systematically employs the relayed phase referencing technique, in which a secondary phase reference source (explained in Chapter 2 of \citealp{Ding22a}) very close to the target on the sky is observed to refine direction-dependent calibration effects.
The observing and correlation tactics are identical to those of the \psrpi\ project \citep{Deller19}.  
All MSPs in the \mspsrpi\ catalogue
(see Table\ref{tab:MSPs}) were observed at L band with the VLBA at 2\,Gbps data rate (256 MHz total bandwidth, dual polarisation) from mid-2015 to no later than early 2018.
To minimise radio-frequency interference (RFI) at L band, we used eight 32\,MHz subbands with central frequencies of 1.41, 1.44, 1.47, 1.50, 1.60, 1.66, 1.70 and 1.73\,GHz, corresponding to an effective central frequency of 1.55\,GHz.
The primary phase calibrators were selected from the Radio Fundamental Catalogue\footnote{\url{astrogeo.org/rfc/}}.
The secondary phase calibrators were identified from 
the FIRST (Faint Images of the Radio Sky at Twenty-cm) catalogue \citep{Becker95} or the NVSS (NRAO VLA sky survey) catalogue \citep{Condon98} (for sky regions not covered by the FIRST survey) using a short multi-field observation.
Normally, more than one secondary phase calibrators were observed together with the target. Among them, a main one that is preferably the brightest and the closest to the target is selected to carry out self-calibration; the other secondary phase calibrators are hereafter referred to as redundant secondary phase calibrators. 
The primary and the main secondary phase calibrators for the astrometry of the 18 MSPs are summarized in Table~\ref{tab:MSPs}, alongside the project codes. 
At correlation time, pulsar gating was applied \citep{Deller11a} to improve the S/N on the target pulsars. The median values of the gating gain, defined as ${(S/N)_\mathrm{gated}}/{(S/N)_\mathrm{ungated}}$, are provided in Table~\ref{tab:MSPs}.

\section{Data Reduction and fiducial systematic errors}
\label{sec:data_reduction}

We reduced all data with the {\tt psrvlbireduce} pipeline\footnote{\label{footnote:parseltongue}available at \url{https://github.com/dingswin/psrvlbireduce}} written in {\tt parseltongue} \citep{Kettenis06}, a {\tt python}-based interface for running functions provided by {\tt AIPS} \citep{Greisen03} and {\tt DIFMAP} \citep{Shepherd94}.
The procedure of data reduction is identical to that outlined in \citet{Ding20}, except for four MSPs --- \psrfa, \psrd, \psro\ and \psrkb.
For \psrfa, the self-calibration solutions acquired with NVSS~J151733$+$491626, a 36-mJy secondary calibrator 13\farcm8 away from the pulsar, are extrapolated to both the pulsar and NVSS~J151815$+$491105 --- a 4.5-mJy source about a factor of two closer to \psrfa\ than NVSS~J151733$+$491626. The  positions relative to NVSS~J151815$+$491105 are used to derive the astrometric parameters of \psrfa.
For the other exceptions, the data reduction procedures as well as fiducial systematics estimation are described in Sections~\ref{subsec:dualphscal} and \ref{subsec:sophisticated_data_reduction}.

At the end of the data reduction, a series of positions as well as their random errors $\sigma_i^\mathcal{R}$ (where $i\!=\!1, 2, 3,...$ refers to right ascension or declination at different epochs) are acquired for each pulsar. 
For each observation, on top of the random errors due to image noise, ionospheric fluctuations would introduce systematic errors that distort and translate the source, the magnitude of which generally increases with the angular separation between a target and its (secondary) phase calibrator \citep[e.g.][]{Chatterjee04,Pradel06,Kirsten15,Deller19}.
We estimate fiducial values for these systematic errors $\sigma_i^\mathcal{S}$ of pulsar positions using the empirical relation (i.e., Equation~1 of \citealp{Deller19}) derived from the whole \psrpi\ sample.
While this empirical relation has proven a reasonable approximation to the actual systematic errors for a large sample of sources, for an individual observational setup $\sigma_i^\mathcal{S}$ may overstate or underestimate the true systematic error (see Section~\ref{sec:parameter_inference}). 
We can account for our uncertainty in this empirical estimator by re-formulating the total positional uncertainty as
\begin{equation}
\label{eq:EFAC}
\begin{split}
    \sigma_{i}\left(\eta_\mathrm{EFAC}\right)=\sqrt{(\sigma_i^\mathcal{R})^2+(\eta_\mathrm{EFAC} \cdot \sigma_i^\mathcal{S})^2} \,,
\end{split}
\end{equation}
where $\eta_\mathrm{EFAC}$ is a positive correction factor on the fiducial systematic errors. In this work, we assume $\eta_\mathrm{EFAC}$ stays the same for each pulsar throughout its astrometric campaign.
The inference of $\eta_\mathrm{EFAC}$ is described in Section~\ref{sec:parameter_inference}. 
We reiterate that the target image frames have been determined by the positions assumed for our reference sources (or virtual calibrators, see Section~\ref{subsec:dualphscal}), and that any change in the assumed reference source position would transfer directly into a change in the recovered position for the target pulsar.  Accordingly, the uncertainty in the reference source position must be accounted for in the pulsar's reference position error budget, after fitting the pulsar's astrometric parameters.

All pulsar positions and their error budgets are provided in the online\footnote{\label{footnote:pulsar_positions}\url{https://github.com/dingswin/publication_related_materials}} ``pmpar.in.preliminary'' and ``pmpar.in'' files. 
The only difference between ``pmpar.in.preliminary'' and ``pmpar.in'' (for each pulsar) files are the position uncertainties: ``pmpar.in.preliminary'' and ``pmpar.in'' offer, respectively, position uncertainties $\sigma_i(0)=\sigma_i^\mathcal{R}$ and $\sigma_i(1)=\sqrt{(\sigma_i^\mathcal{R})^2+(\sigma_i^\mathcal{S})^2}$. 
As an example, the pulsar positions for \psri\ are presented in Table~\ref{tab:pulsar_positions}, where the values on the left and right side of the `` | '' sign stand for, respectively, $\sigma_i(0)$ and $\sigma_i(1)$. 
Additionally, to facilitate reproducibility, the image models for all primary and secondary phase calibrators listed in Table~\ref{tab:MSPs} are released\footnote{\label{footnote:calibrator_models}\url{https://github.com/dingswin/calibrator_models_for_astrometry}} along with this paper. 
Following \citet{Deller19,Ding20}, the calibrator models were made with the calibrator data concatenated from all epochs in an iterative manner.

\begin{table}
    \raggedright
    \caption{An example set of astrometric results for J1738+0333, where the presented uncertainty excludes the calibrator reference point uncertainty as described in the text.}
     
        \resizebox{\columnwidth}{!}{
    	\begin{tabular}{ccc} 
		\hline
		\hline
	obs. date & $\alpha_\mathrm{J2000}$ (RA.)  & $\delta_\mathrm{J2000}$ (Decl.)   \\
	(yr) & &  \\
		\hline
	2015.6166 & $17^{\rm h}38^{\rm m}53\fs 969242(3|5)$ & $03\degr33'10\farcs90430(9|17)$ \\
	2015.8106 & $17^{\rm h}38^{\rm m}53\fs 969329(3|6)$ & $03\degr33'10\farcs90491(9|18)$ \\
	2016.6939 & $17^{\rm h}38^{\rm m}53\fs 969726(5|6)$ & $03\degr33'10\farcs90981(16|21)$ \\
	2017.1304 & $17^{\rm h}38^{\rm m}53\fs 970000(6|7)$ & $03\degr33'10\farcs91262(21|25)$ \\
	2017.2068 & $17^{\rm h}38^{\rm m}53\fs 970040(2|4)$ & $03\degr33'10\farcs91217(7|15)$ \\
	2017.2860 & $17^{\rm h}38^{\rm m}53\fs 970078(3|5)$ & $03\degr33'10\farcs91307(11|17)$ \\
	2017.2997 & $17^{\rm h}38^{\rm m}53\fs 970062(17|17)$ & $03\degr33'10\farcs91272(59|61)$ \\
	2017.7232 & $17^{\rm h}38^{\rm m}53\fs 970208(15|16)$ & $03\degr33'10\farcs91484(64|74)$ \\
	2017.7669 & $17^{\rm h}38^{\rm m}53\fs 970248(7|8)$ & $03\degr33'10\farcs91466(27|33)$ \\
	\hline 
	\multicolumn{3}{l}{$\bullet$ This table is compiled for \psri.}\\
	\multicolumn{3}{l}{$\bullet$ The values on the left and the right side of `` | '' are, respectively,}\\ 
	\multicolumn{3}{l}{\ \ \ statistical errors given in J1738+0333.pmpar.in.preliminary\textsuperscript{\ref{footnote:pulsar_positions}}, and}\\ 
	\multicolumn{3}{l}{\ \ \ systematics-included errors provided in J1738+0333.pmpar.in\textsuperscript{\ref{footnote:pulsar_positions}}.}
	\end{tabular}
	}
    \label{tab:pulsar_positions}
    
\end{table}

\subsection{1D interpolation on \psrd\ and \psro}
\label{subsec:dualphscal}
One can substantially reduce propagation-related systematic errors using 1D interpolation with two phase calibrators quasi-colinear with a target \citep[e.g.][]{Fomalont03,Ding20c}. After 1D interpolation is applied, the target should in effect be referenced to a ``virtual calibrator'' much closer (on the sky) than either of the two physical phase calibrators, assuming the phase screen can be approximated by a linear gradient with sky position \citep{Ding20c}.

According to Table~\ref{tab:MSPs}, 7 secondary phase calibrators (or the final reference sources) are more than 20' away from their targets, 
which would generally lead to relatively large systematic errors \citep[e.g.][]{Chatterjee04,Kirsten15,Deller19}.
Fortunately, there are 3 MSPs --- \psrd, \psro\ and \psrkb, for which the pulsar and its primary and secondary phase calibrators are near-colinear (see online\textsuperscript{\ref{footnote:pulsar_positions}} calibrator plans as well as Figure~\ref{fig:J1939_calibration_plan}). 
Hence, by applying 1D interpolation, each of the 3 ``1D-interpolation-capable'' MSPs can be referenced to a virtual calibrator much closer than the physical secondary phase calibrator (see Table~\ref{tab:MSPs}). 

We implemented 1D interpolation on \psrd\ and \psro\ in the same way as the astrometry of the radio magnetar XTE~J1810$-$197 carried out at 5.7\,GHz \citep{Ding20c}.
Nonetheless, due to our different observing frequency (i.e., 1.55\,GHz), we estimated $\sigma^\mathcal{S}_i$ differently.
The post-1D-interpolation systematic errors should consist of {\bf 1)} first-order residual systematic errors related to the target-to-virtual-calibrator offset $\Delta_\mathrm{psr-VC}$ and {\bf 2)} higher-order terms.
Assuming negligible higher-order terms, we approached post-1D-interpolation $\sigma^\mathcal{S}_i$ with Equation~1 of \citet{Deller19}, using $\Delta_\mathrm{psr-VC}$ as the calibrator-to-target separation. 
The assumption of negligible higher-order terms will be tested later and discussed in Section~\ref{subsubsec:implications_for_1D_interpolation}.

\subsection{Inverse-referenced 1D interpolation on \psrkb}
\label{subsec:sophisticated_data_reduction}
For \psrkb, normal 1D interpolation \citep{Fomalont03,Ding20c}, with respect to the primary phase calibrator ICRF~J193510.4$+$203154 (J1935) and the brightest secondary reference source NVSS~J194104$+$214913 (J194104), is still not the optimal calibration strategy. The $\approx$10-mJy (at 1.55\,GHz) \psrkb\ is the brightest MSP in the northern hemisphere and only second to PSR~J0437$-$4715 in the whole sky.
After pulsar gating, \psrkb\ is actually brighter than J194104.
\psrkb\ is unresolved on VLBI scales, and does not show
long-term radio feature variations (frequently seen in quasars), which makes it an ideal secondary phase calibrator.
Both factors encouraged us to implement the inverse-referenced 1D interpolation (or simply inverse 1D interpolation) on \psrkb, where \psrkb\ is the de-facto secondary phase calibrator and the two ``secondary phase calibrators'' serve as the targets.
To avoid confusion, we refer to the two ``secondary phase calibrators'' for \psrkb\ (see Table~\ref{tab:MSPs}) as secondary reference sources or simply reference sources.

Though inverse phase referencing (without interpolation) has been an observing/calibration strategy broadly used in VLBI astrometry \citep[e.g.][]{Imai12,Yang16,Li18,Deller19}, inverse  interpolation is new, with the 2D approach of \citet{Hyland22} at 8.3 GHz being a recent and independent development.
We implemented inverse 1D interpolation at 1.55\,GHz on \psrkb\ in three steps (in addition to the standard procedure) detailed as follows.

\subsubsection{Tying \psrkb\ to the primary-calibrator reference frame}
\label{subsubsec:tying_J1939_to_J1935}

Inverse 1D interpolation relies on the residual phase solutions $\Delta\phi_{n}(\vec{x},t)$ of self-calibration on \psrkb\ (where $\vec{x}$, $t$ and $n$ refers to, respectively, sky position, time and the $n$-th station in a VLBI array), which, however, change with $\Delta\vec{x}_\mathrm{psr}$ --- the displacement from the ``true'' pulsar position 
to its model position. 
When $|\Delta\vec{x}_\mathrm{psr}|$ is much smaller than the synthesized beam size $\theta_\mathrm{syn}$, the changes in $\Delta\phi_{n}(\vec{x},t)$ would be equal across all epochs, hence not biasing the resultant parallax and proper motion. 
However, if $|\Delta\vec{x}_\mathrm{psr}| \gtrsim \theta_\mathrm{syn}$, then the phase wraps of $\Delta\phi_{n}(\vec{x},t)$ would likely become hard to uncover.
The main contributor of considerable $|\Delta\vec{x}_\mathrm{psr}|$ is an inaccurate pulsar position. The proper motion of the pulsar would also increase $|\Delta\vec{x}_\mathrm{psr}|$ with time, if it is poorly constrained (or neglected). For \psrkb, the effect of proper motion across our observing duration is small ($\lesssim1$\,mas across the nominal observing span of 2.5 years; see the timing proper motion in Section~\ref{sec:inference_with_priors}) compared to $\theta_\mathrm{syn}\sim10$\,mas.

In order to minimize $|\Delta\vec{x}_\mathrm{psr}|$, we shifted the pulsar model position, on an epoch-to-epoch basis, by $\Delta\vec{x}_\mathrm{cor}$ (which ideally should approximate $-\Delta\vec{x}_\mathrm{psr}$), to the position measured in the J1935 reference frame (see Section~4.1 of \citealp{Ding20c} for explanation of ``reference frame'').
This J1935-frame position was derived with the method for determining pulsar absolute position \citep{Ding20} (where J194104 was used temporarily as the secondary phase calibrator) except that there is no need to quantify the position uncertainty.
We typically found $|\Delta\vec{x}_\mathrm{psr}|\sim50$\,mas, which is well above $\theta_\mathrm{syn}\sim10$\,mas.
After the map centre shift, \psrkb\ becomes tied to the J1935 frame.

\subsubsection{1D interpolation on the tied \psrkb}
\label{subsubsec:1D_inter_on_tied_J1939}
The second step of inverse 1D interpolation is simply the normal 1D interpolation on \psrkb\ that has been tied to the J1935 frame as described above (in Section~\ref{subsubsec:tying_J1939_to_J1935}).
When there is only one secondary reference source, optimal 1D interpolation should see the virtual calibrator moved along the interpolation line (that passes through both J1935 and \psrkb) to the closest position to the secondary reference source \citep[e.g.][]{Ding20c}. 
However, there are two reference sources for \psrkb\ (see Table~\ref{tab:MSPs}), 
and the virtual calibrator point can be set at a point that will enable both of them to be used.

\begin{figure*}
    \centering
	\includegraphics[width=18cm]{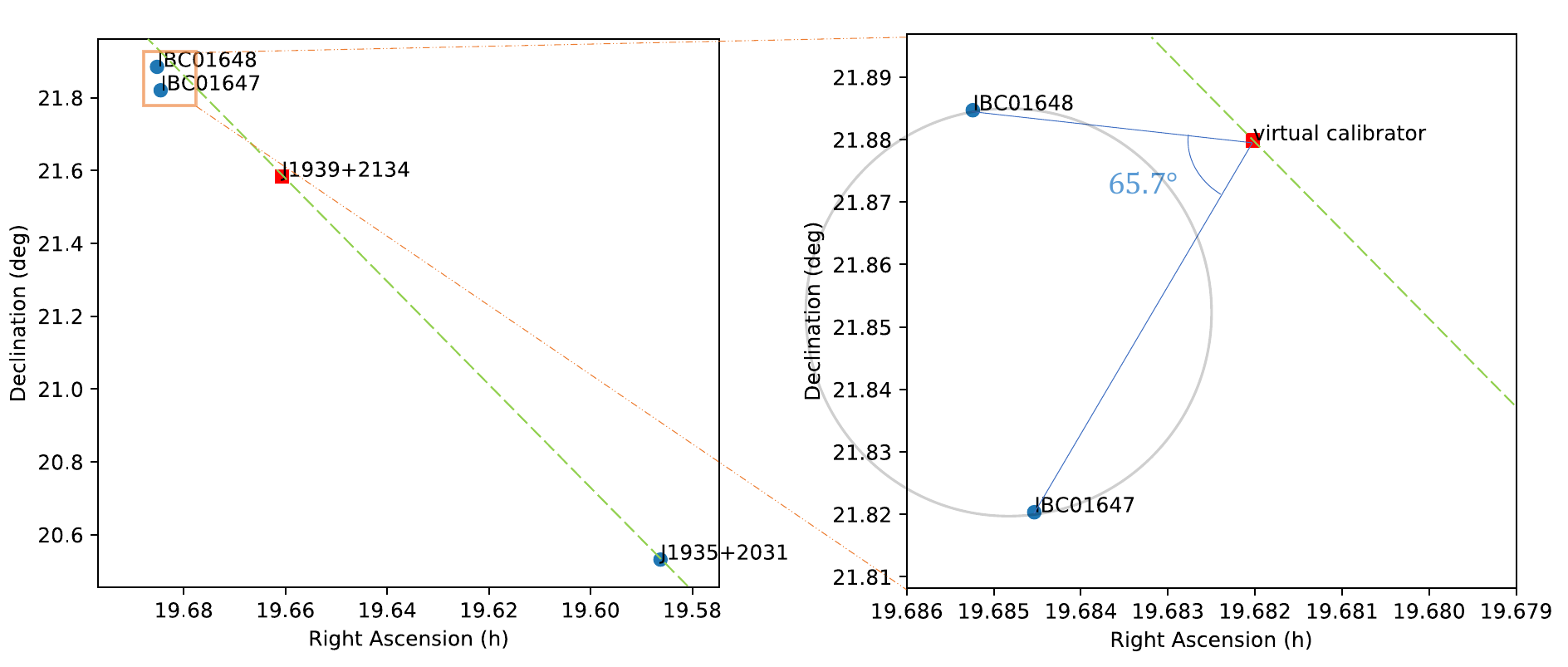}
    \caption{{\bf Left:} The calibrator plan for VLBI astrometry of \psrkb\ (see Table~\ref{tab:MSPs} for full source names), where \psrkb\ serves as the secondary phase calibrator and J1935 is the primary phase calibrator. {\bf Right:} The zoomed-in field for reference sources as well as the virtual calibrator (VC) along the J1935-to-pulsar line. For the inverse 1D interpolation on \psrkb, we used the VC location that forms the largest included angle ($65\fdg7$) with the two reference sources (see Section~\ref{subsec:sophisticated_data_reduction} for explanation), which corresponds to $\Delta_{VC-PC}/\Delta_{PC-psr}$=1.2836 (i.e., the VC-to-J1935 separation is 1.2836 times the J1935-to-pulsar separation). 
    }    
    \label{fig:J1939_calibration_plan}
\end{figure*}

After calibration, a separate position series can be produced for each reference source.
While we used each reference-source position series to infer astrometric parameters separately, we can also directly infer astrometric parameters with the combined knowledge of the two position series (which can be realized with $\tt sterne$\footnote{\label{footnote:sterne}\url{https://github.com/dingswin/sterne}}). 
If the errors in the two position series are (largely) uncorrelated, this can provide superior astrometric precision. 
Since position residuals should be spatially correlated, we would ideally set the virtual calibrator at a location such that the included angle between the two reference sources is $90\degr$.
While achieving this ideal is not possible, we chose a virtual calibrator location that forms the largest possible included angle (65\fdg7) with the two reference sources to minimise spatially correlated errors (see Figure~\ref{fig:J1939_calibration_plan}).
This virtual calibrator is 1.2836 times further away from J1935 than \psrkb. 
Accordingly, the $\Delta\phi_{n}(\vec{x},t)$ solutions (obtained from the self-calibration on the tied \psrkb) were multiplied by 1.2836, and applied to the two reference sources.

\subsubsection{De-shifting reference source positions}
After data reduction involving the two steps outlined in Sections~\ref{subsubsec:tying_J1939_to_J1935} and \ref{subsubsec:1D_inter_on_tied_J1939}, one position series was acquired for each reference source. At this point, however, the two position series are not yet ready for astrometric inference, mainly because both proper motion and parallax signatures have been removed in the first step (see Section~\ref{subsubsec:tying_J1939_to_J1935}) when \psrkb\ was shifted to its J1935-frame position. Therefore, the third step of inverse 1D interpolation is to cancel out the \psrkb\ shift (made in the first step) by moving reference source positions by $-1.2836 \cdot \Delta\vec{x}_\mathrm{cor}$, where the multiplication can be understood by considering Figure~1 of \citealp{Ding20c}.
This de-shifting operation was carried out separately outside the data reduction pipeline\textsuperscript{\ref{footnote:parseltongue}}. After the operation, we estimated $\sigma_{ij}^\mathcal{S}$ of the reference sources (where $j\!=\!1, 2$ refers to an individual reference source) following the method described in Section~\ref{subsec:dualphscal}.
The final position series of the reference sources are available online\textsuperscript{\ref{footnote:pulsar_positions}}. The astrometric parameter inference based on these position series is 
outlined in Section~\ref{sec:parameter_inference}. 

\section{Astrometric inference methods and quasi-VLBI-only astrometric results}
\label{sec:parameter_inference}

After gathering the position series\textsuperscript{\ref{footnote:pulsar_positions}} with basic uncertainty estimation (see Section~\ref{sec:data_reduction}), we proceed to infer the astrometric parameters. 
The inference is made by three different methods: {\bf a)} direct fitting of the position series with {\tt pmpar}\footnote{\label{footnote:pmpar}\url{https://github.com/walterfb/pmpar}}, {\bf b)} bootstrapping (see \citealp{Ding20}) and {\bf c)} Bayesian analysis using {\tt sterne}\textsuperscript{\ref{footnote:sterne}} (see \citealp{Ding21a}).
The two former methods directly adopt $\sigma_i(1)=\sqrt{(\sigma_i^\mathcal{R})^2+(\sigma_i^\mathcal{S})^2}$ as the position errors. 
In Bayesian analysis, however, we inferred $\eta_\mathrm{EFAC}$ along with other model parameters using the likelihood terms
\begin{equation}
\label{eq:probability}
\begin{split}
    P_1 \propto \left(\prod_{i} \sigma_i\right)^{-1} \exp{\left[-\frac{1}{2} \sum_i \left(\frac{\Delta \epsilon_i}{\sigma_i}\right)^2\right]} \,,
\end{split}
\end{equation}
where $\sigma_i=\sigma_i(\eta_\mathrm{EFAC})$ obeys Equation~\ref{eq:EFAC}; $\Delta \epsilon_i$ refers to the model offsets from the measured positions. 
As is discussed in Section~\ref{subsec:Bayesian_as_major}, Bayesian inference outperforms the other two methods, and is hence consistently used to present final results in this work.  In all cases, the uncertainty in the reference source position should be added in quadrature to the uncertainty in the pulsar's reference position acquired with any method (of the three), in order to obtain a final estimate of the absolute positional uncertainty of the pulsar.

To serve different scientific purposes, we present two sets of astrometric results in two sections (i.e., Sections~\ref{sec:parameter_inference} and \ref{sec:inference_with_priors}), which differ in whether timing proper motions and parallaxes are used as prior information in the inference. 


\subsubsection{Priors of canonical model parameters used in Bayesian analysis}
\label{subsubsec:parameter_priors}
To facilitate reproduction of our Bayesian results, the priors (of Bayesian inference) we use for canonical model parameters and $\eta_\mathrm{EFAC}$ are detailed as follows. Priors for the two orbital parameters can be found in Section~\ref{subsec:reflex_motion_inference}. We universally adopt the prior uniform distribution $\mathcal{U}$(0, 15) (i.e., uniformly distributed between 0 and 15) for $\eta_\mathrm{EFAC}$. This prior distribution can be refined for future work with an ensemble of results across many pulsars.
With regard to the canonical astrometric parameters (7 parameters for \psrkb\ and 5 for the other pulsars), we adopt $\mathcal{U}\left(X_0^\mathrm{(DF)}-20~\tilde{\sigma}_X^\mathrm{(DF)},~~ X_0^\mathrm{(DF)}+20~ \tilde{\sigma}_X^\mathrm{(DF)}\right)$ for each $X$, where $X$ refers to one of $\alpha_\mathrm{ref}$, $\delta_\mathrm{ref}$, $\mu_\alpha$, $\mu_\delta$ and $\varpi$. Here, $X_0^\mathrm{(DF)}$ stands for the direct-fitting estimate of $X$; $\tilde{\sigma}_X^\mathrm{(DF)}$ represents the
direct-fitting error corrected by the reduced chi-square $\chi^2_\nu$ (see Table~\ref{tab:astrometric_parameters}) with $\tilde{\sigma}_X^\mathrm{(DF)} \equiv \sigma_{X}^\mathrm{(DF)} \cdot \sqrt{\chi^2_\nu}$.
The calculation of prior range of $X$ is made with the function $\tt sterne.priors.generate\_initsfile$\textsuperscript{\ref{footnote:sterne}}.
We note that the adopted priors are  relaxed enough to ensure robust outcomes: shrinking or enlarging the prior ranges by a factor of two would not change the inferred values. Meanwhile, the specified prior ranges are also reasonably small so that the global minimum of Equation~\ref{eq:probability} can be reached.

\subsection{Astrometric inference disregarding orbital motion}
\label{subsec:non_reflex_motion_inference}

\subsubsection{Single-reference-source astrometric inferences}
All MSPs (in this work) excepting \psrkb\ have only one reference source. For each of these single-reference-source MSPs, we fit for the five canonical astrometric parameters, i.e., reference position ($\alpha_\mathrm{ref}$ and $\delta_\mathrm{ref}$), proper motion ($\mu_\alpha \equiv \dot{\alpha} \cos{\delta}$ and $\mu_\delta$) and parallax ($\varpi$). In the Bayesian analysis alone, $\eta_\mathrm{EFAC}$ is also inferred alongside the astrometric parameters.  At this stage, we neglect any orbital reflex motion for binary pulsars -- the effects of orbital reflex motion are addressed in Section~\ref{subsec:reflex_motion_inference}.
The proper motions and parallaxes derived with single-reference-source astrometry and disregarding orbital motion are summarized in Table~\ref{tab:astrometric_parameters}. 
The reference positions are presented in Section~\ref{subsec:astrometric_results_non_PM_priors}.



\begingroup
\renewcommand{\arraystretch}{1.5} 

\begin{table*}
\raggedright
\caption{Proper motion $\left( \mu_\alpha,\mu_\delta \right)$ and parallax $\varpi$ from astrometry inferences disregarding orbital motion.}
\label{tab:astrometric_parameters}
\resizebox{\textwidth}{!}{
\begin{tabular}{lcccllcccccccc}
\hline
\hline
PSR & $\mu_\alpha^\mathrm{(DF)}$ & $\mu_\delta^\mathrm{(DF)}$ & $\varpi^\mathrm{(DF)}$ & \rcs\ & $\mu_\alpha^\mathrm{(Bo)}$ & $\mu_\delta^\mathrm{(Bo)}$ & $\varpi^\mathrm{(Bo)}$ & $\mu_\alpha^\mathrm{(Ba)}$ & $\mu_\delta^\mathrm{(Ba)}$ & $\varpi^\mathrm{(Ba)}$ & $\eta_\mathrm{EFAC}$ & $\eta_\mathrm{orb}$ & $P_\mathrm{b}$\\
 & (\maspy) & (\maspy) & (mas) & & (\maspy) & (\maspy) & (mas) & (\maspy) & (\maspy) & (mas) & & & (d)\\

\hline
\multicolumn{14}{c}{Non-1D-interpolated results}\\
\hline

{\Psrb}  & -6.13(4) & 0.33(9) & 3.02(4) & 1.4 & -6.1(1) & $0.34^{+0.09}_{-0.08}$ & $3.02^{+0.09}_{-0.08}$ &   -6.13(7) & $0.34^{+0.15}_{-0.16}$ & 3.02(7) & $1.35^{+0.45}_{-0.32}$ & --- & ---\\
\Psrc & 9.11(8) & 15.9(2) & 0.74(7) & 1.0 & $9.10^{+0.06}_{-0.05}$ & 15.9(2) & $0.73^{+0.05}_{-0.04}$ & 9.1(1) &  $15.96^{+0.25}_{-0.24}$ &  0.73(10) & $1.1^{+0.4}_{-0.2}$ & $3\!\times\!10^{-3}$ & 0.29\\ 
\Psrd & 3.51(9) & -1.32(16) & 0.88(7) & 3.7 & $3.5^{+0.3}_{-0.4}$ & -1.3(2) & $0.9^{+0.2}_{-0.3}$ & 3.4(2) & -1.3(5) & 0.85(21) & $2.33^{+0.56}_{-0.44}$ & 0.4 & 8.3\\ 
\Psrea $\,^*$ &  2.68(3) & -25.38(6) & 1.17(2) & 1.9 & $2.67^{+0.13}_{-0.06}$ & -25.39(12) & $1.18^{+0.05}_{-0.06}$ & 2.67(5) & $-25.39^{+0.14}_{-0.15}$ & $1.17^{+0.04}_{-0.05}$ & $1.7^{+0.6}_{-0.4}$ & 0.1 & 0.60 \\

\Psreb & -35.32(4) & -48.2(1) & 0.94(3) & 1.1 & $-35.32^{+0.05}_{-0.04}$ & -48.2(1) & $0.94^{+0.07}_{-0.06}$ &  -35.32(7) & -48.1(2) & 0.94(6) & $1.2^{+0.4}_{-0.3}$ & --- & ---\\

{\Psrfa} & -0.69(2) & -8.54(4) & 1.24(2) & 1.2 & -0.69(3) & $-8.53^{+0.10}_{-0.07}$ & 1.25(3) & -0.69(3) & -8.52(7) & 1.24(3) & $1.2^{+0.4}_{-0.3}$ & 4.9 & 8.6 \\

\Psrfb $\,^*$ & 1.51(2) & -25.31(5) & 1.06(7) & 0.81 & 1.51(2) & $-25.31^{+0.04}_{-0.05}$ & $1.07^{+0.09}_{-0.08}$ & 1.51(3) & $-25.30^{+0.05}_{-0.06}$ & $1.06^{+0.11}_{-0.10}$ & $0.54^{+0.57}_{-0.38}$ & 0.24 & 0.42 \\

\Psrga $\,^*$ & 2.199(56) & -11.29(9) & 0.676(46) & 1.1 & $2.20^{+0.06}_{-0.07}$ & -$11.29^{+0.15}_{-0.17}$ & $0.68^{+0.07}_{-0.06}$ & 2.20(9) & $-11.29^{+0.16}_{-0.14}$ &  0.68(7) & $1.29^{+0.66}_{-0.54}$ & 3.1 & 175 \\

{\Psrgb} & 6.2(2) & 3.3(5) & 1.3(1) & 0.8  & $6.1^{+0.7}_{-0.1}$ & 3.3(4) & $1.3^{+0.2}_{-0.5}$ & 6.2(2) &  3.3(6) & 1.33(18) & $1.0^{+0.3}_{-0.2}$ & 1.1 & 147\\
\Psrha & 2.5(1) & -1.9(3) & 0.02(7) & 4.4  & 2.5(6) & $-1.7^{+0.4}_{-0.7}$ & $0.2^{+0.2}_{-0.3}$ & 2.5(3) &  -1.9(9) & 0.0(2) & $3.1^{+0.8}_{-0.6}$ & --- & ---\\
{\Psrhb} & 20.3(1) & -4.79(26) & 1.56(9) & 1.7 & $20.32^{+0.18}_{-0.15}$ & $-4.80^{+0.33}_{-0.35}$ & $1.56^{+0.15}_{-0.17}$ & 20.3(2) & -4.8(5) & $1.57(18)$ & $1.4^{+0.3}_{-0.2}$ & --- & ---\\

\Psri & 6.97(4) & 5.18(7) & 0.51(3) & 1.8 & $7.00^{+0.06}_{-0.11}$ & 5.2(1) & $0.50^{+0.07}_{-0.06}$ & 6.98(8) & 5.18(16) & 0.50(6) & $1.9^{+0.7}_{-0.6}$ & 0.02 & 0.35\\ 

\Psro & -0.03(49) & -6.6(1.2) & -0.25(37) & 0.8 & $-0.03^{+0.28}_{-0.79}$ & $-6.8^{+1.3}_{-1.5}$ & $-0.20^{+0.54}_{-0.39}$ & -0.1(6) & -6.6(1.5) & -0.22(48) & $0.9^{+0.4}_{-0.3}$ & ---& ---\\

\Psrka & -1.37(9) & -2.8(2) & 0.49(6) & 1.0  & $-1.39^{+0.13}_{-0.28}$ & -2.8(2) & $0.48^{+0.07}_{-0.13}$ &  -1.36(10) & -2.8(2) & 0.50(6) & $0.45^{+0.46}_{-0.30}$ & 1.4 & 116\\

\Psrl & 0.49(8) & -6.85(15) & 0.26(7) & 0.15 & $0.49^{+0.07}_{-0.08}$ & -6.85(6) & $0.26^{+0.06}_{-0.02}$ & 0.50(4) & -6.85(9) & 0.25(3) & $0.19^{+0.15}_{-0.12}$ & 0.87 & 58.5\\
\Psrma & -13.8(1) & -10.3(2) & 0.38(9) & 0.9 & $-13.76^{+0.06}_{-0.08}$ & $-10.3^{+1.1}_{-0.3}$ & 0.36(8) &  -13.8(2) & -10.3(4) & $0.38^{+0.13}_{-0.14}$ & $1.1^{+0.4}_{-0.3}$ & $3\!\times\!10^{-4}$ & 2.7\\
\Psrmb & -7.12(8) & -5.7(2) & 0.60(7) & 1.1 & -7.1(1) & $-5.8^{+0.3}_{-0.2}$ & $0.60^{+0.12}_{-0.08}$ & -7.1(1) & -5.7(3) & 0.60(12) & $1.2^{+0.3}_{-0.2}$ & 0.27 & 10.9\\ 
\Psrkb $^{\,{(i)}}$ & 0.07(14) & -0.24(24) & 0.35(10) & 0.2 &  $0.06^{+0.36}_{-0.45}$ & $-0.24^{+0.10}_{-0.06}$ & $0.34^{+0.08}_{-0.25}$ & 0.08(10) & $-0.23^{+0.17}_{-0.16}$ & 0.34(7) & $0.45^{+0.15}_{-0.11}$ & --- & ---\\

\hline
\multicolumn{14}{c}{Single-reference-source 1D-interpolated results}\\
\hline

\Psrd & 3.68(6) & -1.33(9) & 0.94(4) & 6.3 & $3.53^{+0.36}_{-0.35}$ & -1.3(2) & $0.9^{+0.2}_{-0.3}$ & 3.5(2) &  -1.37(35) & 0.86(15) & $3.7^{+1.0}_{-0.7}$ & 0.4 & 8.3\\
\Psro & 0.3(3) & -3.7(6) & 0.1(3) & 1.5 & $0.4^{+0.5}_{-1.2}$ & $-4^{+1}_{-2}$ & $0.1^{+0.7}_{-0.4}$ & 0.3(6) & $-3.9^{+1.2}_{-1.3}$ & 0.1(5) & $1.6^{+0.8}_{-0.7}$ & ---& ---\\
\Psrkb $^{\,{(ii)}}$ & 0.08(4) & -0.45(6) & 0.38(3) & 1.3 &  $0.08^{+0.34}_{-0.07}$ & $-0.44^{+0.09}_{-0.12}$ & $0.36^{+0.05}_{-0.22}$ & 0.08(7) & -0.44(11) & $0.380^{+0.048}_{-0.049}$ & $1.4^{+0.7}_{-0.5}$ & --- & ---\\
\Psrkb $^{\,{(iii)}}$ & 0.3(3) & -0.3(2) & 0.36(19) & 1.3 & $0.2^{+1.0}_{-0.2}$ & -0.3(4) & $0.36^{+0.35}_{-0.58}$ & 0.3(4) & -0.3(4) & $0.38^{+0.29}_{-0.28}$ & $3.7^{+3.4}_{-2.6}$ & --- & ---\\ 

\hline
\multicolumn{14}{c}{Multi-reference-source 1D-interpolated results}\\
\hline

\Psrkb $^{\,{(iv)}}$ & --- & ---- & --- & --- &  --- & --- & --- & 0.08(7) & -0.43(11) & $0.384^{+0.048}_{-0.046}$ & $1.5^{+0.7}_{-0.6}$ & --- & ---\\

\hline
\multicolumn{14}{l}{$\bullet$ ``DF'', ``Bo'' and ``Ba'' stands for, respectively, direct fitting, bootstrap and Bayesian inference. \rcs\ is the reduced chi-square of direct fitting using {\tt pmpar}\textsuperscript{\ref{footnote:pmpar}}.}\\
\multicolumn{14}{l}{$\bullet$ The middle and top block presents, respectively, 1D-interpolated (see Sections~\ref{subsec:dualphscal} and \ref{subsec:sophisticated_data_reduction}) and non-1D-interpolated results.}\\
\multicolumn{14}{l}{$\bullet$ The bottom entry for \psrkb\ shows the result of multi-reference-source astrometry inference (see Section~\ref{subsubsec:multi-source-inference}).}\\
\multicolumn{14}{l}{$\bullet$ $P_\mathrm{b}$ represents orbital period (see Table~4 for their references). $\eta_\mathrm{orb}$ is defined in Equation~\ref{eq:reflex_motion_measurability}.}\\
\multicolumn{14}{l}{$^*$ Already published in \citet{Vigeland18,Ding20,Ding21a}.}\\
\multicolumn{14}{l}{$^{(i)}$ Based on (non-1D-interpolated) J194104 positions inverse-referenced to \psrkb.}\\
\multicolumn{14}{l}{$^{(ii)}$ Using 1D-interpolated J194104 positions inverse-referenced to \psrkb.}\\
\multicolumn{14}{l}{$^{(iii)}$ Using 1D-interpolated J194106 positions inverse-referenced to \psrkb.}\\ 
\multicolumn{14}{l}{$^{(iv)}$ Based on 1D-interpolated J194104 and J194106 positions inverse-referenced to \psrkb\ (see Section~\ref{subsec:sophisticated_data_reduction}).}\\

\end{tabular}
}
\end{table*}
\endgroup

\subsubsection{Multi-source astrometry inferences}
\label{subsubsec:multi-source-inference}
When multiple sources share proper motion and/or parallax (while each source having its own reference position), a joint multi-source astrometry inference can increase the degrees of freedom of inference (i.e., the number of measurements reduced by the number of parameters to infer), and tighten constraints on the astrometric parameters.
Multi-source astrometry inference has been widely used in maser astrometry (where maser spots with different proper motions scatter around a region of high-mass star formation, \citealp{Reid09a}), but has not yet been used for any pulsar, despite the availability of several bright pulsars with multiple in-beam calibrators (e.g., PSR~J0332$+$5434, PSR~J1136$+$1551) in the \psrpi\ project \citep{Deller19}. 

\psrkb\ is the only source (in this work) that has multiple (i.e., two) reference sources, which provides a rare opportunity to test multi-reference-source astrometry.
We assumed that the position series of J194104 is uncorrelated with that of NVSS~J194106$+$215304 (hereafter J194106), and utilized {\tt sterne}\textsuperscript{\ref{footnote:sterne}} to infer the common parallax and proper motion, alongside two reference positions (one for each reference source). The acquired proper motion and parallax are listed in Table~\ref{tab:astrometric_parameters}. 
As inverse phase referencing is applied for \psrkb, the parallax and proper motion of \psrkb\ are the inverse of the direct astrometric measurements.
For comparison, the proper motion and parallax inferred solely with one reference source are also reported in Table~\ref{tab:astrometric_parameters}. 
Due to the relative faintness of J194106 (see Table~\ref{tab:MSPs}), the inclusion of J194106 only marginally improves the astrometric results (e.g., $\varpi$) over those inferred with J194104 alone. 

\begin{figure}
    \centering
	\includegraphics[width=9cm]{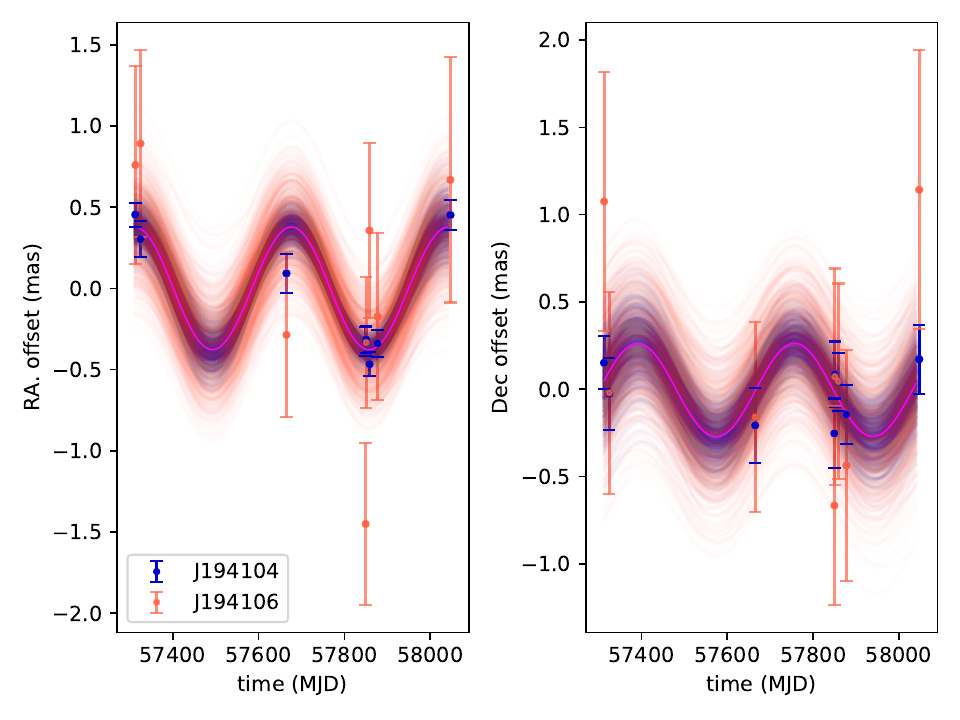}
    \caption{The common parallax signature of \psrkb\ revealed by the position measurements of both reference sources (see Table~\ref{tab:MSPs}). In both panels, the best-fit proper motion has been subtracted. The magenta curve in each panel represents the best-inferred astrometric model. The fuzzy region around the curve consists of various Bayesian simulations, the scatter of which can visualize the uncertainty level of the underlying model parameters (see Section~\ref{subsubsec:multi-source-inference}). As a result of the inverse referencing, the common parallax revealed here is actually the negative of the \psrkb\ parallax presented in Table~\ref{tab:astrometric_parameters}. 
    }    
    \label{fig:J1939_parallax}
\end{figure}

The constraints on the parallax (as well as the proper motion) are visualized in Figure~\ref{fig:J1939_parallax}. The best-inferred model (derived from the J194104 and J194106 positions) is illustrated with a bright magenta curve, amidst two sets of Bayesian simulations --- each set for a reference source.
Each simulated curve is a time series of simulated positions, with the best-inferred reference position ($\alpha_{\mathrm{ref},j}$ and $\delta_{\mathrm{ref},j}$, where $j$ refers to either J194104 or J194106) and proper-motion-related displacements (i.e., $\mu_\alpha \Delta t$ and $\mu_\delta \Delta t$, where $\Delta t$ is the time delay from the reference epoch) subtracted. As the simulated curve depends on the underlying model parameters, the degree of scatter of simulated curves would increase with larger uncertainties of model parameters. 
Though sharing simulated parallaxes and proper motions with J194104, the simulated curves for J194106 exhibits broader scatter (than the J194104 ones) owing to more uncertain reference position (see Section~\ref{subsec:astrometric_results_non_PM_priors} for $\alpha_{\mathrm{ref,J194106}}$ and $\delta_{\mathrm{ref,J194106}}$).
The large scatter implies that the J194106 position measurements impose  relatively limited constraints on the common model parameter (i.e., parallax and proper motion), which is consistent with the findings from Table~\ref{tab:astrometric_parameters}.

\subsubsection{Implications for 1D/2D interpolation}
\label{subsubsec:implications_for_1D_interpolation}
On the three 1D-interpolation-capable MSPs, we compared astrometric inference with both the 1D-interpolated and non-1D-interpolated position series (one at a time).
For \psrkb, the $\eta_\mathrm{EFAC}$ of the three 1D-interpolated realizations are consistent with each other, but larger than the non-1D-interpolated counterpart. This post-1D-interpolation inflation of $\eta_\mathrm{EFAC}$ also occurs to the other two 1D-interpolation-capable pulsars (see Table~\ref{tab:astrometric_parameters}), which suggests the post-1D-interpolation fiducial systematic errors $\sigma^\mathcal{S}_i$ might be systematically under-estimated. One obvious explanation for this under-estimation is that the higher-order terms of systematic errors are non-negligible (as opposed to the assumption we started with in Section~\ref{subsec:dualphscal}): they might be actually comparable to the first-order residual systematic errors (that are related to $\Delta_\mathrm{psr-VC}$) at the $\sim1.55$\,GHz observing frequencies.

On the other hand, the astrometric results based on the non-1D-interpolated J194104 positions inverse-referenced to \psrkb\ are less precise than the 1D-interpolated counterpart by $\approx40$\% , as is also the case for \psrd\ (see Table~\ref{tab:astrometric_parameters}). Moreover, the post-1D-interpolation parallax of \psro\ becomes relatively more accurate than the negative parallax obtained without applying 1D interpolation. All of these demonstrate the utility of 1D/2D interpolation, even in the scenario of in-beam astrometry that is already precise. 
In the remainder of this paper, we only focus on the 1D-interpolated astrometric results for the three 1D-interpolation-capable MSPs.

\subsection{Bayesian inference as the major method for \mspsrpi}
\label{subsec:Bayesian_as_major}
We now compare the three sets of astrometric parameters (in Table~\ref{tab:astrometric_parameters}) obtained with different inference methods, and seek to proceed with only one set in order to simplify the structure of this paper.
Among the three inference methods we use in this work, direct least-square fitting is the most time-efficient, but is also the least robust against improperly estimated positional uncertainties. Conversely, the other two methods (i.e., bootstrap and Bayesian methods) do not rely solely on the input positional uncertainties, and can still estimate the model parameters and their uncertainties $\sigma_X^{(Y)}$ ($X\!=\!\mu_{\alpha}, \mu_{\delta}$ or $\varpi$; $Y\!=$\,``Bo'' or ``Ba'') more robustly in the presence of incorrectly estimated positional errors. 

\begin{figure*}
    \centering
	\includegraphics[width=13cm]{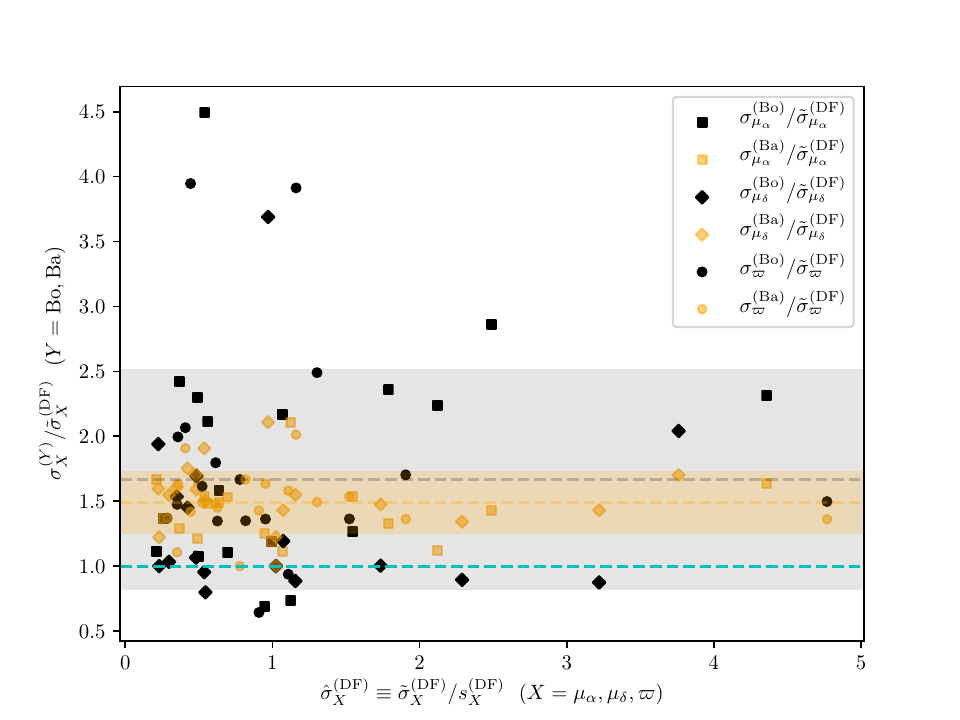}
    \caption{Bootstrap (denoted as ``Bo'') and Bayesian (``Ba'') errors (of three inferred parameters) divided by the corresponding \rcs-corrected direct-fitting errors. Here, $\tilde{\sigma}_X^{(DF)} \equiv \sigma_X^{(DF)} \cdot \sqrt{\chi^2_\nu}$ represents the $\chi^2_\nu$-corrected errors of direct fitting, where $X$ stands for one of the $\mu_\alpha$, $\mu_\delta$ and $\varpi$ groups. The dimensionless $\hat{\sigma}_X^{(DF)}$ is defined as an individual $\tilde{\sigma}_X^{(DF)}$ divided by the standard deviation $s_X^{(DF)}$ for all $\tilde{\sigma}_X^{(DF)}$ of the group $X$. The grey and orange shaded regions show, respectively, the standard deviation of $\sigma_X^\mathrm{(Bo)}/\tilde{\sigma}_X^\mathrm{(DF)}$ and $\sigma_X^\mathrm{(Ba)}/\tilde{\sigma}_X^\mathrm{(DF)}$ across all of the three groups (i.e., $\mu_\alpha$, $\mu_\delta$ and $\varpi$) around the respective mean value outlined with the grey and orange dashed lines.
    Both bootstrap and Bayesian errors are generally slightly higher than the level of direct-fitting errors illustrated with the cyan dashed line, and are well consistent with each other as anticipated. Despite the consistency, bootstrap errors show larger scatter than Bayesian ones. 
    }    
    \label{fig:Bo_errs_vs_Ba_errs}
\end{figure*}

Generally speaking, $\sigma_X^{(Y)}$ inferred from a pulsar position series are expected to change with the corresponding \rcs-corrected direct-fitting error $\tilde{\sigma}_X^\mathrm{(DF)} \equiv \sigma_{X}^\mathrm{(DF)} \cdot \sqrt{\chi^2_\nu}$.
In order to investigate the relation between $\sigma_X^{(Y)}$ and $\tilde{\sigma}_{X}^\mathrm{(DF)}$, we divided $\sigma_X^{(Y)}$ by $\tilde{\sigma}_{X}^\mathrm{(DF)}$ for each pulsar entry in the top block of Table~\ref{tab:astrometric_parameters}. 
The results are displayed in Figure~\ref{fig:Bo_errs_vs_Ba_errs}. 
For the convenience of illustration, we calculated the dimensionless $\tilde{\sigma}_{X}^\mathrm{(DF)}$ defined as $\tilde{\sigma}_{X}^\mathrm{(DF)}/s_X^{(DF)}$ (where $s_X^{(DF)}$ represents  the standard deviation of $\tilde{\sigma}_{X}^\mathrm{(DF)}$ over the group $X$), which allows all the three sets (i.e., $\mu_{\alpha}$, $\mu_{\delta}$ and $\varpi$) of dimensionless $\tilde{\sigma}_{X}^\mathrm{(DF)}$ to be horizontally more evenly plotted in Figure~\ref{fig:Bo_errs_vs_Ba_errs}. 

Across the entire \mspsrpi\ sample, we see that $\sigma_X^{(Y)}$ scales with $\tilde{\sigma}_X^\mathrm{(DF)}$ in a near-linear fashion.
The mean scaling factors across all of the three parameter groups (i.e., $\mu_{\alpha}$, $\mu_{\delta}$ and $\varpi$) are $\left<\sigma^\mathrm{(Bo)}_X / \tilde{\sigma}_X^\mathrm{(DF)}\right>=1.67\pm 0.85$ and $\left<\sigma^\mathrm{(Ba)}_X / \tilde{\sigma}_X^\mathrm{(DF)}\right>=1.49\pm0.24$ (see Figure~\ref{fig:Bo_errs_vs_Ba_errs}). The two mean scaling factors show that parameter uncertainties inferred using either a bootstrap or Bayesian approach will be slightly higher (and on average, consistent between the two approaches) than would be obtained utilising direct-fitting (illustrated with the cyan dashed line in Figure~\ref{fig:Bo_errs_vs_Ba_errs}).

The more optimistic uncertainty predictions of $\tilde{\sigma}_X^\mathrm{(DF)}$ can be understood as resulting from two causes: first, it neglects both the finite width and the skewness of the $\chi^2$ distribution, and second, to achieve the expected $\chi^2$ it scales the {\em total} uncertainty contribution at each epoch, rather than the systematic uncertainty contribution alone.  When (as is typical for pulsar observations) the S/N and hence statistical positional precision can vary substantially between observing epochs, this simplified approach preserves the relative weighting between epochs, whereas increasing the estimated systematic uncertainty contribution acts to equalise the weighting between epochs (by reducing the position precision more for epochs where the pulsar was bright and the statistical precision high, than for epochs where the pulsar was faint and the statistical precision is already low).  

While the consistency between $\left<\sigma^\mathrm{(Bo)}_X / \tilde{\sigma}_X^\mathrm{(DF)}\right>$ and $\left<\sigma^\mathrm{(Ba)}_X / \tilde{\sigma}_X^\mathrm{(DF)}\right>$ suggests that both approaches can overcome this shortcoming in the direct fitting method, $\sigma^\mathrm{(Bo)}_X / \tilde{\sigma}_X^\mathrm{(DF)}$ shows a much larger scatter (3.5 times) compared to $\sigma^\mathrm{(Ba)}_X / \tilde{\sigma}_X^\mathrm{(DF)}$ (see Figure~\ref{fig:Bo_errs_vs_Ba_errs}). To determine which approach best represents the true (and unknown) parameter uncertainties, it is instructive to consider the outliers in the bootstrap distribution results.  

First, consider cases where the bootstrap results in a lower uncertainty than $\tilde{\sigma}_X^\mathrm{(DF)}$.  
For the reasons noted above, we expect $\tilde{\sigma}_X^\mathrm{(DF)}$ to yield the most optimistic final parameter uncertainty estimates, and yet the bootstrap returns a lower uncertainty than $\tilde{\sigma}_X^\mathrm{(DF)}$ in a number of cases.
Second, the cases with the highest values of $\sigma^\mathrm{(Bo)}_X / \tilde{\sigma}_X^\mathrm{(DF)}$ reach $\gtrsim$3 on a number of occasions, which imply an extremely large (or very non-Gaussian) systematic uncertainty contribution, which would lead (in those cases) to a surprisingly low reduced $\chi^2$ for the best-fitting model.
Given the frequency with which these outliers arise, we regard it likely that bootstrap approach mis-estimates parameter uncertainties at least occasionally, likely due to the small number of observations available.
Therefore, we consider the Bayesian method described in this paper as the preferred inference method for the \mspsrpi\ sample, and consistently use the Bayesian results in the following discussions.
We note that as continued VLBI observing campaigns add more results, the systematic uncertainty estimation scheme applied to Bayesian inference can be further refined in the future.

\subsection{Astrometric inference accounting for orbital motion}
\label{subsec:reflex_motion_inference}
For some binary pulsars, VLBI astrometry can also refine parameters related to the binary orbit, on top of the canonical astrometric parameters. The orbital inclination $i$ and the orbital ascending node longitude $\Omega_\mathrm{asc}$ have been previously constrained for a few nearby pulsars, such as PSR~J1022$+$1001, PSR~J2145$-$0750 and PSR~J2222$-$0137 \citep{Deller13,Deller16,Guo21}.
To assess the feasibility of detecting orbital reflex motion with VLBI, we computed
\begin{equation}
\label{eq:reflex_motion_measurability}
    \eta_\mathrm{orb}  \equiv \frac{2a_1}{1\,\mathrm{AU}} \cdot \frac{\varpi}{\sigma_{\varpi}} 
    = 2a_1 \cdot \left( \frac{1\,\mathrm{AU}}{\varpi}\right)^{-1} \cdot \frac{1}{\sigma_{\varpi}} = \frac{2a_1}{D} \cdot \frac{1}{\sigma_\varpi} \,,
\end{equation}
where $D$ and $a_1 \equiv a \sin{i}$ stands for, respectively, the distance (to the pulsar) and the orbital semi-major axis projected onto the sightline. 
On the other hand, $\tilde{\theta}_\mathrm{orb} \equiv 2a/D$ reflects the apparent angular size of orbit. Provided the parallax uncertainty $\sigma_{\varpi}$, $\tilde{\theta}_\mathrm{orb} / \sigma_{\varpi}$ quantifies the detectability of orbital parameters using VLBI astrometry. Hence,
\begin{equation}
\label{eq:theta_orb}
   \frac{\tilde{\theta}_\mathrm{orb}}{\sigma_{\varpi}} \equiv \frac{2a}{D}\cdot \frac{1}{\sigma_{\varpi}} \geq  \eta_\mathrm{orb}
    \,.
\end{equation}
Since $i$ is usually unknown, the $\eta_\mathrm{orb}$ defined in Equation~\ref{eq:reflex_motion_measurability} serves as a lower limit for $\tilde{\theta}_\mathrm{orb} / \sigma_{\varpi}$, and is used in this work to find out pulsar systems with $i$ and $\Omega_\mathrm{asc}$ potentially measurable with VLBI observations. 
In general, the orbital reflex motion should be negligible when $\eta_\mathrm{orb} \ll 1$, easily measurable when $\eta_\mathrm{orb} \gg 1$, and difficult to constrain (but non-negligible) when $\eta_\mathrm{orb} \sim 1$. By way of comparison, \citet{Guo21} were able to firmly constrain $\Omega_\mathrm{asc}$ and $i$ for PSR~J2222$-$0137 ($\eta_\mathrm{orb} = 10.2$), while \citet{Deller16} could place weak constraints for PSR~J1022$+$1001 and PSR~J2145$-$0750 ($\eta_\mathrm{orb} = 3.2$ and 1.6, respectively)

Accordingly, in this work, we fit for orbital reflex motion if all the following conditions are met:
\begin{enumerate}[label=(\roman*), leftmargin=*,align=left]
    \item $a_1$ is well determined with pulsar timing;
    \item $\eta_\mathrm{orb} > 1$;
    \item the orbital period $P_\mathrm{b} < 2$\,yr, where 2\,yr is the nominal time span of an \mspsrpi\ astrometric campaign.
\end{enumerate}
For the calculation of $\eta_\mathrm{orb}$, we simply use the direct-fitting parallax $\varpi^{\mathrm{(DF)}}$ for $\varpi$, and its \rcs-corrected uncertainty $\sigma_{\varpi}^\mathrm{(DF)} \cdot \sqrt{\chi^2_\nu}$ for $\sigma_{\varpi}$ (see Table~\ref{tab:astrometric_parameters}). 
We note that this choice of parallax and its uncertainty would generally lead to slightly larger $\eta_\mathrm{orb}$ compared to using $\varpi^{\mathrm{(Ba)}}$ and $\sigma_{\varpi}^\mathrm{(Ba)}$, according to Figure~\ref{fig:Bo_errs_vs_Ba_errs} and the discussion in Section~\ref{subsec:Bayesian_as_major}.  
Nevertheless, the choice {\bf 1)} enables the comparison with $\eta_\mathrm{orb}$ of the historically published pulsars (that do not have $\varpi^{\mathrm{(Ba)}}$ and $\sigma_{\varpi}^\mathrm{(Ba)}$), {\bf 2)} simplifies the procedure of analysis, {\bf 3)} facilitates the reproduction of $\eta_\mathrm{orb}$ by other researchers, and {\bf 4)} is more conservative in the sense that more candidates with $\eta_\mathrm{orb} > 1$ would be found.
The calculated $\eta_\mathrm{orb}$ as well as $P_\mathrm{b}$ are summarized in Table~\ref{tab:astrometric_parameters}. 
Among the 18 \mspsrpi\ pulsars, \psrfa, \psrga, \psrgb\ and \psrka\ meet our criteria (see Table~\ref{tab:astrometric_parameters}), where \psrfa\ is a DNS system and the others are pulsar-WD binaries.
Hereafter, the 4 pulsars are referred to as the ``8P'' pulsars for the sake of brevity, as we would perform 8-parameter (i.e., the 5 canonical astrometric parameters and $\eta_\mathrm{EFAC}$ plus $i$ and $\Omega_\mathrm{asc}$) inference on them. 


For the 8-parameter inference, prior probability distributions of the canonical parameters and $\eta_\mathrm{EFAC}$ are described in Section~\ref{subsubsec:parameter_priors}. 
Both $i$ and $\Omega_\mathrm{asc}$ are defined in the TEMPO2 \citep{Edwards06} convention.
The prior probability distribution of $\Omega_\mathrm{asc}$ follows $\mathcal{U}$(0, 360\degr).
Sine distribution $\mathcal{S}$(0, 180\degr) is used for $i$ of the four 8P pulsars (i.e., the probability density $p(i) \propto \sin{i}$, $i \in \left[0, 180\degr\right]$). 
Where available, tighter constraints are applied to $i$ in accordance with Table~\ref{tab:7_parameter_inference} (also see the descriptions in Section~\ref{sec:individual_pulsars}).

Moreover, extra prior constraints can be applied to $i$ and $\Omega_\mathrm{asc}$ based on $\dot{a}_1$, the time derivative of $a_1$ \citep[e.g.][]{Nice01,Deller16,Reardon21}.
As $a_1 \equiv a\sin{i}$, 
\begin{equation}
\label{eq:a1dot}
\begin{split}
    \frac{\dot{a}_1}{a_1} = \frac{\dot{a}}{a} + \frac{\partial i}{\partial t} \cot{i}  \approx \frac{\partial i}{\partial t} \cot{i}\,.
\end{split}
\end{equation}
Here, the $\dot{a}/a$ term reflects the intrinsic variation of the semi-major axis $a$ due to GR effects \citep{Peters64}, which is however $\sim$8 and $\sim$5 orders smaller than $\dot{a}_1/a_1$ for the 8P WD-pulsar systems and the DNS system \psrfa, respectively (see \citealp{Nice01} for an analogy).
Accordingly, the apparent $\dot{a}_1/a_1$ is predominantly caused by apparent $i$ change as a result of the sightline shift \citep{Kopeikin96}. When proper motion contributes predominantly to the sky position shift (as is the case for the 8P pulsars), 
\begin{equation}
\label{eq:i_dot}
\begin{split}
    \frac{\partial i}{\partial t} = \mu \sin{\left(\theta_\mu-\Omega_\mathrm{asc}\right)}\,,
\end{split}
\end{equation}
where $\theta_\mu$ refers to the position angle (east of north) of the proper motion $\mu$ \citep{Kopeikin96,Nice01}.
We incorporated the $\dot{a}_1/a_1$ measurements (with Equations~\ref{eq:a1dot} and \ref{eq:i_dot}) on top of other prior constraints,
and inferred $i$, $\Omega_\mathrm{asc}$, $\eta_\mathrm{EFAC}$ and the canonical five astrometric parameters for the 8P pulsars with {\tt sterne}\textsuperscript{\ref{footnote:sterne}}, following similar approaches taken by \citet{Deller16,Guo21}.

While we ultimately did not significantly constrain $i$ or $\Omega_\mathrm{asc}$ for any pulsar, including their non-negligible reflex motion in the inference is still necessary for correctly inferring the uncertainties of the non-orbital model parameters. 
The non-orbital inferred parameters are provided in Section~\ref{subsec:astrometric_results_non_PM_priors} below, along with all the non-8P pulsars.
As we found minimal differences between the constraints obtained on orbital parameters with or without the adoption of priors based on pulsar timing, we defer the presentation of the posterior constraints on orbital inclinations and ascending node longitudes (of the 8P pulsars) to Section~\ref{sec:inference_with_priors} in order to avoid repetition.

\begingroup
\renewcommand{\arraystretch}{1.4} 

\begin{table}
\raggedright
\caption{Prior constraints on $i$ and $\Omega_\mathrm{asc}$}
\label{tab:7_parameter_inference}
\resizebox{\columnwidth}{!}{
\begin{tabular}{lcccc}
\hline
\hline
PSR & $\dot{a}_1$ & $\dot{a}_1/a_1$ & $i$ & $\Omega_\mathrm{asc}$ \\
 & ($10^{-15}~\text{lt-s~s}^{-1}$) & ($10^{-15}~\mathrm{s}^{-1}$) &  & (deg)  \\
\hline

\Psrfa & -11(3) $^{a_1}$ & -0.55(15) & $\sin{i}\leq0.73$ $^{a_2}$ & ---   \\

\Psrga & 12(1) $^b$ & 0.22(2) & $\sin{i}=0.973(9)$ $^b$  & ---   \\

\Psrgb & -49.7(7) $^c$ & -1.98(3) & ---  & ---   \\
\Psrka & 14(2) $^b$ & 0.34(5) & 85(14)\degr\ $^d$ & ---  \\
\hline

\multicolumn{5}{l}{
$^{a_1}$\citet{Janssen08};
$^{a_2}$ inferred from the non-detection of Shapiro delay effects.}\\
\multicolumn{5}{l}{ $^b$\citet{Perera19}; $^c$\citet{Reardon21}.}\\
\multicolumn{5}{l}{
$^d$ based on Shapiro delay measurements \citep{Faisal-Alam20}.}\\
\end{tabular}
}
\end{table}
\endgroup

\subsection{The quasi-VLBI-only astrometric results}
\label{subsec:astrometric_results_non_PM_priors}
To wrap up this section, we summarize in Table~\ref{tab:models_no_pm_prior} the full (including $\alpha_\mathrm{ref}$ and $\delta_\mathrm{ref}$) final astrometric results obtained with no exterior prior proper motion or parallax constraints, which we simply refer to as quasi-VLBI-only astrometric results (we add ``quasi'' because timing constraints on two orbital parameters, i.e., $i$ and $\dot{a}_1$, have already been used for the 8P pulsars).
These quasi-VLBI-only results are mainly meant for independent checks of timing results (which would enable the frame connection mentioned in Section~\ref{subsec:MSP_VLBI_astrometry}), or as priors for future timing analyses.
For the most precise possible pulsar parallaxes and hence distances, we recommend the use of the ``VLBI~$+$~timing'' results presented in Section~\ref{sec:inference_with_priors}.

The reference positions $\alpha_\mathrm{ref}$ and $\delta_\mathrm{ref}$ we provide in Table~\ref{tab:models_no_pm_prior} are precisely measured, but only with respect to the assumed location of the in-beam calibrator source for each pulsar. In all cases, the uncertainties on the in-beam source locations (also shown in Table~\ref{tab:models_no_pm_prior}) dominate the total uncertainty in the pulsar's reference position.  A future work, incorporating additional multi-frequency observations of the in-beam calibrations, will enable significantly more precise pulsar reference positions to be obtained, as is discussed in Section~\ref{subsec:mspsrpi}.

\begingroup
\renewcommand{\arraystretch}{1.4} 

    \begin{table*}
    \raggedright
    \caption{Final astrometric models inferred without using timing proper motions as priors.}
     
        \resizebox{\textwidth}{!}{
    	\begin{tabular}{lccccccccccc} 
		\hline
		\hline
	PSR & $t_\mathrm{ref}$ & $\alpha_\mathrm{ref}$ (J2000) $^*$  & $\sigma_\mathrm{\alpha_\mathrm{ref}}$ & $\delta_\mathrm{ref}$ (J2000) $^*$ & $\sigma_\mathrm{\delta_\mathrm{ref}}$ & $\mu_\alpha$ & $\mu_\delta$ & $\varpi$ & $\eta_\mathrm{EFAC}$ &   $\rho_{\mu_{\alpha},\varpi}$  & $\rho_{\mu_{\delta},\varpi}$ \\
	 & (MJD) & & (mas) & & (mas) & (\maspy) & (\maspy) & (mas) &   & & \\
		\hline
	\Psrb\ & 57849 & $00^{\rm h}30^{\rm m}27\fs 42502$ & $0.06[\pm0.3\pm0.2\pm0.8]$ & $04\degr51'39\farcs7159$ & $0.2[\pm0.8\pm0.3\pm0.8]$ & -6.13(7) & $0.34^{+0.15}_{-0.16}$ &  3.02(7)  & $1.35^{+0.45}_{-0.32}$  & 0.39 & 0.08 \\
	\Psrc & 57757& $06^{\rm h}10^{\rm m}13\fs 60053$ & $0.08[\pm0\pm3.5\pm0.8]$ & $-21\degr00'27\farcs7923$ & $0.2[\pm0\pm15.2\pm0.8]$ & 9.1(1) &  $15.96^{+0.25}_{-0.24}$ &  0.73(10)   & $1.1^{+0.4}_{-0.2}$  & 0.51 & 0.05\\
	\Psrd & 57685& $06^{\rm h}21^{\rm m}22\fs11617$ & $0.12[\pm2.1\pm0.5\pm0.8]$ & $10\degr02'38\farcs7261$ & $0.3[\pm2.8\pm0.7\pm0.8]$ & 3.5(2) &  -1.37(35) &  0.86(15)   & $3.7^{+1.0}_{-0.7}$ & 0.60 & -0.01 \\
	\Psrea & 57700 & $10^{\rm h}12^{\rm m}33\fs 43991$ & $0.04[\pm1.2\pm0.3\pm0.8]$ & $53\degr07'02\farcs1110$ & $0.1[\pm2.8\pm0.3\pm0.8]$ & 2.67(5) & $-25.39^{+0.14}_{-0.15}$ & $1.17^{+0.04}_{-0.05}$   & $1.7^{+0.6}_{-0.4}$  & 0.37 & -0.06 \\
	\Psreb & 57797& $10^{\rm h}24^{\rm m}38\fs65725$ & $0.06[\pm0.6\pm0.8\pm0.8]$ & $-07\degr19'19\farcs8014$ & $0.2[\pm1.4\pm1.5\pm0.8]$ & -35.32(7) & -48.1(2) &  0.94(6)  & $1.2^{+0.4}_{-0.3}$ & 0.23 & $2\!\times\!10^{-3}$ \\
	\Psrfa\ $^b$ & 57795 & $15^{\rm h}18^{\rm m}16\fs79817$ & $0.04[\pm0.5\pm0.4\pm0.8]$ & $49\degr04'34\farcs1132$ & $0.1[\pm1.4\pm0.4\pm0.8]$ & $-0.69^{+0.04}_{-0.03}$ & $-8.53^{+0.07}_{-0.09}$ & 1.238(36) & $1.4^{+0.6}_{-0.4}$ & -0.19 & -0.68 \\
	\Psrfb & 57964 & $15^{\rm h}37^{\rm m}09\fs96347$ & $0.06[\pm1.2\pm0.1\pm0.8]$ & $11\degr55'55\farcs0274$ & $0.1[\pm2.7\pm0.1\pm0.8]$ & 1.51(3) & $-25.30^{+0.05}_{-0.06}$ & $1.06^{+0.11}_{-0.10}$ & $0.5^{+0.6}_{-0.4}$ & -0.17 & 0.04\\
	\Psrga\ $^b$ & 57500 & $16^{\rm h}40^{\rm m}16\fs74587$ & $0.07[\pm0.4\pm0.3\pm0.8]$ & $22\degr24'08\farcs7642$ & $0.1[\pm0.3\pm0.6\pm0.8]$ & 2.19(9) & $-11.30^{+0.16}_{-0.13}$ & 0.68(8) & $1.3^{+0.7}_{-0.5}$ & -0.57 & -0.02\\
	\Psrgb\ $^b$ & 57700 & $16^{\rm h}43^{\rm m}38\fs16407$ & $0.1[\pm2.0\pm0.1\pm0.8]$ & $-12\degr24'58\farcs6531$ & $0.4[\pm6.2\pm0.1\pm0.8]$ & 6.2(2) & 3.3(6) & $1.31^{+0.17}_{-0.18}$ & $1.0^{+0.3}_{-0.2}$ & -0.43 & 0.04\\
	\Psrha & 57820 & $17^{\rm h}21^{\rm m}05\fs49936$ & $0.2[\pm3.0\pm0.3\pm0.8]$ & $-24\degr57'06\farcs2210$ & $0.6[\pm7.3\pm0.6\pm0.8]$ & 2.5(3) & -1.9(9) & 0.0(2) & $3.1^{+0.8}_{-0.6}$ & -0.13 & -0.01\\
	\Psrhb & 57821 & $17^{\rm h}30^{\rm m}21\fs 67969$ & $0.2[\pm0.5\pm0.3\pm0.8]$ & $-23\degr04'31\farcs1749$ & $0.5[\pm1.1\pm0.6\pm0.8]$ & 20.3(2) & -4.8(5) & $1.57(18)$ & $1.4^{+0.3}_{-0.2}$ & -0.38 & 0.01 \\
	\Psri & 57829 & $17^{\rm h}38^{\rm m}53\fs 97001$ & $0.06[\pm0.2\pm0.3\pm0.8]$ & $03\degr33'10\farcs9124$ & $0.1[\pm0.6\pm0.6\pm0.8]$ & 6.98(8) & 5.18(16) & 0.50(6) & $1.9^{+0.7}_{-0.6}$ & -0.53 & $4\!\times\!10^{-4}$ \\
	\Psro & 57836 & $18^{\rm h}24^{\rm m}32\fs00791$ & $0.4[\pm2.0\pm0.1\pm0.8]$ & $-24\degr52'10\farcs912$ & $1[\pm5.2\pm0.2\pm0.8]$ & 0.3(6) & $-3.9^{+1.2}_{-1.3}$ &  0.1(5) & $1.6^{+0.8}_{-0.7}$ & -0.65 & $-3\!\times\!10^{-3}$ \\
	\Psrka\ $^b$ & 57846 & $18^{\rm h}53^{\rm m}57\fs31785$ & $0.06[\pm0.8\pm0.2\pm0.8]$ & $13\degr03'44\farcs0471$ & $0.1[\pm1.6\pm0.4\pm0.8]$ & -1.4(1) & -2.8(2) & 0.49(7) & $0.5^{+0.6}_{-0.3}$ & -0.37 & 0.26 \\
	\Psrl & 57847 & $19^{\rm h}10^{\rm m}09\fs 70165$ & $0.03[\pm1.0\pm0.1\pm0.8]$ & $12\degr56'25\farcs4316$ & $0.06[\pm2.5\pm0.1\pm0.8]$ & 0.50(4) & -6.85(9) &  0.254(35) & $0.19^{+0.15}_{-0.12}$ & -0.48 & 0.03 \\
	\Psrma & 57768 & $19^{\rm h}11^{\rm m}49\fs 27544$ & $0.1[\pm0.9\pm0.3\pm0.8]$ & $-11\degr14'22\farcs5547$ & $0.3[\pm2.1\pm0.5\pm0.8]$ & -13.8(2) & -10.3(4) & $0.38^{+0.13}_{-0.14}$ & $1.1^{+0.4}_{-0.3}$ & -0.39 & -0.02\\
	\Psrmb & 57768 & $19^{\rm h}18^{\rm m}48\fs 02959$ & $0.1[\pm0.9\pm0.2\pm0.8]$ & $-06\degr42'34\farcs9335$ & $0.2[\pm2.2\pm0.4\pm0.8]$ & -7.1(1) & -5.7(3) & 0.60(12) & $1.2^{+0.3}_{-0.2}$ & -0.39 & -0.01 \\
	\Psrkb\ $^a$ & 57850 & $19^{\rm h}39^{\rm m}38\fs56134$  
	& $[0.07[\pm0.8\pm0.1\pm0.8]$ & $21\degr34'59\farcs1233$ & $0.2[\pm1.9\pm0.1\pm0.8]$ & 0.08(7) & -0.43(11) & $0.384^{+0.048}_{-0.046}$ & $1.5^{+0.7}_{-0.6}$ & -0.62 & -0.06 \\
	\hline 
	\multicolumn{12}{l}{$^*$ $\alpha_\mathrm{ref}$ and $\delta_\mathrm{ref}$ refer to the reference position at reference epoch $t_\mathrm{ref}$. The error budgets of the reference positions are provided in the adjacent columns, which include, from left to right, the error of relative}\\
	\multicolumn{12}{l}{\ \ \   reference position with respect to the reference point, the uncertainty of the reference point with regard to the main phase calibrator (estimated with Equation~1 of \citealp{Deller19}), the position uncertainty}\\
	\multicolumn{12}{l}{\ \ \  of the main phase calibrator, and the typical (0.8\,mas in each direction, \citealp{Sokolovsky11}) frequency-dependent core shift \citep[e.g.][]{Bartel86,Lobanov98} between 1.55\,GHz and $\sim8$\,GHz.}\\
	\multicolumn{12}{l}{\ \ \   We note that the errors outside ``[ ]'' are obtained with Bayesian inference, while the errors inside ``[ ]'' are only indicative. To properly determine the absolute pulsar position and its uncertainty requires}\\
	\multicolumn{12}{l}{\ \ \   the procedure described in Section~3.2 of \citet{Ding20}. This analysis will be made and presented in an upcoming paper.}\\
	\multicolumn{12}{l}{$\bullet$  $\rho_{\mu_{\alpha},\varpi}$ and $\rho_{\mu_{\delta},\varpi}$ stand for correlation coefficients between $\varpi$ and the two proper motion components.}\\
	\multicolumn{12}{l}{$\bullet$ The special parameter $\eta_\mathrm{EFAC}$ (that has been provided in Tables~\ref{tab:astrometric_parameters} and \ref{tab:7_parameter_inference}) is not reiterated in this table.}\\
	\multicolumn{12}{l}{$^a$ 
	Since inverse referencing is applied for \psrkb, the two reference sources are the de-facto targets. Accordingly, the proper motion and parallax are the negative values of the direct measurements out }\\
	\multicolumn{12}{l}{\ \ \ of inverse referencing. For the original astrometric model, the reference positions for the two reference sources J194104 and J194106 are, respectively, $19^{\rm h}41^{\rm m}04\fs319769(2)+21\degr49'13\farcs19731(7)$ and}\\
	\multicolumn{12}{l}{\ \ \  $19^{\rm h}41^{\rm m}06\fs86774(1)+21\degr53'04\farcs9594(2)$, where the uncertainties do not contain those of the reference point (i.e. the inside-the-bracket terms of $\sigma_\mathrm{\alpha_\mathrm{ref}}$ and $\sigma_\mathrm{\delta_\mathrm{ref}}$). The reference position of \psrkb}\\
	\multicolumn{12}{l}{\ \ \ presented in the table is estimated using normal  phase referencing with respect to J194104.}\\
	\multicolumn{12}{l}{$^b$ Results of the 8-parameter Bayesian inference are reported here; the constraints on $i$ and $\Omega_\mathrm{asc}$ are described in Section~\ref{sec:inference_with_priors} (see Section~\ref{subsec:reflex_motion_inference} for explanations).}\\

	\end{tabular}
	}
    \label{tab:models_no_pm_prior}
    \end{table*}
\endgroup

\section{VLBI+timing astrometric results}
\label{sec:inference_with_priors}
In Bayesian inference, the output of a model parameter $X_j$ (where $j$ refers to various model parameters) hinges on its prior probability distribution: generally speaking, tighter prior constraints (on $X_j$) that are consistent with data (in the sense of Bayesian analysis) would sharpen the output $X_j$. In cases where a strong correlation between $X_j$ and another model parameter $X_k$ is present, tighter prior $X_j$ constraints that are consistent with the data would potentially sharpen both the output $X_j$ and the output $X_k$.

As noted in Section~\ref{subsec:MSP_VLBI_astrometry}, VLBI astrometry serves as the prime method to measure parallaxes of Galactic pulsars. A VLBI astrometric campaign (on a Galactic pulsar) normally spans $\sim2$ years, as a substantial parallax can likely be achieved in this timespan. On the other hand, most \mspsrpi\ pulsars have been timed routinely for $\gtrsim10$ years, which allows their proper motions to be precisely determined, as the precision on proper motion grows with $t^{3/2}$ (see, e.g., Section~4.4 of \citealp{Ding21}) for a regularly observed pulsar. In Table~\ref{tab:VLBI_timing_results}, we collect one timing proper motion (denoted as $\mu_\alpha^\mathrm{(Ti)}$ and $\mu_\delta^\mathrm{(Ti)}$) and one timing parallax ($\varpi^\mathrm{(Ti)}$) for each \mspsrpi\ pulsar. 
Among the published timing results, we select the timing proper motions measured over the longest timespan, and the $\varpi^\mathrm{(Ti)}$ having the smallest uncertainties.
According to Tables~\ref{tab:models_no_pm_prior} and \ref{tab:VLBI_timing_results}, most timing proper motions are more precise than the quasi-VLBI-only counterparts. On the other hand, timing parallaxes are mostly less precise than the quasi-VLBI-only counterparts.
Nevertheless, adopting appropriate timing parallaxes as priors can still effectively lower parallax uncertainties.

The precisely measured $\mu_\alpha^\mathrm{(Ti)}$ and $\mu_\delta^\mathrm{(Ti)}$ provide the opportunity to significantly refine the quasi-VLBI-only proper motions. 
Furthermore, as shown with the Pearson correlation coefficients \citep{Pearson95} $\rho_{\mu_\alpha,\varpi}$ and $\rho_{\mu_\delta,\varpi}$ that we summarized in Table~\ref{tab:models_no_pm_prior}, large correlation between parallax and proper motion is not rare for VLBI astrometry. Therefore, using the $\mu_\alpha^\mathrm{(Ti)}$ and $\mu_\delta^\mathrm{(Ti)}$ measurements as the prior proper motion constraints in Bayesian inference can potentially refine both proper motion and parallax determination.

The astrometric results inferred with timing priors, hereafter referred to as VLBI+timing results, are reported in Table~\ref{tab:VLBI_timing_results}. To differentiate from the notation of quasi-VLBI-only astrometric parameter $Y$, we denote a VLBI+timing model parameter in the form of $Y'$.
Comparing Tables~\ref{tab:models_no_pm_prior} and \ref{tab:VLBI_timing_results}, we find almost all VLBI+timing proper motions and parallaxes more precise than the quasi-VLBI-only counterparts; 
the most significant parallax precision enhancement occurs to \psrmb\ (by 42\%), followed by \psrkb\ (by 36\%) and \psrfb\ (by 33\%). 
Hence, we use the VLBI+timing results in the remainder of this paper.

In 7 cases (i.e., \psrc, \psrgb, \psrhb, \psri, \psrka, \psro, \psrl), one of $\mu_\alpha^\mathrm{(Ti)}$, $\mu_\delta^\mathrm{(Ti)}$ or $\varpi^\mathrm{(Ti)}$ is more than 2\,$\sigma$ discrepant from the quasi-VLBI-only counterpart. Using such timing priors may widen the uncertainties of resultant model parameters, as $\eta_\mathrm{EFAC}$ would be lifted to counter-balance the increased \rcs. Without any indication that the discrepant timing values are less reliable, we use them as priors regardless. 
However, we caution the use of these 7 sets of VLBI+timing results, and would recommend the quasi-VLBI-only results to be considered if our adopted timing priors are proven inaccurate in future.

We also now consider any  possible effects that could, despite our best efforts to characterise all sources of position noise, bias the fitted VLBI positions.  For any given VLBI calibrator source, evolution in the source structure can lead to a detectable position offset \citep[e.g.][]{Perger18,Zhang20c} that is then transferred to the target pulsar.  Due to the long timescales of AGN structure evolution, over the $\sim2$-year timescale of the \mspsrpi\ observations, this error may be quasi-linear in time and be absorbed into the pulsar proper motion \citep[e.g.][]{Deller13}.
Redundant secondary calibrators can be used to probe the astrometric effect of structure evolution. However, with small numbers of redundant calibrator sources, such probes are hardly conclusive, as the structure evolution of the redundant calibrators would also be involved.
Among the 7 pulsars showing $>2\,\sigma$ discrepancy  between quasi-VLBI-only and timing results (see Table~\ref{tab:VLBI_timing_results}), 
\psrb, \psrgb, \psrhb, \psri\ and \psro\ either display no relative motion between the redundant secondary calibrators and the main secondary calibrators or do not have any redundant calibrator (i.e. \psrgb), although the sub-optimal main secondary calibrators of \psrgb\ and \psro\ (see Sections~\ref{subsec:J1643} and \ref{subsec:J1824}) may likely affect the astrometric performance.
For \psrka, the main secondary calibrator has a clear jet aligned roughly with the right ascension (RA) direction, and thus source structure evolution is potentially significant.  
The two redundant calibrators for \psrka\ do display a relative proper motion of up to 0.2 mas/yr with respect to the main secondary calibrator, so while the mean relative motion seen between the two redundant secondary calibrators is small, calibrator structure evolution remains a possible explanation for the VLBI-timing discrepancy.
Finally, the main secondary calibrator of \psrl\ also exhibits a jet structure at a position angle of $\sim45$\degr. When using the only redundant calibrator of \psrl\ as the reference source, we obtained the VLBI-only result $\mu_\alpha=0.25\pm0.06$\,\maspy, $\mu_\delta=-7.3\pm0.1$\,\maspy\ and $\varpi=0.61\pm0.05$\,mas with Bayesian inference, where $\mu_\alpha$ becomes consistent with $\mu_\alpha^\mathrm{(Ti)}$ but $\mu_\delta$ and $\varpi$ are further away from the timing counterparts. 
The $\mu_\alpha$ consistency between VLBI and timing indicates that  structure evolution in our chosen calibrator is likely contributing to the VLBI-timing discrepancy. However, as the redundant calibrator is both fainter and further away from \psrl\ (compared to the main secondary calibrator), we do not use this source as the final reference source.

\begingroup
\renewcommand{\arraystretch}{1.4} 
\begin{table*}
\raggedright
\caption{{\it Left of the dashed line:} proper motions $\{ \mu'_\alpha,\mu'_\delta \}$ and parallaxes $\varpi'$ inferred with timing proper motion and parallax priors. {\it Right:} distances $D$ and transverse space velocities $v_\perp$ based on $ \mu'_\alpha$, $\mu'_\delta$ and $\varpi'$.}
\label{tab:VLBI_timing_results}
\resizebox{\textwidth}{!}{
\begin{tabular}{lccccccc:cccccc}
\hline
\hline
PSR &  $\mu_\alpha^\mathrm{(Ti)}$ & $\mu_\delta^\mathrm{(Ti)}$ &
$\varpi^\mathrm{(Ti)}$ & $\mu_{\alpha}'$ & $\mu_\delta'$ & $\varpi'$  & $\eta'_\mathrm{EFAC}$  & DM  &
$d_\mathrm{DM}^\mathrm{(NE)}$ $^*$ & $d_\mathrm{DM}^\mathrm{(YMW)}$ $^*$ & $b$ & $D$ & $v_\perp$ $^{**}$ \\
 & (\maspy) & (\maspy) & (mas) & (\maspy) & (\maspy) & (mas) &   & ($\mathrm{pc~{cm}^{-3}}$) & (kpc) & (kpc) & (deg) & (kpc) & (\kmps) \\
\hline
\Psrb &  -6.2(1)  & 0.5(3) $^d$ & 3.08(8) $^d$ & -6.15(5) & 0.37(14) & 3.04(5) &  $1.3^{+0.4}_{-0.3}$  & 4.3 & 0.32(6) & 0.35(7) & -57.6 & $0.329^{+0.006}_{-0.005}$ & 15.4(2) \\
\Psrc  & 9.04(8)  & $^{!!}$16.7(1) $^c$ & --- & 9.06(7) & 16.6(1) & 0.72(11)  &  $1.4^{+0.4}_{-0.3}$  & 60.7 & 3.5(7) & 3.3(7) & -18.2 & $1.5^{+0.3}_{-0.2}$ & $120^{+25}_{-17}$ \\
\Psrd &  $^!3.2(1)$  & $^!-0.6(5)$ $^c$ & ---  & 3.27(9) & -1.1(3) & 0.74(14) &  $3.9^{+1.1}_{-0.8}$  & 36.5 & 1.4(3) & 0.42(8) & -2.0 & $1.6^{+0.5}_{-0.3}$ & $20^{+8}_{-5}$ \\ 
\Psrea & $^!2.61(1)$  & $-25.49(1)$ $^c$ & $^!$0.9(2) $^c$ & 2.61(1) & -25.49(1) & 1.14(4) &  $1.7^{+0.5}_{-0.4}$  & 9.02 & 0.41(8) & 0.8(2) & 50.9 & 0.877(35) & 95(4) \\
\Psreb & -35.270(17)  & -48.22(3) $^b$ & 0.83(13) $^b$ & -35.27(2) & -48.22(3) & 0.93(5) &  $1.2^{+0.4}_{-0.3}$  & 6.5 & 0.39(8) & 0.38(8) & 40.5 & 1.08(6) & 300(20) \\
\Psrfa & $-0.67(4)$  & $-8.53(4)$ $^h$ & --- & -0.683(26) & -8.528(36) & $1.237^{+0.035}_{-0.031}$ &  $1.2^{+0.6}_{-0.4}$  & 11.61 & 0.6(1) & 1.0(2) & 54.3 & 0.81(2) & 16.0(6) \\
\Psrfb &  1.482(7)  & -25.285(12) $^g$ & 0.86(18) $^g$  & 1.484(7) & -25.286(11)  & 1.07(7)  & $0.5^{+0.5}_{-0.3}$   & 11.62 & 1.0(2) & 0.9(2) & 48.3 & $0.94^{+0.07}_{-0.06}$ & $102^{+8}_{-7}$ \\
\Psrga &  2.08(1) & -11.34(2) $^c$ & 0.6(4) $^c$ & 2.08(1) & -11.34(2) & 0.73(6) &  $1.3^{+0.7}_{-0.6}$  & 18.43 & 1.2(2) & 1.5(3) & 38.3 & $1.39^{+0.13}_{-0.11}$ & $53^{+5}_{-4}$ \\
\Psrgb &  $^!5.970(18)$  & 3.77(8) $^b$ & $^{!!}$0.82(17) $^b$ & 5.97(2) & 3.76(8) & 1.1(1) &  $1.0^{+0.3}_{-0.2}$  & 62.3 & 2.4(5) & 0.8(2) & 21.2 & $0.95^{+0.15}_{-0.11}$ & $41^{+5}_{-4}$ \\
\Psrha &  1.9(1.2)  & $^!-25(16)$ $^f$ & --- & 2.5(3) & -1.9(9) & 0.0(2) &  $3.1^{+0.8}_{-0.6}$  & 48.3 & 1.3(3) & 1.4(3) & 6.8 & $>1.5$ $^{\blacktriangle}$ & --- \\
\Psrhb & $^!20.06(12)$ & -4(2) $^b$ & $^{!!}$2.11(11) $^b$ & 20.1(1) & -4.7(6) & 2.0(1) &  $1.6^{+0.4}_{-0.3}$  & 9.62 & 0.5(1) & 0.5(1) & 6.0 & 0.51(3) & $54^{+3}_{-2}$ \\
\Psri &  7.037(5)  & 5.073(12) $^e$ & $^{!!}$0.68(5) $^e$ & 7.036(5) & 5.07(1) & 0.589(46) & $2.3^{+0.8}_{-0.6}$  & 33.8 & 1.4(3) & 1.5(3) & 17.7 & $1.74^{+0.15}_{-0.13}$ & $90^{+7}_{-6}$\\
\Psro &  -0.25(4)  & $^{!!!}-8.6(8)$ $^b$ & --- & -0.25(4) & -7.8(8) & 0.4(5)  & $2.5^{+0.8}_{-0.6}$  & 119.9 & 3.1(6) & 3.7(7) & -5.6 & $>0.5$ $^{\blacktriangle}$ & --- \\
\Psrka &  $^{!!}-1.63(2)$  & -2.96(4) $^c$ & 0.48(14) $^a$ & -1.62(2) & -2.96(4) & 0.53(7) &  0.9(5)  & 30.6 & 2.1(4) & 1.3(3) & 5.4 & 2.0(3) & $14^{+3}_{-2}$ \\
\Psrl &  $^{!!!!!}0.21(3)$  & $^!-7.04(5)$ $^c$ & 0.1(3) $^c$ & 0.24(3) & -7.03(5) & 0.36(6) &  $0.8^{+0.3}_{-0.2}$  & 38.1 & 2.3(5) & 1.5(3) & 1.8 & $3.4^{+1.0}_{-0.6}$ & $55^{+16}_{-9}$ \\
\Psrma &  -13.7(2)  & $^!-9.3(9)$ $^c$ & --- & -13.7(1) & $-10.2^{+0.4}_{-0.3}$ & $0.37^{+0.13}_{-0.14}$  &  $1.1^{+0.4}_{-0.3}$  & 31.0 & 1.2(2) & 1.1(2) & -9.6 & $>1.3$ $^{\blacktriangle}$ & --- \\
\Psrmb &  -7.15(2)  & -5.94(5) $^c$ & $^{!}$0.8(1) $^c$ & -7.15(2) & -5.93(5) & 0.71(7) &  $1.1^{+0.3}_{-0.2}$  & 6.1 & 1.2(2) & 1.0(2) & -9.1 & $1.48^{+0.19}_{-0.14}$ & $44^{+6}_{-5}$ \\
\Psrkb &  0.074(2)  & -0.410(3) $^c$ & $^{!}$0.28(5) $^a$ & 0.074(2) & -0.410(3) & 0.35(3)  & 1.2(6)  & 71.1 & 3.6(7) & 2.9(6) & -0.3 & $2.9^{+0.3}_{-0.2}$ & $80^{+9}_{-8}$ \\

\hline
\multicolumn{14}{l}{$\bullet$ ``Ti'' denotes historical pulsar timing results. $\rho$ stand for correlation coefficients of Bayesian inference without using timing proper motion priors.}\\
\multicolumn{14}{l}{$\bullet$ The ``!''s before $\mu_\alpha^\mathrm{(Ti)}$, $\mu_\delta^\mathrm{(Ti)}$ and $\varpi^\mathrm{(Ti)}$ convey the significance of the offset from the quasi-VLBI-only counterparts. In specific, the ``!'' repetition number $N_!$ means the offset significance is between}\\
\multicolumn{14}{l}{\ \ \ $N_!\,\sigma$ and $(N_!+1)\,\sigma$. VLBI+timing results obtained with timing proper motion or parallax priors that are more than 2\,$\sigma$ away from the quasi-VLBI-only counterparts (i.e., $\mu_\alpha^\mathrm{(Ti)}$,  $\mu_\delta^\mathrm{(Ti)}$ and $\varpi^\mathrm{(Ti)}$ }\\
\multicolumn{14}{l}{\ \ \ marked with $\geq2$ ``!''s) should be used with caution.}\\
\multicolumn{14}{l}{$\bullet$ $\mu_\alpha''$, $\mu_\delta''$ and $\varpi''$ designate results of Bayesian inference using timing proper motion priors.}\\
\multicolumn{14}{l}{$\bullet$ For each pulsar, we present the most precise timing estimates published in $^a$\citet{Faisal-Alam20},  $^b$\citet{Reardon21}, $^c$\citet{Perera19}, $^d$\citet{Arzoumanian18a}, $^e$\citet{Freire12},}\\
\multicolumn{14}{l}{\ \ \ $^f$\citet{Desvignes16}, $^g$\citet{Fonseca14} and $^h$\citet{Janssen08}.}\\ 
\multicolumn{14}{l}{$\bullet$ For comparison, we list $d_\mathrm{DM}^\mathrm{(NE)}$ and $d_\mathrm{DM}^\mathrm{(YMW)}$ derived with {\tt pygedm}\textsuperscript{\ref{footnote:pygedm}} given the sky position and the DM of a pulsar, based on the two latest $n_\mathrm{e}(\vec{x})$ models \citep{Cordes02,Yao17}.}\\
\multicolumn{14}{l}{$^*$20\% relative uncertainties are assumed for all DM-based distances (i.e., $d_\mathrm{DM}^\mathrm{(NE)}$ and $d_\mathrm{DM}^\mathrm{(YMW)}$).}\\ 
\multicolumn{14}{l}{$^{**}$Tangential space velocities corrected for the differential rotation of the Galaxy (see \ref{subsec:v_t}).}\\ 
\multicolumn{14}{l}{$^{\blacktriangle}$The reciprocal of the 3\,$\sigma$ upper limit of the parallax is adopted as the lower limit of the distance.}\\ 

\end{tabular}
}
\end{table*}
\endgroup

\subsection{The posterior orbital inclinations and ascending node longitudes}
\label{subsec:i_and_Omega_asc}
For the four 8P pulsars, orbital inclinations $i'$ and ascending node longitudes $\Omega'_\mathrm{asc}$ are also inferred alongside the five canonical parameters and $\eta'_\mathrm{EFAC}$ (see Section~\ref{subsec:reflex_motion_inference}). 
The full 8D corner plots out of the 8-parameter inferences are available online\textsuperscript{\ref{footnote:pulsar_positions}}.
Prior constraints on $i'$ and $\Omega'_\mathrm{asc}$ have been provided in Section~\ref{subsubsec:parameter_priors}.
Owing to bi-modal features of all 1D histograms of $i'$, no likelihood component is substantially favored over the other. Hence, no tight posterior constraint on $i'$ is achieved for any 8P pulsar. Likewise, all 1D histograms of $\Omega'_\mathrm{asc}$ show multi-modal features, which precludes stringent constraints on $\Omega'_\mathrm{asc}$. 

\multilinecomment{
\begin{figure*}
    \centering
	\includegraphics[width=12cm]{Figures/orbital_constraints.pdf}
    \caption{The VLBI+timing posterior constraints on the orbital inclination $i$ (i.e., $i'$ elsewhere in this paper) and the ascending node longitude $\Omega_\mathrm{asc}$ (i.e., $\Omega'_\mathrm{asc}$ elsewhere in this paper). 
    Both $i$ and $\Omega_\mathrm{asc}$ are defined in the TEMPO2 \citep{Edwards06} convention.
    The corner plots are made with {\tt corner.py}, where the $n$-th contour in each 2D histogram comprises $1-\exp\left(-n^2/2\right)$ of the simulated data chain \citep{Foreman-Mackey16}.
    }    
    \label{fig:orbital_constraints}
\end{figure*}
}

\subsection{Comparison with Gaia results}
\label{subsec:Gaia_results}

From the Gaia Data Release 2 \citep{Gaia-Collaboration18a}, Gaia counterparts for pulsars with optically bright companions have been identified and studied by \citet{Jennings18,Mingarelli18,Antoniadis21}. In the \mspsrpi\ sample, \psrea\ and \psreb\ have secure Gaia counterparts, while \psrl\ has a proposed Gaia counterpart candidate \citep{Mingarelli18}. 
In Table~\ref{tab:Gaia_results}, we updated the Gaia results for these three Gaia sources to the Gaia Data Release 3 (DR3, \citealp{Gaia-Collaboration22}).

For \psreb, the Gaia  proper motion \{$\mu_\alpha^\mathrm{(G)}$, $\mu_\delta^\mathrm{(G)}$\} and parallax $\varpi_1^\mathrm{(G)}$ are highly consistent with the VLBI+timing ones, which further strengthens the proposal that \psreb\ is in an ultra-wide orbit with a companion star (\citealp{Bassa16,Kaplan16}, also see Sections~\ref{subsec:v_t} and \ref{subsec:A_Gal}).
The Gaia proper motion and parallax of \psrea\ is largely consistent with the VLBI+timing counterparts. The $>1\,\sigma$ discrepancy between $\mu_\delta^\mathrm{(G)}$ and $\varpi_1^\mathrm{(G)}$ and the respective VLBI+timing counterparts can be explained by non-optimal goodness of (Gaia astrometric) fitting (GoF) (see Table~\ref{tab:Gaia_results}).
On the other hand, the Gaia counterpart candidate for \psrl\ (proposed by \citealp{Mingarelli18}) possesses a $\mu_\alpha^\mathrm{(G)}$ 4\,$\sigma$ discrepant from the VLBI+timing one. Though this discrepancy is discounted by the relatively bad GoF by roughly a factor of 1.9 (see Table~\ref{tab:Gaia_results}), the connection between the Gaia source and \psrl\ remains inconclusive.
We note that the parallax zero-points $\varpi_0^\mathrm{(G)}$ \citep{Lindegren21} of the three Gaia sources are negligible and hence not considered, as $\varpi_0^\mathrm{(G)}$ is small ($|\varpi_0^\mathrm{(G)}|\lesssim0.02$\,mas, \citealp{Ding21}) compared to the uncertainty of $\varpi_1^\mathrm{(G)}$ (see Table~\ref{tab:Gaia_results}). 

\begingroup
\renewcommand{\arraystretch}{1.4} 

\begin{table}
\raggedright
\caption{Gaia astrometric results}
\label{tab:Gaia_results}
\resizebox{\columnwidth}{!}{
\begin{tabular}{lccccc}
\hline
\hline
PSR & Gaia DR3 & $\mu_\alpha^\mathrm{(G)}$ & $\mu_\delta^\mathrm{(G)}$ & $\varpi_1^\mathrm{(G)}$ & GoF.$^*$\\
 & source ID & (\maspy) & (\maspy) & (mas) & \\
\hline

\Psrea & 851610861391010944 & 2.7(3) & $^!$-25.9(3) & $^!$1.7(3) & -1.5\\

\Psreb & 3775277872387310208 & -35.5(3) & -48.35(36) & 0.86(28) & 0.4\\

\Psrl & 4314046781982561920$^?$ & $^{!!!!}$-2.3(6) & $^!$-6.1(6) & -0.1(8) & 1.9\\
\hline

\multicolumn{6}{l}{$\bullet$ Sources marked with ``?'' are tentative Gaia counterpart candidates.}\\
\multicolumn{6}{l}{$\bullet$ Values marked with $N$ ``!''s are $N \sigma - (N+1) \sigma$ offset from the VLBI+timing counterparts}\\
\multicolumn{6}{l}{$^*$ Goodness of fitting, a parameter (of Gaia data releases) approximately following $\mathcal{N}(0,1)$ }\\
\multicolumn{6}{l}{\ \ \ distribution. A GoF closer to zero indicates better fitting performance.}\\

\end{tabular}
}
\end{table}
\endgroup

\section{Distances and Space velocities}
\label{sec:distances_and_velocities}
In this section, we derive pulsar distances $D$ from parallaxes $\varpi'$ (see Section~\ref{sec:inference_with_priors}), and compare them to the dispersion-measure-based distances.
Incorporating the proper motions $\{\mu'_\alpha, \mu'_\delta\}$ (see Section~\ref{sec:inference_with_priors}), we infer the transverse space velocity $v_\perp$ (i.e., the velocity with respect to the stellar neighbourhood) for each pulsar, in an effort to enrich the sample of $\sim$40 MSPs with precise $v_\perp$ \citep{Hobbs05,Gonzalez11} and refine the $v_\perp$ distributions of MSP subgroups such as binary MSPs and solitary MSPs.

\subsection{Parallax-based distances}
\label{subsec:px_based_D}
Inferring a source distance from a measured parallax requires assumptions about the source properties, for which a simple inversion implicitly makes unphysical assumptions \citep[e.g.][]{Bailer-Jones21}. 
Various works \citep[e.g.][]{Lutz73,Verbiest12,Bailer-Jones15,Igoshev16} have contributed to developing and consolidating the mathematical formalism of parallax-based distance inference, which we briefly recapitulate as follows, in order to facilitate comprehension and ready the mathematical formalism for further discussion.

A parallax-based distance $D$ can be approached from the conditional probability density function (PDF) 
\begin{equation}
\label{eq:px_based_D}
\begin{split}
    p(D|\varpi', l, b) \propto p(\varpi'|D) p(D, l, b),
\end{split}
\end{equation}
where $l$ and $b$ stands for Galactic longitude and latitude, respectively; 
$\varpi'=\varpi'_0 \pm \sigma_{\varpi'}$.
The first term on the right takes the form of 
\begin{equation}
\label{eq:gaussian_distribution}
\begin{split}
    p(\varpi'|D) \propto \exp\left[{-\frac{1}{2}\left(\frac{1/D-\varpi'_0}{\sigma_{\varpi'}}\right)^2}\right],
\end{split}
\end{equation}
assuming $\varpi_0'$ is Gaussian-distributed, or more specifically, $\varpi_0' \sim \mathcal{N}\left(1/D, \sigma_{\varpi'}^2\right)$. 
The second term on the right side of Equation~\ref{eq:px_based_D} can be approximated as $p(D,l,b) \propto D^2$, when the parent population $\Psi$ of the target celestial body is uniformly distributed spatially \citep{Lutz73}. Given a postulated (Galactic) spatial distribution $\rho(D,l,b)$ of $\Psi$, $p(D,l,b) \propto D^2 \rho(D,l,b)$.
Hence, 
\begin{equation}
\label{eq:p_D_extended}
\begin{split}
    p(D|\varpi',l,b) \propto D^2 \rho(D,l,b) \exp\left[{-\frac{1}{2}\left(\frac{1/D-\varpi'_0}{\sigma_{\varpi'}}\right)^2}\right] \,.
\end{split}
\end{equation}
We join \citet{Verbiest12} and \citet{Jennings18} to adopt the $\rho(D,l,b)$ (of the ``Model C'') determined by \citet{Lorimer06a} for Galactic pulsars. While calculating the $\rho(D,l,b)$ with Equations~10 and 11 of \citet{Lorimer06a}, we follow \citet{Verbiest12} and \citet{Jennings18} to increase the scale height (i.e., the parameter ``$E$'' of \citealp{Lorimer06a}) to 0.5\,kpc to accommodate the MSP population. In addition, the distance to the Galactic centre (GC) in Equation~10 of \citealp{Lorimer06a} is updated to $d_\odot = 8.12\pm0.03$\,kpc \citep{Gravity-Collaboration18}. 
We do not follow \citet{Verbiest12,Igoshev16} to use pulsar radio fluxes to constrain pulsar distances, as pulsar luminosity is relatively poorly constrained.

\begin{figure}
    \centering
	\includegraphics[width=9cm]{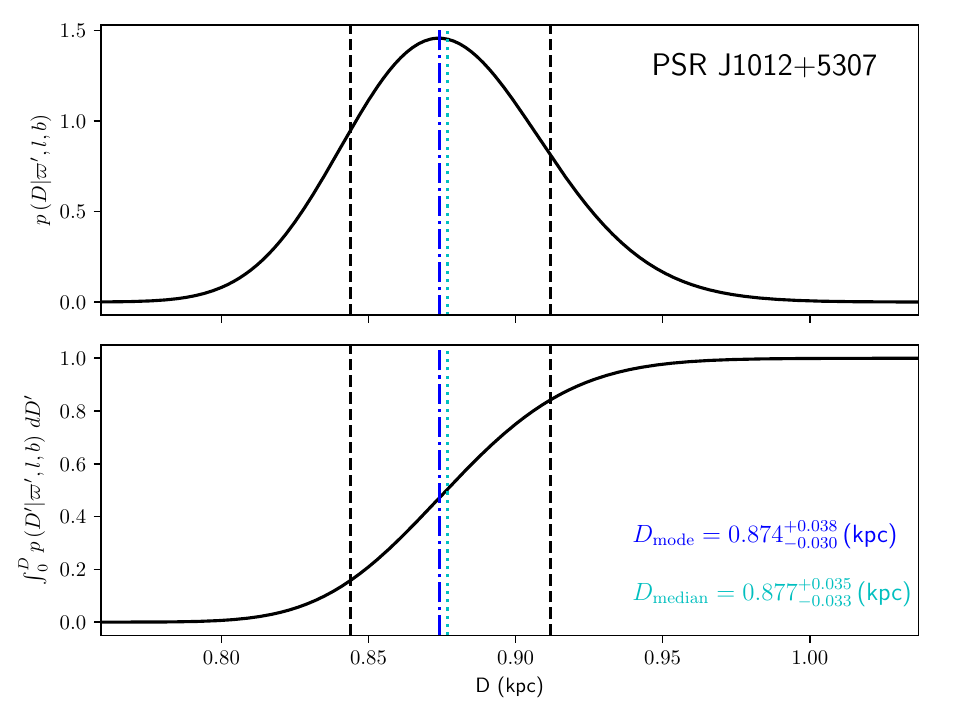}
    \caption{An example posterior probability density function $p(D|\varpi',l,b)$ (of distance) and its cumulative distribution function $\Phi(D|\varpi',l,b)=\int^{D}_{\,0} p(D'|\varpi',l,b)\,dD'$. The vertical dashed lines correspond to $\Phi(D|\varpi',l,b)\!=\!0.16$ and $\Phi(D|\varpi',l,b)\!=\!0.84$, which are respectively used as the lower and upper bound of the 1\,$\sigma$ uncertainty interval. The mode distance $D_\mathrm{mode}$ and median distance $D_\mathrm{median}$ are marked with dot-dashed blue line and dotted cyan line, respectively. Plots of this kind are also made for other \mspsrpi\ pulsars, and made available online\textsuperscript{\ref{footnote:pulsar_positions}}. Staying in line with the norm (see Section~\ref{subsec:MSP_VLBI_astrometry}) of this paper, we universally adopt $D_\mathrm{median}$ as the distances (i.e., $D$ in Table~\ref{tab:VLBI_timing_results}) for all \mspsrpi\ pulsars in this paper.
    }    
    \label{fig:J1012_dist}
\end{figure}

Using the aforementioned mathematical formalism, we calculated $p(D|\varpi',l,b)$ for each \mspsrpi\ pulsar, and integrated it into the cumulative distribution function (CDF) $\Phi(D|\varpi',l,b)=\int^{D}_{\,0} p(D'|\varpi',l,b)\,dD'$. 
The $p(D|\varpi',l,b)$ and $\Phi(D|\varpi',l,b)$ is plotted for each pulsar and made available online\textsuperscript{\ref{footnote:pulsar_positions}}. An example of these plots are presented in Figure~\ref{fig:J1012_dist}. The median distances $D_\mathrm{median}$ corresponding to $\Phi(D|\varpi',l,b)\!=\!0.5$ are taken as the pulsar distances, and summarized in Table~\ref{tab:VLBI_timing_results}. The distances matching $\Phi(D|\varpi',l,b)\!=\!0.16$ and $\Phi(D|\varpi',l,b)\!=\!0.84$ are respectively used as the lower and upper bound of the 1\,$\sigma$ uncertainty interval.

\subsubsection{Comparison with DM distances}
\label{subsubsec:DM_distances}
As mentioned in Section~\ref{subsec:MSP_VLBI_astrometry}, the precise DM measured from a pulsar can be used to assess the pulsar distance, provided an $n_\mathrm{e}(\vec{x})$ model. Using {{\tt pygedm}\footnote{\label{footnote:pygedm}\url{https://github.com/FRBs/pygedm}}}, we compile into Table~\ref{tab:VLBI_timing_results} the DM distances (i.e., $d^\mathrm{(NE)}_\mathrm{DM}$ and $d^\mathrm{(YMW)}_\mathrm{DM}$) of each pulsar based on the two latest realisations of $n_\mathrm{e}(\vec{x})$ model --- the NE2001 model \citep{Cordes02} and the YMW16 model \citep{Yao17}. 
For all the DM distances, we adopt typical 20\% fractional uncertainties.
We have obtained significant ($\geq3\,\sigma$) parallax-based distances $D$ for 15 out of 18 \mspsrpi\ pulsars. These distances enable an independent quality check of both $n_\mathrm{e}(\vec{x})$ models.

Among the 15 pulsars with parallax-based distance measurements, YMW16 is more accurate than NE2001 in three cases (i.e., \psrea, \psrgb\ and \psrkb), but turns out to be the other way around in four cases (i.e. \psrd, \psrka, \psrl\ and \psrmb). In other 8 cases, the $D$ cannot discriminate between the two models.
The small sample of 15 $D$ measurements shows that NE2001 and YMW16 remain comparable in terms of outliers.
In 2 (out of the 15) cases (i.e., \psrc, \psreb), $D$ is about $2.6\,\sigma$ and $6.8\,\sigma$ away from either DM distance, which reveals the need to further refine the $n_\mathrm{e}(\vec{x})$ models. Such a refinement can be achieved with improved pulsar distances including the ones determined in this work.

\subsection{Transverse space velocities}
\label{subsec:v_t}

Having determined the parallax-based distances $D$ and the proper motions $\{\mu'_\alpha, \mu'_\delta\}$, we proceed to calculate transverse space velocities $v_\perp$ for each pulsar, namely the transverse velocity with respect to the neighbouring star field of the pulsar. In estimating the transverse velocity of a pulsar neighbourhood, we assume the neighbourhood observes circular motion about the axis connecting the North and South Galactic Poles, which is roughly valid given that all \mspsrpi\ pulsars with significant ($>3\,\sigma$) $D$ share a median $|z|=D\sin{|b|}$ of 0.3\,kpc. Using the Galactic rotation curve from \citet{Reid19} and the full circular velocity of the Sun $247\pm1$\,\kmps, we derived the apparent transverse velocity of the neighbourhood $v_{\perp,N}$, thus obtaining $v_\perp$ by subtracting the apparent transverse velocity of the pulsar by $v_{\perp,N}$. Here, the full circular velocity (denoted as $\Theta_0+\Vsun$ in \citealp{Reid19}) is calculated with $d_\odot = 8.12\!\pm\!0.03$\,kpc \citep{Gravity-Collaboration18} and the proper motion of Sgr~$\mathrm{A^*}$ from \citet{Reid19}.

To estimate the uncertainty of $v_\perp$, we simulated a chain of 50,000 distances for each pulsar based on the $p(D|\varpi',l,b)$ that we have obtained in Section~\ref{subsec:px_based_D}. Besides, we also acquired chains of 50,000 $\mu'_\alpha$ and $\mu'_\delta$ given the VLBI+timing proper motions of Table~\ref{tab:VLBI_timing_results}, assuming $\mu'_\alpha$ and $\mu'_\delta$ follow Gaussian distributions. 
With these chains of $D$, $\mu'_\alpha$ and $\mu'_\delta$, we calculated 50,000 $v_\perp$ values, which form a PDF of $v_\perp$ for each pulsar. The $v_\perp$ inferred from the PDFs are summarized in Table~\ref{tab:VLBI_timing_results}.

\begin{figure*}
    \centering
	\includegraphics[width=19cm]{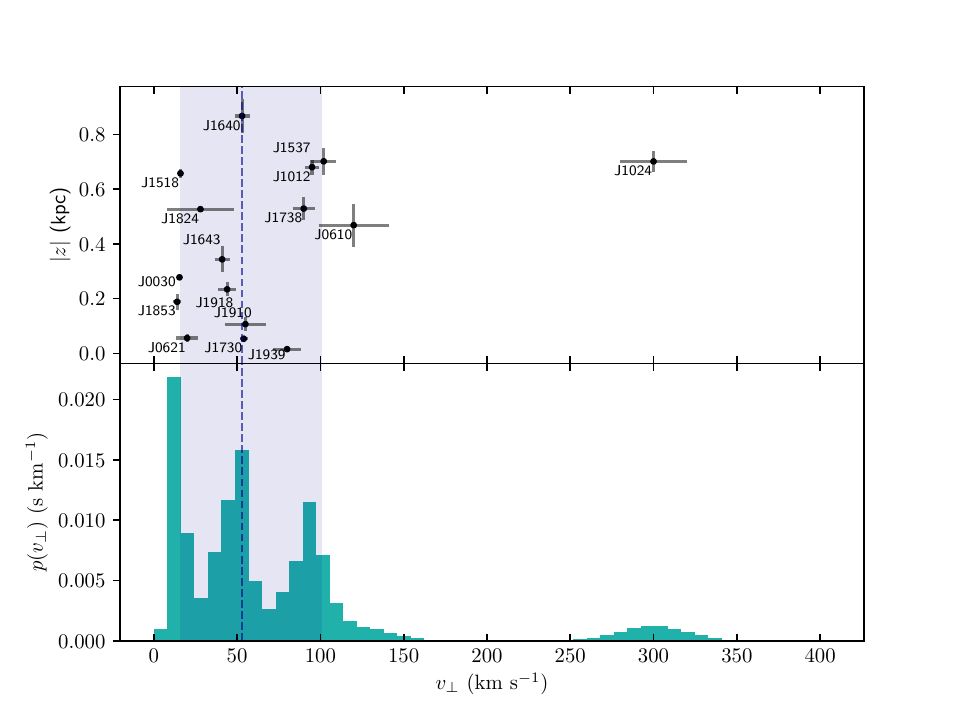}
    \caption{{\bfseries Upper:} The transverse space velocities $v_\perp$ versus the Galactic vertical heights $|z|=D\sin{|b|}$ of the 16 \mspsrpi\ pulsars with significant ($>3\,\sigma$) distance measurements (including 15 parallax-based distances and a globular cluster distance). {\bfseries Lower:} The probability density function (PDF) of $v_\perp$ for the 16 MSPs.
    The median of the $v_\perp$ PDF is marked with the dashed line, while the 1\,$\sigma$ error interval is shown with the shaded region.
    }    
    \label{fig:v_t}
\end{figure*}

In Figure~\ref{fig:v_t}, we illustrate the $v_\perp$ in relation to $|z|$ for 16 pulsars with precise distance estimates. Among the 16 pulsars, only \psro\ does not have a significant parallax-based distance. Nevertheless, its $v_\perp$ can be inferred by incorporating its proper motion with the astrometric information (i.e., distance and proper motion) of its host globular cluster (see Section~\ref{subsec:J1824}).
No clear correlation is revealed between $v_\perp$ and $|z|$, which reinforces our decision to treat all \mspsrpi\ pulsars across the $|z|\lesssim1$\,kpc regime equally. 
By concatenating the simulated $v_\perp$ chains, we acquired the PDF for the 16 MSPs (see Figure~\ref{fig:v_t}), which gives $v^\mathrm{(MSP)}_\perp=53^{+48}_{-37}$\,\kmps.

Amongst the \mspsrpi\ sources, \psreb\ is an obvious outlier, with a velocity of $\sim$300 km s$^{-1}$ that is 3$\sigma$ above the mean. As proposed by \citet{Bassa16} and \citet{Kaplan16}, \psreb\ is theorized to have been ejected from a dense stellar region, thus possibly following a different $v_\perp$ distribution from typical field MSPs (isolated along with their respective companions throughout their lives). In this regard, we turn our attention to the binary sample of pulsars with well determined orbital periods $P_\mathrm{b}$ (see $P_\mathrm{b}$ of Table~\ref{tab:astrometric_parameters}), and obtain $v_\perp^\mathrm{(BI)}=50^{+49}_{-34}$\,\kmps\ for field binary MSPs.
Based on this small sample, we do not find the $v_\perp$ of the three solitary MSPs (i.e., \psrb, \psrhb\ and \psrkb) to be inconsistent with $v_\perp^\mathrm{(BI)}$.  
Neither are the two DNSs (i.e., \psrfa\ and \psrfb).
If we exclude the two DNSs from the binary sample, we would come to $v_\perp^\mathrm{(WD)}=50^{+46}_{-31}$\,\kmps\ for the \mspsrpi\ pulsars with WD companions, which is highly consistent with  $v_\perp^\mathrm{(BI)}$ and $v^\mathrm{(MSP)}_\perp$.

Compared to $113\pm17$\,\kmps\ previously estimated for a sample of $\sim40$ MSPs \citep{Gonzalez11}, our $v_\perp^\mathrm{(MSP)}$ is largely consistent but on the smaller side.
\citet{Boodram22} recently shows that MSP space velocities have to be near zero to explain the Galactic Centre $\gamma$-ray excess \citep[e.g.][]{Abazajian12}.
Interestingly, the $v_\perp$ PDF based on our small sample of 16 shows a multi-modal feature, with the lowest mode consistent with zero.
Specifically, the 7 \mspsrpi\ pulsars with the smallest $v_\perp$ share an equally weighted mean $v_\perp$ of only 25\,\kmps, which suggests MSPs with extremely low space velocities are not uncommon. 
Accordingly, we suspect the MSP origin of the GC $\gamma$-ray excess can still not be ruled out based on our sample of $v_\perp$.


\section{Radial accelerations of pulsars and orbital-decay tests of gravitational theories}
\label{sec:orbital_decay_tests}

As described in Section~\ref{subsec:MSP_VLBI_astrometry}, 
VLBI astrometry of pulsars, in conjunction with pulsar timing, can enhance the orbital-decay tests of gravitational theories.
For binary systems involved in this work, the observed orbital decay has three significant components:
\begin{equation}
\label{eq:Pb_budget}
\begin{split}
    \dot{P}_\mathrm{b}^\mathrm{obs} = \dot{P}_\mathrm{b}^\mathrm{GW} + \dot{P}_\mathrm{b}^\mathrm{Shk} + \dot{P}_\mathrm{b}^\mathrm{Gal}\,,
\end{split}
\end{equation}
where $\dot{P}_\mathrm{b}^\mathrm{GW}$ reflects the effect of gravitational-wave damping intrinsic to a binary system, while $\dot{P}_\mathrm{b}^\mathrm{Shk}$ and $\dot{P}_\mathrm{b}^\mathrm{Gal}$ are both extrinsic contributions caused, respectively, by relative line-of-sight accelerations (of pulsars) $\mathcal{A}_\mathrm{Shk}$ and $\mathcal{A}_\mathrm{Gal}$. 
Specifically, $\dot{P}_\mathrm{b}^\mathrm{Shk}=\mathcal{A}_\mathrm{Shk}/c \cdot P_\mathrm{b} =\mu^2 D/c \cdot P_\mathrm{b}$ (where $\mu^2={\mu'_\alpha}^2+{\mu'_\delta}^2$) is the radial acceleration caused by the tangential motion of pulsars \citep{Shklovskii70}, which becomes increasingly crucial for pulsars with larger $\mu$ (e.g. \psrfb, \citealp{Ding21a}), as $\mathcal{A}_\mathrm{Shk} \propto \mu^2$. 
On the other hand, 
\begin{equation}
\label{eq:PbGal}
\begin{split}
    \dot{P}_\mathrm{b}^\mathrm{Gal} &= \frac{\mathcal{A}_\mathrm{Gal}}{c} P_\mathrm{b} 
    = \frac{\left[-\nabla \varphi\left(\vec{x}\right)\right]\big|^{\vec{x}_\mathrm{target}}_{\vec{x}_{\odot}} \cdot \vec{e}_{r}}{c} P_\mathrm{b}
\end{split}
\end{equation}
is a consequence of the gravitational pull (or push) exerted by the Galaxy.
Here, $\varphi(\vec{x})$ and $\vec{e}_{r}$ are, respectively, the Galactic gravitational potential (as a function of Galactic position $\vec{x}$) and the unit vector in the Earth-to-pulsar direction. 

In order to test any theoretical prediction of $\dot{P}_\mathrm{b}^\mathrm{GW}$, it is necessary to estimate $\mathcal{A}_\mathrm{Shk}$ and $\mathcal{A}_\mathrm{Gal}$ and remove their effect on $\dot{P}_\mathrm{b}^\mathrm{obs}$. Besides this impact, the radial accelerations $\mathcal{A}_\mathrm{Shk}$ and $\mathcal{A}_\mathrm{Gal}$ would, more generally, affect the time derivative of all periodicities intrinsic to a pulsar system, which include the pulsar spin period derivative $\dot{P}_\mathrm{s}$. 
Similar to $\dot{P}_\mathrm{b}^\mathrm{Shk}$ and $\dot{P}_\mathrm{b}^\mathrm{Gal}$, $\dot{P}_\mathrm{s}^\mathrm{Shk}=\mathcal{A}_\mathrm{Shk}/c \cdot P_\mathrm{s}$ and $\dot{P}_\mathrm{s}^\mathrm{Gal}=\mathcal{A}_\mathrm{Gal}/c \cdot P_\mathrm{s}$ (where ${P}_\mathrm{s}$ stands for the spin period of a pulsar).
As MSPs consist of nearly half of the $\gamma$-ray pulsar population, determining the extrinsic terms of $\dot{P}_\mathrm{s}$ and the intrinsic spin period derivative $\dot{P}_\mathrm{s}^\mathrm{int}=\dot{P}_\mathrm{s}^\mathrm{obs}-\dot{P}_\mathrm{s}^\mathrm{Shk}-\dot{P}_\mathrm{s}^\mathrm{Gal}$ is essential for exploring the ``death line'' (i.e., the lower limit) of high-energy emissions from pulsars \citep[e.g.][]{Guillemot16}. 
In Sections~\ref{subsec:Shk} and \ref{subsec:A_Gal}, we evaluate $\mathcal{A}_\mathrm{Shk}$ and $\mathcal{A}_\mathrm{Gal}$ one after another.
The evaluation only covers pulsars with significant $D$, as both $\mathcal{A}_\mathrm{Shk}$ and $\mathcal{A}_\mathrm{Gal}$ are distance-dependent.

\subsection{Shklovkii effects}
\label{subsec:Shk}
We estimate the model-independent $\mathcal{A}_\mathrm{Shk}$ in a way similar to the estimation of $v_\perp$ (see Section~\ref{subsec:v_t}). Three chains of 50,000 $\mu'_\alpha$, $\mu'_\delta$ and $D$ were simulated from their respective PDFs. Using the relation $\mathcal{A}_\mathrm{Shk} = \left({\mu'_\alpha}^2+{\mu'_\delta}^2\right) D$, 50,000 $\mathcal{A}_\mathrm{Shk}$ were calculated to assemble the PDF of $\mathcal{A}_\mathrm{Shk}$ for each pulsar with significant $D$. The $\mathcal{A}_\mathrm{Shk}$ inferred from the PDFs are compiled in Table~\ref{tab:Shk} along with their resultant $\dot{P}_\mathrm{s}^\mathrm{Shk}$ and $\dot{P}_\mathrm{b}^\mathrm{Shk}$.

\begingroup
\renewcommand{\arraystretch}{1.4} 

\begin{table*}
\raggedright
\caption{Extrinsic terms of $\dot{P}_\mathrm{s}$ and $\dot{P}_\mathrm{b}$ for 15 pulsars with significant $D$.}
\label{tab:Shk}
\resizebox{\textwidth}{!}{
\begin{tabular}{lcc:ccccc:ccccc}
\hline
\hline
PSR & $\mathcal{A}_\mathrm{Shk}$ & $\mathcal{A}_\mathrm{Gal}$ & $P_\mathrm{s}$ & $\dot{P}_\mathrm{s}^\mathrm{Shk}$ & $\dot{P}_\mathrm{s}^\mathrm{Gal}$ & $\dot{P}_\mathrm{s}^\mathrm{obs}$ & $\dot{P}_\mathrm{s}^\mathrm{int}$ & $\dot{P}_\mathrm{b}^\mathrm{Shk}$ & $\dot{P}_\mathrm{b}^\mathrm{Gal}$ & $\dot{P}_\mathrm{b}^\mathrm{obs}$ & $\dot{P}_\mathrm{b}^\mathrm{int}$ & $\dot{P}_\mathrm{b}^\mathrm{GW}$ \\
 & ($\mathrm{pm~s^{-2}}$) $^*$ & ($\mathrm{pm~s^{-2}}$)  &  (ms) & ($\mathrm{zs~s^{-1}}$) $^*$ & ($\mathrm{zs~s^{-1}}$) & ($\mathrm{zs~s^{-1}}$) & ($\mathrm{zs~s^{-1}}$) & ($\mathrm{fs~s^{-1}}$) & ($\mathrm{fs~s^{-1}}$) $^*$ & ($\mathrm{fs~s^{-1}}$) & ($\mathrm{fs~s^{-1}}$) & ($\mathrm{fs~s^{-1}}$)\\
\hline
\Psrb & 9.1(2) & -33.0(3.7) & 4.87 & 0.148(3) & -0.54(6) & 10.2 & 10.59(6) &  --- & --- & --- & --- & ---\\
\Psrc & $3.9^{+0.8}_{-0.6}\!\times\!10^2$ & -9(2) & 3.86 & $5.0^{+1.0}_{-0.7}$ & -0.12(3) & 12.3 $^a$ & 7.4(9) &  $32^{+6}_{-5}$ & -0.72(17) & $-70(30)$ $^a$ & -101(31) & $\sim-4.6$ $^a$ \\
\Psrd & $14^{+4}_{-3}$ & 23.8(4.5) & 28.85 & $1.3^{+0.4}_{-0.3}$ & 2.3(4) & 47.3 & 43.7(6) &  $33^{+11}_{-7}$ & 57(11) & ---& ---&---\\
\Psrea & $419^{+17}_{-15}$ & -23.5(2.4) & 5.26 & 7.3(3) & -0.41(4) & 17.1 & 10.2(3) &  73(3) & -4.1(4) & 61(4) & -7.9(5.0) & -13(1) $^b$ \\
\Psreb & $2.8(2)\!\times\!10^3$ & -40(3) & 5.16 & 48(3) & -0.69(5) & 18.6 & -29(3) $^{**}$ & --- & ---& --- & ---&---\\
\Psrfa & 43(1) & -48.5(3.2) & 40.93 & 5.9(2) & -6.6(4) & 27.2 & 27.9(5) &  107(3) & -120(8) & $2.4(2.2)\!\times\!10^2$ & $2.6(2.2)\!\times\!10^2$ & $\sim-1.2$ $^e$ \\
\Psrfb & $4.4(3)\!\times\!10^2$ & -42(3) & 37.90 & $55.6^{+4.0}_{-3.5}$ & -5.3(4) & 2422.5 & 2372(4) & $53.3^{+3.8}_{-3.3}$ & -5.1(4) & -136.6(3) & -185(4) & -192.45(6) $^c$ \\
\Psrga & $1.3(1)\!\times\!10^2$ & -48.5(4.3) & 3.16 & 1.4(1) & -0.51(5) & 2.8 & 1.9(1) & $6.8^{+0.6}_{-0.5}\!\times\!10^3$ & $-2.45(22)\!\times\!10^3$ & ---& --- &---\\
\Psrgb & $35^{+5}_{-4}$ & 1.0(1.7) & 4.62 & $0.53^{+0.08}_{-0.06}$ & $2(3)\!\times\!10^{-2}$ & 18.5 & 17.95(7) & $1.5(2)\!\times\!10^3$ & 41(72) & ---& ---&---\\
\Psrhb & $158^{+9}_{-8}$ & 11.7(9) & 8.12 & 4.3(2) & 0.32(2) & 20.2 & 15.6(2) & --- & ---& --- & --- &---\\
\Psri & $95^{+8}_{-7}$ & -6.2(1.6) & 5.85 & $1.86^{+0.16}_{-0.14}$ & -0.12(3) & 24.1 & 22.4(2) & $9.7^{+0.9}_{-0.7}$ & -0.64(16) & -17(3) & -26.1(3.1) & $-27.7^{+1.5}_{-1.9}$ $^d$\\
\Psrka & $17^{+3}_{-2}$ & -16(5) & 4.09 & $0.23^{+0.04}_{-0.03}$ & -0.22(7) & 8.7 & 8.69(8) &  $5.6^{+0.9}_{-0.7}\!\times\!10^2$ & $-5(2)\times10^2$ & --- & --- &---\\
\Psrl & $1.2^{+0.4}_{-0.2}\!\times\!10^2$ & -35(17) & 4.98 & $2.0^{+0.6}_{-0.4}$ & -0.6(3) & 9.7 & 8.3(6) & $2.0^{+0.6}_{-0.4}\!\times\!10^3$ & $-6(3)\times10^2$ & ---& --- &---\\
\Psrmb & $93^{+12}_{-9}$ & 6.8(1.5) & 7.65 & $2.4^{+0.3}_{-0.2}$ & 0.17(4) & 25.7 & 23.1(3) & $2.9^{+0.4}_{-0.3}\!\times\!10^2$ & 21(5) & ---& --- &---\\
\Psrkb & $0.37^{+0.04}_{-0.03}$ & -60(10) & 1.56 & $1.9(2)\!\times\!10^{-3}$ & -0.31(5) & 105.1 & 105.41(5) & --- & --- & --- & --- &---\\

\hline

\multicolumn{13}{l}{$\bullet$ Unless otherwise specified, we adopted $P_\mathrm{s}$, $\dot{P}^\mathrm{obs}_\mathrm{s}$, $P_\mathrm{b}$ and $\dot{P}^\mathrm{obs}_\mathrm{b}$ from the ATNF Pulsar Catalogue\textsuperscript{\ref{footnote:PSRCAT}}.}\\ 
\multicolumn{13}{l}{$\bullet$ Other references: $^a$\citet{van-der-Wateren22}; $^b$\citet{Ding20} and references therein; $^c$\citet{Ding21a} and references therein; $^d$\citet{Freire12}; $^e$\citet{Janssen08}.}\\
\multicolumn{13}{l}{$^{*}$ ``zs'' means zeptosecond, which is $10^{-21}$\,s. Besides, ``pm'' and ``fs'' represent $10^{-12}$\,m and $10^{-15}$\,s, respectively.}\\
\multicolumn{13}{l}{$^{**}$ The negative $\dot{P}_\mathrm{s}^\mathrm{int}$ is the result of the radial acceleration caused by a far-away companion \citep{Bassa16,Kaplan16}.}\\

\end{tabular}
}
\end{table*}
\endgroup

\subsection{Relative radial accelerations due to Galactic gravitational pull}
\label{subsec:A_Gal}
We estimate $\mathcal{A}_\mathrm{Gal}$  in the same way as \citet{Ding21a}, following the pioneering work of \citet{Zhu19}. 
To briefly demonstrate this method, we present, in Table~\ref{tab:A_Gal}, the $\mathcal{A}_\mathrm{Gal}$ based on five different $\varphi(\vec{x})$ models for the 15 pulsars with significant $D$ measurements. 
The five $\varphi(\vec{x})$ models are denoted as NT95 \citep{Nice95}, DB98 \citep{Dehnen98}, BT08 \citep{Binney11}, P14 \citep{Piffl14} and M17 \citep{McMillan17} in this paper.
The results obtained with NT95, which uses a simple analytical approach, are frequently discrepant compared to the other 4 $\varphi(\vec{x})$ models. Accordingly, and following \citet{Ding21a}, we exclude it and use the remaining 4 models to derive the estimate for $\mathcal{A}_\mathrm{Gal}$ and its uncertainty, which we present in Table~\ref{tab:Shk} (along with $\dot{P}_\mathrm{b}^\mathrm{Gal}$ and $\dot{P}_\mathrm{s}^\mathrm{Gal}$).

Incorporating the $\dot{P}_\mathrm{s}^\mathrm{Shk}$ derived in Section~\ref{subsec:Shk}, we calculated the intrinsic spin period derivative $\dot{P}_\mathrm{s}^\mathrm{int}=\dot{P}_\mathrm{s}^\mathrm{obs}-\dot{P}_\mathrm{s}^\mathrm{Shk}-\dot{P}_\mathrm{s}^\mathrm{Gal}$.
We note that the negative $\dot{P}_\mathrm{s}^\mathrm{int}$ of \psreb\ is probably the consequence of radial acceleration induced by a putative companion in an extremely wide orbit with \psreb\ (\citealp{Bassa16,Kaplan16}, also see Section~\ref{subsec:Gaia_results}).
In addition to $\dot{P}_\mathrm{s}^\mathrm{int}$, $\dot{P}_\mathrm{b}^\mathrm{int}=\dot{P}_\mathrm{b}^\mathrm{obs}-\dot{P}_\mathrm{b}^\mathrm{Shk}-\dot{P}_\mathrm{b}^\mathrm{Gal}$ are estimated for four pulsar systems with reported $\dot{P}_\mathrm{b}^\mathrm{obs}$. 
The improved \psri\ parallax as well as the re-assessed \psrea\ parallax calls for an update to the constraint on alternative theories of gravity \citep[e.g.][]{Freire12,Zhu19,Ding20}, which is discussed in Section~\ref{subsec:alternative_gravity}.

While performing the $\mathcal{A}_\mathrm{Gal}$ analysis, we found an error in the code that had been used to implement the calculation of Equation~\ref{eq:PbGal} for the \citet{Ding21a} work (which, to be clear, is not an error in the {\tt GalPot}\footnote{\label{footnote:galpot}\url{https://github.com/PaulMcMillan-Astro/GalPot}} package that provides the $\varphi(\vec{x})$ models). Therefore, we note that the $\dot{P}_\mathrm{b}^\mathrm{Gal}$ of \psrfb\ in Table~\ref{tab:Shk} is a correction to the \citet{Ding21a} counterpart. 
Further discussions on \psrfb\ can be found in Section~\ref{subsec:J1537}.

Last but not least, assuming GR is correct, the approach taken above can be inverted to infer
$\mathcal{A}_\mathrm{Gal}^\mathrm{(GR)}\!=\!(\dot{P}_\mathrm{b}^\mathrm{obs} - \dot{P}_\mathrm{b}^\mathrm{GW} - \dot{P}_\mathrm{b}^\mathrm{Shk})c/P_\mathrm{b}$, which can be used to constrain Galactic parameters for the local environment (of the Solar system) \citep{Chakrabarti21,Bovy20}, or probe the Galactic dark matter distribution in the long run \citep{Phillips21}. The  $\mathcal{A}_\mathrm{Gal}^\mathrm{(GR)}$ for the three viable pulsars are listed in Table~\ref{tab:A_Gal}. 


\begingroup
\renewcommand{\arraystretch}{1.4} 

\begin{table}
\raggedright
\caption{Radial accelerations due to Galactic gravitational pull based on different models of Galactic gravitational potential}
\label{tab:A_Gal}
\resizebox{\columnwidth}{!}{
\begin{tabular}{lcccccc}
\hline
\hline
PSR   & $\mathcal{A}_\mathrm{Gal}^\mathrm{(NT95)}$ & $\mathcal{A}_\mathrm{Gal}^\mathrm{(DB98)}$ & $\mathcal{A}_\mathrm{Gal}^\mathrm{(BT08)}$ & $\mathcal{A}_\mathrm{Gal}^\mathrm{(P14)}$ & $\mathcal{A}_\mathrm{Gal}^\mathrm{(M17)}$ & $\mathcal{A}_\mathrm{Gal}^\mathrm{(GR)}$ $^{*}$ \\
 & ($\mathrm{pm~s^{-2}}$) &  ($\mathrm{pm~s^{-2}}$) & ($\mathrm{pm~s^{-2}}$) & ($\mathrm{pm~s^{-2}}$) & ($\mathrm{pm~s^{-2}}$) & ($\mathrm{pm~s^{-2}}$)\\
\hline
\Psrb & -29(3) & $^!$-37.0(2) & $^!$-27.3(2) & -35.3(2) & -32.5(3) & --- \\
\Psrc & $^!$-12(1) & $-8.6^{+1.2}_{-0.8}$ & $^!$ $-6.0^{+1.0}_{-0.5}$ & $-10.9^{+0.7}_{-0.3}$ & $-8.8^{+1.0}_{-0.4}$ & --- \\
\Psrd & 24(4) & 23(5) & 24(5) & 24(5) & 25(5) & --- \\
\Psrea & $^{!!!}$-32.0(6) & -24.0(2) & $^!$-19.44(9) & -24.80(9) & -25.80(6) & 05(29) \\
\Psreb & $^!$-45.1(9) & -38.5(6) & $^!$-35.6(9) & -42(1) & -43(1) & --- \\
\Psrfa & -47.5(5) & -48.1(5) & $^!$-44.8(7) & -50.4(7) & $^!$-51.9(7) & --- \\
\Psrfb & $^{!!!!}$-29(1) & -42(1) & -39(2) & -43(2) & -45(2) & $21^{+28}_{-31}$ \\
\Psrga & $^{!!!}$-33(1) & -46(3) & -45(3) & -50(3) & -52(4) & --- \\
\Psrgb & $^{!!}$10(3) & $^!$ $-1.2^{+0.8}_{-0.6}$ & $^!$ $3.2^{+0.7}_{-0.5}$ & $0.6^{+0.7}_{-0.6}$ & $1.3^{+0.6}_{-0.4}$ & --- \\
\Psrhb & 13.2(8) & 10.8(6) & 12.1(6) & 11.5(6) & 12.5(7) & --- \\
\Psri  & $^{!!!!!!}$10.1(1.8) & $-6.4^{+0.4}_{-0.6}$ & $-4.2^{+0.8}_{-1.1}$ & $-7.5^{+0.7}_{-1.0}$ & $-6.9^{+0.8}_{-1.2}$ & 9(35) \\
\Psrka & -13(3) & $-13^{+3}_{-4}$ & $-13^{+3}_{-5}$ & $-19^{+4}_{-5}$ & $-16^{+4}_{-5}$ & --- \\
\Psrl  & -35(13) & $-29^{+8}_{-16}$ & $-31^{+10}_{-18}$ & $-42^{+12}_{-21}$ & $-36^{+10}_{-20}$ & --- \\
\Psrmb & $^{!!}$14(2) & 5.9(5) & $^!$8.7(5) & $^!$5.0(2) & 7.4(3) & --- \\
\Psrkb & -64(8) & -53(8) & -56(8) & -67(9) & -63(9) & --- \\

\hline

\multicolumn{7}{l}{$\bullet$ NT95, DB98, BT08, P14 and M17 refer to five different $\varphi(\vec{x})$ models (see Section~\ref{subsec:A_Gal} for}\\
\multicolumn{7}{l}{\ \ \ the references).}\\
\multicolumn{7}{l}{$\bullet$ The ``!''s indicate, in the same way as Table~\ref{tab:VLBI_timing_results}, the significance of the offset between the $\mathcal{A}_\mathrm{Gal}$}\\
\multicolumn{7}{l}{\ \ \ in Table~\ref{tab:Shk} and that of each $\varphi(\vec{x})$ model.}\\
\multicolumn{7}{l}{$^*$ $\mathcal{A}_\mathrm{Gal}^\mathrm{(GR)}\!=\!(\dot{P}_\mathrm{b}^\mathrm{obs} - \dot{P}_\mathrm{b}^\mathrm{GW} - \dot{P}_\mathrm{b}^\mathrm{Shk})c/P_\mathrm{b}$ is based on the assumption that GR is correct.}\\

\end{tabular}
}
\end{table}
\endgroup

\subsection{New constraints on alternative theories of gravity}
\label{subsec:alternative_gravity}
In the GR framework, the excess orbital decay $\dot{P}_\mathrm{b}^\mathrm{ex}=\dot{P}_\mathrm{b}^\mathrm{int}-\dot{P}_\mathrm{b}^\mathrm{GW}$ is expected to agree with zero. However, some alternative theories of gravity expect otherwise due to their predictions of non-zero dipole gravitational radiation and time-varying Newton's gravitational constant $G$. Both phenomena are prohibited by GR. Namely, in GR, the dipole gravitational radiation coupling constant $\kappa_D=0$, and $\dot{G}/G=0$. The large asymmetry of gravitational binding energy of pulsar-WD systems makes them ideal testbeds for dipole gravitational emissions \citep[e.g.][]{Eardley75}. 
In an effort to test (and possibly eliminate) alternative theories of gravity, increasingly tight constraints on $\kappa_D$ and $\dot{G}/G$ have been placed using multiple pulsar-WD systems \citep{Deller08,Freire12,Zhu19,Ding20}.

With the reassessed astrometric results of \psrea, the $\dot{P}_\mathrm{b}^\mathrm{ex}$ of \psrea\ changes from $10.6\pm6.1\,\mathrm{fs~s^{-1}}$ in \citet{Ding20} to $5.1\pm5.1\,\mathrm{fs~s^{-1}}$. This change is mainly caused by three reasons: {\bf 1)} priors are placed on the proper motion during inference in this work (but not in \citealp{Ding20}); {\bf 2)} a Bayesian framework is applied in this work (while \citealp{Ding20} reported bootstrap results); {\bf 3)} this work adopts PDF medians as the estimates (while \citealp{Ding20} used PDF modes). Though barely affecting this work (see Figure~\ref{fig:J1012_dist}), the choice between PDF mode and median makes a difference to \citet{Ding20} given that their parallax PDF is more skewed (see Figure~4 of \citealp{Ding20}).
After employing the new VLBI+timing distance, the $\dot{P}_\mathrm{b}^\mathrm{ex}$ of \psri\ has shifted from $2.0\pm3.7\,\mathrm{fs~s^{-1}}$ \citep{Freire12} to $1.6\pm3.5\,\mathrm{fs~s^{-1}}$. More discussions on \psri\ can be found in Section~\ref{subsec:J1738}.

With the new $\dot{P}_\mathrm{b}^\mathrm{ex}$ of \psrea\ and \psri, we updated the constraints on $\kappa_D$ and $\dot{G}/G$ in exactly the same way as \citet{Ding20}. The prerequisites of this inference are reproduced in Table~\ref{tab:PbdotEx}, where the two underlined $\dot{P}_\mathrm{b}^\mathrm{ex}$ are the only difference from the Table~6 of \citet{Ding20}.  
We obtained 
\begin{equation}
\label{eq:gdot_kD}
\begin{split}
    \dot{G}/G &= -1.6^{\,+5.3}_{\,-4.8}\times10^{-13}\,\mathrm{yr^{-1}}, \\
    \kappa_D &= -1.1^{\,+2.4}_{\,-0.9}\times10^{-4}.
\end{split}
\end{equation}
Compared to \citet{Ding20}, $\kappa_D$ becomes more consistent with zero, while the new uncertainties of $\kappa_D$ and $\dot{G}/G$ remain at the same level.

\begingroup
\renewcommand{\arraystretch}{1.4} 

\begin{table}
\raggedright
\caption{Excess orbital decay $\dot{P}_\mathrm{b}^\mathrm{ex}=\dot{P}_\mathrm{b}^\mathrm{obs}-\dot{P}_\mathrm{b}^\mathrm{Shk}-\dot{P}_\mathrm{b}^\mathrm{Gal}$ and other prerequisites for constraining $\dot{G}/G$ and $\kappa_D$}
\label{tab:PbdotEx}
\resizebox{\columnwidth}{!}{
\begin{tabular}{lccccc}
\hline
\hline
PSR   & $P_\mathrm{b}$ & $\dot{P}_\mathrm{b}^\mathrm{ex}$ & $m_\mathrm{p}$ & $m_\mathrm{c}$ & $q$  \\
 & (d) &  ($\mathrm{fs~s^{-1}}$) & (\msun) & (\msun) &   \\
\hline
\Psrna & 5.74 & 12(32) & 1.44(7) & 0.224(7) & --- \\
\Psrea & 0.60 & \underline{5.1(5.1)} & --- & 0.174(11) & 10.44(11) \\
\Psrnb & 67.83 & 30(150) & 1.33(10) & 0.290(11) & --- \\
\Psri & 0.35 & \underline{1.6(3.5)} & 1.46(6) & --- & 8.1(2) \\

\hline

\multicolumn{6}{l}{$\bullet$ $m_\mathrm{p}$, $m_\mathrm{c}$ and $q$ stand for, respectively, pulsar mass, companion mass and}\\
\multicolumn{6}{l}{\ \ \ mass ratio (i.e., $m_\mathrm{p}/m_\mathrm{c}$).
See \citet{Ding20} for their references.}\\

\end{tabular}
}
\end{table}
\endgroup


\section{Individual pulsars}
\label{sec:individual_pulsars}

In this section, we discuss the impacts of the new astrometric measurements (particularly the new distances) on the scientific studies around individual pulsars. Accordingly, special attention is paid to the cases where there is no published timing parallax $\varpi^\mathrm{(Ti)}$.
In addition, we also look into the two pulsars (i.e. \psrha\ and \psro) that have $\varpi'$ consistent with zero, in an effort to understand the causes of parallax non-detection.


\subsection{\psrc}
\label{subsec:J0610}
\psrc\ is the third black widow pulsar discovered \citep{Burgay06}, which is in a 7-hr orbit with an extremely low-mass ($\approx0.02$\,\msun, \citealp{Pallanca12}) star.
Adopting a distance of around 2.2\,kpc, \citet{van-der-Wateren22} obtained a $\gamma$-ray emission efficiency $\eta_\gamma \equiv 4 \pi F_\mathrm{\gamma} D^2 / \dot{E}^\mathrm{int}$ in the range of 0.5--3.7, where $\dot{E}^\mathrm{int}$ and $F_\mathrm{\gamma}$ are, respectively, the intrinsic NS spin-down power and the $\gamma$-ray flux above 100\,MeV.

In addition, \citet{van-der-Wateren22} estimated a mass function
\begin{equation}
\label{eq:mass_func}
f(m_\mathrm{p},q)=m_\mathrm{p} \frac{\sin^3{i}}{q(q+1)^2}=\frac{4\pi^2 a_1^3}{G P_\mathrm{b}^2}
\end{equation}
of $5.2\times10^{-6}$\,\msun\ for the \psrc\ system (where $q \equiv m_\mathrm{p}/m_\mathrm{c}$). Besides, they determined the irradiation temperature (of the companion) $T_\mathrm{irr}=2820\pm190$\,K as well as the projected orbital semi-major axis $a_1=7.3\times10^{-3}$\,lt-s. 
Combining these three estimates, we calculated the heating luminosity
\begin{equation}
\label{eq:L_irr}
\begin{split}
   L_\mathrm{irr} &\equiv 4 \pi \left[\frac{a_1 (1+q)}{\sin{i}}\right]^2 \sigma_\mathrm{SB} T_\mathrm{irr}^4 \\
   &\approx 4 \pi a_1^2 \left[ \frac{m_\mathrm{p}}{f(m_\mathrm{p},q)} \right]^{2/3} \sigma_\mathrm{SB} T_\mathrm{irr}^4 \\
   &\sim  9.1\times10^{32} \left( \frac{m_\mathrm{p}}{1.4\,\msun}\right)^{2/3} \mathrm{erg~s^{-1}}, 
\end{split}
\end{equation}
where $\sigma_\mathrm{SB}$ represents the Stefan-Boltzmann constant.

Our new distance $D=1.5^{+0.3}_{-0.2}$\,kpc to \psrc\ is less than half the DM-based distances (see Table~\ref{tab:VLBI_timing_results}), and significantly below that assumed by \citet{van-der-Wateren22}. 
Assuming a NS moment of inertia $I_\mathrm{NS}=10^{45}\,\mathrm{g~cm^2}$,
the $\dot{P}_\mathrm{s}^\mathrm{int}$ of \psrc\ (see Table~\ref{tab:Shk}) corresponds to an intrinsic spin-down power 
\begin{equation}
\label{eq:Edot}
    \dot{E}^\mathrm{int} \equiv 4 \pi^2 I_\mathrm{NS} \dot{P}_\mathrm{s}^\mathrm{int}/P_\mathrm{s}^3
\end{equation}
of $(5.1\pm0.5)\times10^{33}\,\mathrm{erg~s^{-1}}$, which is roughly twice as large as the $\dot{E}^\mathrm{int}$ range calculated by \citet{van-der-Wateren22}.
In conjunction with a smaller $\gamma$-ray luminosity $L_\gamma=4 \pi F_\mathrm{\gamma} D^2$ (due to closer distance), the $\dot{E}^\mathrm{int}$ reduced $\eta_\gamma$ to around 0.37 (from $0.5 < \eta_\gamma < 3.7$ estimated by \citealp{van-der-Wateren22}), disfavoring unusually high $\gamma$-ray beaming towards us. 
Moreover, the heating efficiency $\epsilon_T$ drops to $\sim0.17$ (from $0.15<\epsilon_T<0.77$ evaluated by \citealp{van-der-Wateren22}), disfavoring the scenario where the NS radiation is strongly beamed towards the companion.





\subsubsection{On the DM discrepancy}
\label{subsubsec:J0610_DM}
In Section~\ref{subsubsec:DM_distances}, we noted that our VLBI parallax-derived distance and the DM model-inferred distance to this pulsar differed 
substantially.
Specifically, PSR~J0610$-$2100 has a measured $\mathrm{DM} = 60.7$~pc~cm${}^{-3}$ while the NE2001 model predicts 27.5~pc~cm${}^{-3}$ for a line of sight of length 1.5\,kpc.  We attribute this discrepancy to thermal plasma or ``free electrons’’ along the line of sight that is not captured fully by a ``clump’’ in the NE2001 model.  The NE2001 model includes this ``clump’’ to describe the effects due to the Mon~R2 \ion{H}{2} region, centered at a Galactic longitude and latitude of (214\degr, -12.6\degr), 
located at an approximate distance of $\sim0.9$\,kpc \citep{Herbst75}.  However, the WHAM survey shows considerable H$\alpha$ in this direction, extending over tens of degrees.  Lines of sight close to the pulsar show changes in the H$\alpha$ intensity by factors of two, but an approximate value toward the pulsar is roughly 13~Rayleighs, equivalent to an emission measure $\mathrm{EM} = 29$~pc~cm${}^{-6}$ (for a temperature $T = 8000$~K).  Using standard expressions, as provided in the NE2001 model, to convert EM to \hbox{DM}, there is sufficient H$\alpha$ intensity along the line of sight to account for the excess DM that we infer from the difference between our parallax-derived distance and the NE2001 model distance.

\subsection{\psrfa}
\label{subsec:J1518}
The 41-ms \psrfa, discovered by \citet{Sayer97}, is one of the only two DNSs in the current sample.
According to \citet{Janssen08}, the non-detection of Shapiro delay effects suggests $\sin{i} \leq 0.73$ at 99\% confidence level. 
Accordingly, we adopted 0.73 as the upper limit of $\sin{i}$, and carried out 8-parameter Bayesian inference, which led to a bi-modal posterior PDF on $i'$ and a multi-modal PDF on $\Omega'_\mathrm{asc}$ (see the online corner plot\textsuperscript{\ref{footnote:pulsar_positions}}). The predominant constraints on both $i'$ and $\Omega'_\mathrm{asc}$ come from the $\dot{a}_1$ measurement (\citealp{Janssen08} or see Table~\ref{tab:7_parameter_inference}). 
Though there are 3 major likelihood peaks for the $\Omega'_\mathrm{asc}$, two of them gather around 171\degr, making the PDF relatively more concentrated.
When a much more precise $\dot{a}_1$ measurement is reached with new timing observations, the existing VLBI data will likely place useful constraints on $i'$ and $\Omega'_\mathrm{asc}$. So will additional VLBI observations.

In addition to $i'$ and $\Omega'_\mathrm{asc}$, the 8-parameter Bayesian inference also renders a $40\,\sigma$ parallax $\varpi'$, which becomes the most significant parallax achieved for a DNS. 
In contrast, to detect a timing parallax $\varpi^\mathrm{(Ti)}$ for \psrfa\ would take $\gtrsim600$ years \citep{Janssen08}, due to its relatively high ecliptic latitude of 63\degr.

\subsection{\psrfb}
\label{subsec:J1537}
\psrfb, also known as PSR~B1534$+$12, is the second discovered DNS \citep{Wolszczan91}. The DNS displays an exceptionally high proper motion amongst all Galactic DNSs (see Table~3 of \citealp{Tauris17}), leading to an unusually large Shklovskii contribution to observed timing parameters. Therefore, precise astrometry of the DNS plays an essential role in its orbital decay test of GR.
The most precise astrometric parameters of \psrfb\ are provided by \citet{Ding21a} based on the same dataset used in this work, which result in $\dot{P}_\mathrm{b}^\mathrm{Shk}=53\pm4$\,\fsps. 
Subsequently, \citet{Ding21a} estimated $\dot{P}_\mathrm{b}^\mathrm{Gal}=-1.9\pm0.2$\,\fsps, and concluded $\dot{P}_\mathrm{b}^\mathrm{int}/\dot{P}_\mathrm{b}^\mathrm{GW}=0.977\pm0.020$.

In this work, we inferred $\eta_\mathrm{EFAC}$ on top of the canonical astrometric parameters, which is the only difference from the Bayesian method of \citet{Ding21a}. Despite this difference, our astrometric results of \psrfb\ remain almost the same as \citet{Ding21a}. So is our re-derived $\dot{P}_\mathrm{b}^\mathrm{Shk}=53.3^{+3.8}_{-3.3}$\,\fsps. 
However, as is mentioned in Section~\ref{subsec:A_Gal}, the $\dot{P}_\mathrm{b}^\mathrm{Gal}$ estimated by \citet{Ding20c} is incorrect due to a coding error. After correction, $\dot{P}_\mathrm{b}^\mathrm{Gal}$ drops to $-5.1\pm0.4$\,\fsps\ (see Table~\ref{tab:Shk}). Consequently, we obtained $\dot{P}_\mathrm{b}^\mathrm{int}/\dot{P}_\mathrm{b}^\mathrm{GW}=0.96\pm0.02$. 

As \citet{Ding21a} have pointed out, the limiting factor of the GR orbital decay test using \psrfb\ remains the distance precision, which generally improves as $t^{-1/2}$ with additional observations, but can be accelerated if more sensitive instrumentation can be deployed.
On the other hand, the extremely high braking index of 157 (two orders higher than the normal level) calculated from the rotational frequency $\nu_\mathrm{s} \equiv 1/P_\mathrm{s}$, its first derivative $\dot{\nu}_\mathrm{s}$ and its second derivative $\ddot{\nu}_\mathrm{s}$ \citep{Fonseca14}  indicate likely timing noise contributions that may affect the observed orbital period derivative to some degree.
This will be clarified with continued timing observations and refined timing analysis.

\subsection{\psrga}
\label{subsec:J1640}

\psrga\ is a 3.2-ms MSP \citep{Foster95} in a wide ($P_\mathrm{b}=175$\,d) orbit with a WD companion \citep{Lundgren96}. 
The \mspsrpi\ results for \psrga\ have been determined using bootstrap and published in \citet{Vigeland18}, which are highly consistent with our re-assessed quasi-VLBI-only results (see Table~2 of \citealp{Vigeland18} and Table~\ref{tab:models_no_pm_prior}), and also agree with the VLBI+timing ones (see Table~\ref{tab:VLBI_timing_results}).
Our 8-parameter Bayesian inference renders 
a 1D histogram of $\Omega'_\mathrm{asc}$ with 4 likelihood components at 0\degr, 140\degr, 200\degr\ and 320\degr, which is predominantly shaped by the prior on $\dot{a}_1$ from pulsar timing (see Table~\ref{tab:7_parameter_inference}).

\subsection{\psrgb}
\label{subsec:J1643}

\psrgb\ is a 4.6-ms pulsar in a 147-d orbit with a WD companion \citep{Lorimer95a}. 
As a result of multi-path propagation due to inhomogeneities in the ionised interstellar medium (IISM), the pulse profiles of \psrgb\ are temporally broadened \citep[e.g.][]{Lentati17}.   
As the Earth-to-pulsar sightline moves through inhomogeneous scattering ``screen(s)'' (in the IISM), the temporal broadening $\tau_\mathrm{sc}$ varies with time; at 1\,GHz, $\tau_\mathrm{sc}$ fluctuates up and down by $\lesssim 5\,\mu$s on a yearly time scale \citep{Lentati17}.
Meanwhile, the moving scattering ``screen(s)'' would also change the radio brightness of the pulsar. This effect, as known as pulsar scintillation, is used to constrain the properties of both \psrgb\ and the scattering screen(s) between the Earth and the pulsar \citep{Mall22}. The scintillation of \psrgb\ has previously been modelled with both isotropic and anisotropic screens \citep{Mall22}. The isotropic model renders a pulsar distance  $D=1.0\pm0.3$\,kpc and locates the main scattering screen at $D_\mathrm{sc}=0.21\pm0.02$\,kpc; in comparison, the anisotropic model yields a pulsar distance $D=1.2\pm0.3$\,kpc, and necessitates a secondary scattering screen $0.34\pm0.09$\,kpc away (from the Earth) in addition to a main scattering screen at $0.13\pm0.02$\,kpc distance \citep{Mall22}.
On the other hand, the HII region Sh~2-27 in front of \psrgb\ is suspected to be associated with the main scattering screen of the pulsar. 
This postulated association is strengthened by the agreement between the distance to the main scattering screen (based on the two-screen anisotropic model, \citealp{Mall22}) and the distance to the HII region (i.e., $112\pm3$\,pc, \citealp{Ocker20}).

\subsubsection{Independent check on the postulated association between the HII region Sh~2-27 and the main scattering screen}
\label{subsubsec:HII_region}

Besides the pulse broadening effect, the scattering by the IISM would lead to apparent angular broadening of the pulsar, which has been detected with VLBI at $\gtrsim8$\,GHz \citep[e.g.][]{Bower14}.
By the method outlined in Appendix~A of \citet{Ding20c}, we measured a semi-angular-broadened size $\theta_\mathrm{sc}=3.65\pm0.43$\,mas for \psrgb, which is the only significant $\theta_\mathrm{sc}$ determination in the \mspsrpi\ catalogue. 
Likewise, the secondary in-beam calibrator of \psrgb\ is also scatter-broadened, which may likely introduce additional astrometric uncertainties (see more explanation in Section~\ref{subsec:J1721}).

As both pulse broadening and angular broadening are caused by the IISM deflection, $\theta_\mathrm{sc}$, $\tau_\mathrm{sc}$, the pulsar distance $D$, and the distance(s) $D_\mathrm{sc}$ to the scattering screen(s) are geometrically related. Assuming there is one dominant thin scattering screen, we make use of following approximate relation 
\begin{equation}
\label{eq:tau_sc}
\begin{split}
   \frac{\theta_\mathrm{sc}^2}{2c \tau_\mathrm{sc}} = \frac{1}{D_\mathrm{sc}} - \frac{1}{D}\,\,\,\,\,\,\,\,\,\, (\mathrm{when}\,\, \theta_\mathrm{sc} \lesssim 1\degr),
\end{split}
\end{equation}
where $c$ stands for the speed of light in vacuum.

\begin{figure}
    \centering
	\includegraphics[width=9cm]{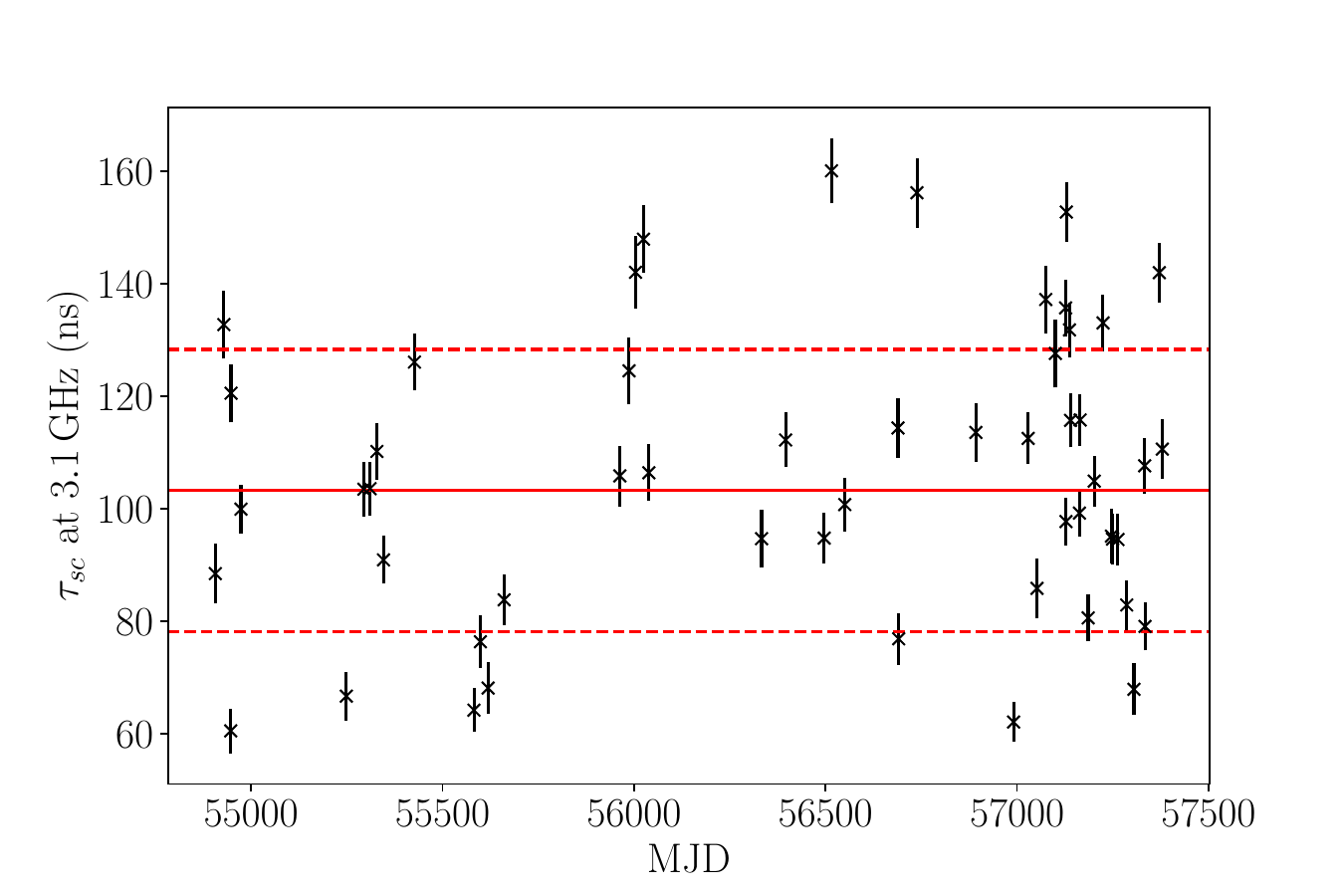}
    \caption{Temporal broadening $\tau_\mathrm{sc}$ of \psrgb\ at 3.1\,GHz. The solid red line and the dashed red line show the mean temporal broadening and a 68\% confidence interval, respectively.
    }    
    \label{fig:tau_sc}
\end{figure}

To estimate the unknown $\tau_\mathrm{sc}$ at our observing frequency of $\sim$1.55\,GHz, we used the data spanning MJD~54900---57500 from the PPTA second data release \citep{Kerr20}. 
We analysed the dynamic spectra of observations centred around 3.1\,GHz and recorded with the PDFB4 processor, using the scintools\footnote{\url{https://github.com/danielreardon/scintools}} package \citep{Reardon20}. A model was fit to the auto-correlation function of each dynamic spectrum, which has an exponential decay with frequency \citep{Reardon19}. The characteristic scintillation scale (in frequency) $\Delta\nu_d$ is related to the scattering timescale with $\tau_{sc} = 1/(2\pi\Delta\nu_{d})$. 
We found the mean temporal broadening $\tau_\mathrm{sc}=103\pm25$\,ns at 3.1\,GHz, with fluctuations of $\lesssim60$\,ns (see Figure~\ref{fig:tau_sc}).
To convert this $\tau_\mathrm{sc}$ to our observing frequency 1.55\,GHz, we compare the maximum degree (i.e., 60\,ns) of fluctuations at 3.1\,GHz to that (i.e., $5\,\mu$s, \citealp{Lentati17}) at 1\,GHz, and acquired an indicative scaling relation 
\begin{equation}
    \label{eq:tau_sc_scaling}
    \tau_\mathrm{sc} \propto \nu^{-3.9},
\end{equation}
where $\nu$ is the observing frequency. This relation reasonably agrees with the scaling relation $\tau_\mathrm{sc} \propto \nu^{-11/3}$ associated with the Kolmogorov turbulence \citep[e.g.][]{Armstrong95}. 
With the indicative scaling relation, we calculated $\tau_\mathrm{sc}=1.54\pm0.37\,\mu$s.
It is timely to note that $\theta_\mathrm{sc}^2 / \tau_\mathrm{sc}$ (on the left side of Equation~\ref{eq:tau_sc}) is frequency-independent. By combining Equations~\ref{eq:tau_sc} and \ref{eq:tau_sc_scaling}, we come to another equivalent indicative scaling relation
\begin{equation}
\label{eq:theta_scaling}
    \theta_\mathrm{sc}\propto\nu^{-1.95}.
\end{equation}

Substituting $\tau_\mathrm{sc}=1.54\pm0.37\,\mu$s, $\theta_\mathrm{sc}=3.65\pm0.43$\,mas and $D=0.95^{+0.15}_{-0.11}$\,kpc into Equation~\ref{eq:tau_sc}, we obtained $D_\mathrm{sc}=86^{+30}_{-24}$\,pc, where the uncertainty is derived with a Monte-Carlo simulation. This $D_\mathrm{sc}$ is consistent with the distance to the HII region Sh~2-27 \citep{Ocker20}, hence independently supporting the association between the HII region and the main scattering screen of \psrgb.

\subsubsection{Probing scintillation models}
\label{subsubsec:scintillation_models}

Apart from the above check on the connection between the HII region Sh~2-27 and the main scattering screen, the angular broadening of \psrgb\ also promises a test of the aforementioned isotropic scintillation model proposed by \citet{Mall22}. 
To do so, we changed the pulsar distance to the one inferred with the model (i.e., $1.0\pm0.3$\,kpc). 
With this change, we derive $D_\mathrm{sc}=86^{+30}_{-24}$\,pc, which disagrees with $0.21\pm0.02$\,kpc based on the isotropic model. 
%
To investigate the impact of a different scaling relation $\tau_\mathrm{sc} \propto \nu^{-\alpha_\mathrm{sc}}$, we inferred $\tau_\mathrm{sc}=4.3\,\mu$s with both $D$ and $D_\mathrm{sc}$ based on the isotropic model \citep{Mall22}, which corresponds to an unreasonably large $\alpha_\mathrm{sc}=5.4$. 
Hence, we conclude that our $\theta_\mathrm{sc}$ and $\tau_\mathrm{sc}$ cannot easily reconcile with the one-screen isotropic model proposed by \citet{Mall22}.

Fundamentally, the irreconcilability implies a one-screen model might be incapable of describing both scintillation and angular broadening effects of \psrgb.
In principle, it is possible to analyse a multi-screen model with a $\theta_\mathrm{sc}(t)$ series (at various time $t$) and its associated $\tau_\mathrm{sc}(t)$, instead of only using their mean values. 
However, this analysis is not feasible for this work, as $\tau_\mathrm{sc}$ and $\theta_\mathrm{sc}$ were not measured on the same days.
Nonetheless, we can still investigate whether our observations of \psrgb\ can reconcile with the scintillation observations \citep{Mall22} in the context of a two-screen model.

In the scenario of two thin scattering screens, we derived the more complicated relation
\begin{equation}
\label{eq:tau_sc2}
   \begin{cases}
   2c\tau_\mathrm{sc} &= k_1 \beta_\mathrm{sc}^2+k_2\beta_\mathrm{sc}\theta_\mathrm{sc}+k_3\theta_\mathrm{sc}^2\,\,\,\,\,\,\,\, (\theta_\mathrm{sc} \lesssim 1\degr \text{ and } \beta_\mathrm{sc} \lesssim 1\degr)\\[5pt]
   k_1 &= \frac{(D-D_\mathrm{sc2})(D-D_\mathrm{sc1})}{D_\mathrm{sc2}-D_\mathrm{sc1}} \\[5pt]
   k_2 &= -\frac{2D_\mathrm{sc1}(D-D_\mathrm{sc2})}{D_\mathrm{sc2}-D_\mathrm{sc1}}\\[5pt]
   k_3 &= \frac{D_\mathrm{sc1} D_\mathrm{sc2}}{D_\mathrm{sc2}-D_\mathrm{sc1}},
   \end{cases}
\end{equation}
where $D_\mathrm{sc1}$ and $D_\mathrm{sc2}$ are the distance to the first and the second scattering screen, respectively; $\beta_\mathrm{sc}$ represents a half of the opening angle of the second scattering screen (closer to the pulsar) as seen from the pulsar.
As a side note, Equation~\ref{eq:tau_sc} can be considered a special case (i.e., $D=D_\mathrm{sc2}$) of Equation~\ref{eq:tau_sc2}.
In Equation~\ref{eq:tau_sc2}, all parameters except $\beta_\mathrm{sc}$ are known, either determined with the anisotropic two-screen model or obtained in this work. Hence, it is feasible to constrain the geometric parameter $\beta_\mathrm{sc}$ with the known parameters.

However, Equation~\ref{eq:tau_sc2} can yield unphysical solutions (i.e., $\beta_\mathrm{sc}>0$). We applied the simple condition 
\begin{equation}
\label{eq:two_screen_condition1}
\begin{split}
    \frac{\theta_\mathrm{sc}^2}{2c \tau_\mathrm{sc}} \leq \frac{1}{D_\mathrm{sc1}} - \frac{1}{D} 
\end{split}
\end{equation}
to ensure Equation~\ref{eq:tau_sc2} gives physical solutions of $\beta_\mathrm{sc}$. 
This equation is equivalent to $D_\mathrm{sc1} \leq D_\mathrm{sc}$, 
where $D_\mathrm{sc}$ corresponds to the one-screen solution of Equation~\ref{eq:tau_sc}. This is because $D_\mathrm{sc1} > D_\mathrm{sc}$ would always lead to longer routes, thus exceeding the $\tau_\mathrm{sc}$ budget. It is important to note that Equation~\ref{eq:two_screen_condition1} is valid for a model with any number of scattering screens. Hence, we recommend to use Equation~\ref{eq:two_screen_condition1} in scintillation model inferences as a prior condition, to cater for the constraints imposed by $\theta_\mathrm{sc}$ and $\tau_\mathrm{sc}$ (and thereby truncate the parameter space of a scintillation model).



To test the anisotropic two-screen model \citep{Mall22} with our $\theta_\mathrm{sc}$ and $\tau_\mathrm{sc}$, we calculated $D_\mathrm{sc}=89^{+33}_{-26}$\,pc with the pulsar distance (i.e., $D=1.2\pm0.3$\,kpc) based on the anisotropic two-screen model. This $D_\mathrm{sc}$ is consistent with $D_\mathrm{sc1}=129\pm15$\,pc \citep{Mall22} (therefore not ruling out $D_\mathrm{sc1} < D_\mathrm{sc}$). That is to say, our $\theta_\mathrm{sc}$ and $\tau_\mathrm{sc}$ measurements can loosely reconcile with the anisotropic two-screen model \citep{Mall22}.
In comparison, we reiterate our finding that a one-screen model is difficult to describe both scintillation and angular broadening effects of \psrgb.


\subsection{\psrha}
\label{subsec:J1721}
\psrha\ is a 3.5-ms solitary MSP discovered at intermediate Galactic latitudes \citep{Edwards01}.
The main secondary phase calibrator of \psrha\ (and indeed, all the sources near it on the plane of the sky) is heavily resolved due to IISM scattering, which leads to non-detections on the longest baselines and a lack of calibration solutions for some antennas, reducing the spatial resolution of the VLBI observations. The non-uniform IISM distribution also leads to refractive image wander as the line-of-sight to the pulsar changes \citep[e.g.,][]{Kramer21a}, which is most pronounced for heavily scatter-broadened sources such as the calibrator for \psrha. In conjunction with the lower spatial resolutions, which reduces positional precision to begin with, this additional noise term likely results in the parallax non-detection (see Table~\ref{tab:models_no_pm_prior}).

\subsection{\psrhb}
\label{subsec:J1730}
\psrhb\ is a solitary pulsar spinning at $P_\mathrm{s}=8.1$\,ms \citep{Lorimer95a}. Being so far the least-energetic (in terms of $\dot{E}^\mathrm{int}$) $\gamma$-ray pulsar \citep{Guillemot16}, the pulsar plays a key role in exploring the death line of NS high-energy radiation. 
Substituting $\dot{P}_\mathrm{s}^\mathrm{int}$ and $P_\mathrm{s}$ of Equation~\ref{eq:Edot} with values listed in Table~\ref{tab:Shk}, we substantially refined the $\dot{E}^\mathrm{int}$ death line (of all $\gamma$-ray-emitting pulsars) to 
\begin{equation}
    \dot{E}_\mathrm{death} \leq \dot{E}^\mathrm{int}_\mathrm{J1730} = (1.15\pm0.01)\times10^{33}\left(\frac{I_\mathrm{NS}}{10^{45}~\mathrm{g~cm^2}}\right)\,\mathrm{erg~s^{-1}},
\end{equation}
which is consistent with (but on the higher side of) the previous estimate $(8.4\pm2.2)\times10^{32}\,\mathrm{erg~s^{-1}}$ by \citet{Guillemot16} (assuming the same $I_\mathrm{NS}$).
On the other hand, we updated the $\gamma$-ray luminosity (above 100\,MeV) to $L_\gamma=4 \pi D^2 F_\mathrm{\gamma}=(3.1\pm1.6)\times10^{32}\,\mathrm{erg~s^{-1}}$, where the precision is limited by the less precise $F_\gamma$ \citep{Guillemot16}. Accordingly, we obtained $\eta_\gamma = 0.27\pm0.14$.

\subsection{\psri}
\label{subsec:J1738}
\psri, discovered from a 1.4-GHz high-Galactic-latitude survey with the 64-m Parkes radio telescope \citep{Jacoby09}, is a 5.85-ms pulsar in a 8.5-hr orbit with a WD companion. Thanks to the relatively short $P_\mathrm{b}$, the WD-pulsar system plays a leading role in constraining alternative gravitational theories that predict dipole gravitational radiation \citep{Freire12,Zhu15}.

Our VLBI-only $\varpi$ is $2.3\,\sigma$ away from the most precise $\varpi^\mathrm{(Ti)}$ (see Table~\ref{tab:models_no_pm_prior} and \ref{tab:VLBI_timing_results}). After adopting timing priors, $\varpi'=0.589\pm0.046$\,mas becomes closer to $\varpi^\mathrm{(Ti)}=0.68\pm0.05$\,mas \citep{Freire12}, meaning that $\dot{P}^\mathrm{Shk}_\mathrm{b}$ is only 1.2 times larger than the previous estimate.
On the other hand, our re-assessed $\dot{P}^\mathrm{Gal}_\mathrm{b}$, based on more realistic $\varphi(\vec{x})$ models, is smaller than that estimated with the NT95 $\varphi(\vec{x})$ model \citep{Freire12}.
Combining the unchanged  $\dot{P}^\mathrm{obs}_\mathrm{b}=-17\pm3$\,\fsps\, the re-derived  $\dot{P}^\mathrm{Int}_\mathrm{b}=-26.1\pm3.1$\,\fsps\ is almost the same as the previous estimate, as the change of $\dot{P}^\mathrm{Gal}_\mathrm{b}$ happens to nearly cancels out that of $\dot{P}^\mathrm{Shk}_\mathrm{b}$.

Future pulsar timing or VLBI investigation into the discrepancy between $\varpi^\mathrm{(Ti)}$ \citep{Freire12} and $\varpi$ is merited by the importance of the pulsar system. Specifically, if the true parallax turns out to be around 0.5\,mas, it would not only mean that $\dot{P}^\mathrm{Shk}_\mathrm{b}$ is 1.4 times higher than the estimate by \citet{Freire12}, but also suggest the WD radius $R_\mathrm{WD}$ to be 1.4 larger (as $R_\mathrm{WD} \propto D$ according to Equation~1 of \citealp{Antoniadis12}). A higher $R_\mathrm{WD}$ would lead to  lighter WD and NS (as the mass ratio is well determined), thus smaller $\dot{P}^\mathrm{GW}$.


\subsection{\psro}
\label{subsec:J1824}
\psro\ is a 3-ms solitary pulsar discovered in the Globular cluster M28 (NGC~6626) \citep{Lyne87a}. 
The calibration configuration for this pulsar was sub-optimal, as the best in-beam phase calibrator for \psro\ was both resolved and faint (3.3\,mJy, see Table~\ref{tab:MSPs}), leading to noisy solutions, especially on the longest baselines.  This is likely responsible for the parallax non-detection (see Table~\ref{tab:models_no_pm_prior}), and indicates that higher sensitivity to enable improved calibration solutions would be advantageous in any future VLBI campaign.

The proper motion of M28 is estimated to be $\mu_\alpha^\mathrm{M28}=-0.28\pm0.03$\,\maspy\ and $\mu_\delta^\mathrm{M28}=-8.92\pm0.03$\,\maspy\ \citep{Vasiliev21} with Gaia Early Data Release 3 (EDR3). Hence, the relative proper motion of \psro\ with respect to M28 is $\Delta\mu_\alpha=0.03\pm0.05$\,\maspy\ and $\Delta\mu_\delta=1.1\pm0.8$\,\maspy.
Combining the M28 distance $D=5.4\pm0.1$\,kpc estimated by \citet{Baumgardt21}, we obtained the transverse space velocity $v_\perp=28\pm20$\,\kmps\ for \psro, which is smaller than the typical escape velocity ($\approx50$\,\kmps) of a globular cluster. Therefore, the pulsar is probably (as expected) bound to M28.






\section{Summary and Future prospects}
\label{sec:summary}

In this \mspsrpi\ release paper, we have presented VLBI astrometric results for 18 MSPs, including a re-analysis of three previously published sources.  From the 18 sources, we detect significant parallaxes for all but three.
For each \mspsrpi\ pulsar, at least one self-calibratable in-beam calibrator was identified to serve as the reference source of relative astrometry. In three cases, 1D interpolation, a more complex observing and data reduction strategy, is adopted to further suppress propagation-related systematic errors. Among the three pulsars, \psrkb\ is the brightest MSP in the northern hemisphere. Hence, we took one step further to perform inverse-referenced 1D interpolation using \psrkb\ as the in-beam calibrator. 
Compared to the pioneering Multi-View study of \citet{Rioja17} at 1.6\,GHz, the larger number of observations and targets here provides more opportunities to characterise the interpolation performance, which is crucial for ultra-precise astrometric calibration schemes proposed for future VLBI arrays incorporating the Square Kilometre Array (SKA).
Based on a small sample of three, we found that $\eta_\mathrm{EFAC}$ has consistently inflated after applying 1D interpolation (see Section~\ref{subsubsec:implications_for_1D_interpolation}). This inflation implies that the higher-order terms of in the phase screen approximation may not be negligible, and could become the limiting factor of the ultra-precise SKA-based astrometry using the Multi-View strategy. Further investigations of the same nature, especially using temporally simultaneous (in-beam) calibrators, at low observing frequencies are merited and encouraged.

In this paper, we present two sets of astrometric results -- the quasi-VLBI-only results (see Section~\ref{sec:parameter_inference}) and the VLBI+timing results (see Section~\ref{sec:inference_with_priors}).
Both sets of astrometric results are inferred with the astrometry inference package {\tt sterne}\textsuperscript{\ref{footnote:sterne}}. 
The former set of results is largely independent of any input based on pulsar timing, making use only of orbital parameters as priors in the inference of orbital reflex motion, which affects only four pulsars from our sample and is near-negligible in any case.  The latter, however, additionally adopts the latest available timing parallaxes and proper motions as priors of inference wherever possible, affecting all pulsars in our sample. While the latter approach typically gives more precise results, we note that this is dependent on the accuracy of the timing priors, and identify seven pulsars (\psrc, \psrgb, \psrhb, \psri, \psro, \psrka\ and \psrl) for which disagreement between the quasi-VLBI-only and the timing priors mean that the VLBI+timing results should be treated with caution. 
From the VLBI perspective, we looked into possible causes of additional astrometric uncertainties, including non-optimal calibrator quality (see Sections~\ref{subsec:J1721} and \ref{subsec:J1824}) and calibrator structure evolution (see Section~\ref{sec:inference_with_priors}).
In future, proper motion uncertainties (including any unaccounted systematic error due to calibrator structure evolution) can be greatly reduced with only $\lesssim2$ extra observations per pulsar. For example, a 10-yr time baseline can improve the current VLBI-only proper motion precision by roughly a factor of 8.

From the VLBI+timing parallaxes $\varpi'$, we derived distances $D$  using Equation~\ref{eq:p_D_extended}. Incorporating the PDFs of $D$ and proper motions \{$\mu'_\alpha$, $\mu'_\delta$\}, we estimated the transverse space velocities $v_\perp$ for 16 pulsars with significant distance determination, and found their $v_\perp$ to be generally on the smaller side of the previous estimates \citep{Hobbs05,Gonzalez11}.
\citet{Boodram22} propose that MSPs must have near-zero space velocities in order to explain the Fermi Galactic centre excess.
Our relatively small space velocities inferred for 16 MSPs suggest that MSPs may not be ruled out as the source of the Galactic $\gamma$-ray centre excess.
If the multi-modal feature of the $v_\perp$ is confirmed with a sample of $\sim50$ MSPs, it may serve as a kinematic evidence for alternative formation channels of MSPs (\citealp{Bailyn90,Gautam22}, also see \citealp{Ding22} as an analogy). 

In addition, we estimated the radial accelerations of pulsars with their distances and proper motions (see Section~\ref{sec:orbital_decay_tests}), which allows us to constrain the intrinsic spin period derivative $\dot{P}_\mathrm{s}^\mathrm{int}$ and the intrinsic orbital decay $\dot{P}_\mathrm{b}^\mathrm{int}$ (see Table~\ref{tab:Shk}). We used the improved $\dot{P}_\mathrm{s}^\mathrm{int}$ of \psrhb\ to place a refined upper limit to the death line of $\gamma$-ray pulsars (see Section~\ref{subsec:J1730}), and the $\dot{P}_\mathrm{b}^\mathrm{int}$ (of \psrea\ and \psri) to constrain alternative theories of gravity (see Section~\ref{subsec:alternative_gravity}).
As already noted by \citet{Ding20}, the orbital decay tests (of gravitational theories) with the three viable \mspsrpi\ systems (i.e., \psrea, \psrfb\ and \psri) will be limited by the distance uncertainties, as parallax precision improves much slower than the $\dot{P}_\mathrm{b}^\mathrm{obs}$ precision \citep{Bell96}. 


Moreover, we detected significant angular broadening of \psrgb, which we used to {\bfseries 1)} provide an independent check of the postulated connection between the HII region Sh~2-27 and the main scattering screen of \psrgb, and {\bfseries 2)} test the scintillation models proposed by \citet{Mall22}.
In future scintillation model inferences, angular broadening and temporal broadening measurements, where available, are suggested to be used as priors using Equation~\ref{eq:two_screen_condition1}, in order to achieve more reliable (and potentially more precise) scintillation model parameters. 
Such an inference would also complete the geometric information of the deflection routes (using Equation~\ref{eq:tau_sc2}, for example, in the two-screen case).

\section*{Acknowledgements}

The authors thank Y. Kovalev, L. Petrov and N. Wex for useful discussions.
HD is supported by the OzGrav scholarship through the Australian Research Council project number CE170100004.
PF thanks the continued support by the Max-Planck-Gesellschaft. SC, JMC, EF, DK and TJWL are members of the NANOGrav Physics Frontiers Center, which is supported by NSF award PHY-1430284.
This work is based on observations with the Very Long Baseline Array (VLBA), which is operated by the National Radio Astronomy Observatory (NRAO). The NRAO is a facility of the National Science Foundation operated under cooperative agreement by Associated Universities, Inc.
Pulsar research at Jodrell Bank Observatory is supported by a consolidated grant from STFC.
Pulsar research at UBC is supported by an NSERC Discovery Grant and by the Canadian Institute for Advanced Research.
The Nan\c{c}ay Radio Observatory is operated by the Paris Observatory, associated with the French Centre National de la Recherche Scientifique (CNRS). We acknowledge financial support from the ``Programme National de Cosmologie et Galaxies'' (PNCG) and ``Programme National Hautes Energies'' (PNHE) of CNRS/INSU, France.
The Parkes radio telescope (Murriyang) is part of the Australia Telescope, which is funded by the Commonwealth Government for operation as a National Facility managed by CSIRO.
The Wisconsin $\mathrm{H\alpha}$ Mapper and its $\mathrm{H\alpha}$ Sky Survey have been funded primarily by the National Science Foundation. The facility was designed and built with the help of the University of  Wisconsin Graduate School, Physical Sciences Lab, and Space Astronomy Lab. NOAO staff at Kitt Peak and Cerro Tololo provided on-site support for its remote operation.
Data reduction and analysis was performed on OzSTAR, the Swinburne-based supercomputer.
This work made use of the Swinburne University of Technology software correlator, developed as part of the Australian Major National Research Facilities Programme and operated under license. {\tt sterne} as well as other data analysis involved in this work made use of {\tt numpy} \citep{Harris20}, {\tt scipy} \citep{Virtanen20}, {\tt astropy} \citep{The-Astropy-Collaboration18} and the {\tt bilby} package \citep{Ashton19}.

\section*{Data and Code Availability}
\label{sec:data_availability}
\begin{itemize}
    \item The VLBA data can be downloaded from the NRAO Archive Interface at \url{https://data.nrao.edu} with the project codes in Table~\ref{tab:MSPs}.
    \item Image models of phase calibrators are provided at \url{https://github.com/dingswin/calibrator_models_for_astrometry}.
    \item Supplementary materials supporting this paper can be found at \url{https://github.com/dingswin/publication_related_materials}.
    \item The data reduction pipeline {\tt psrvlbireduce} is available at \url{https://github.com/dingswin/psrvlbireduce} (version ID: b8ddafd).
    \item The astrometry inference package {\tt sterne} can be accessed at \url{https://github.com/dingswin/sterne}.
\end{itemize}



\bibliographystyle{mnras}
\bibliography{haoding} 

\begin{thebibliography}{}
\makeatletter
\relax
\def\mn@urlcharsother{\let\do\@makeother \do\$\do\&\do\#\do\^\do\_\do\%\do\~}
\def\mn@doi{\begingroup\mn@urlcharsother \@ifnextchar [ {\mn@doi@}
  {\mn@doi@[]}}
\def\mn@doi@[#1]#2{\def\@tempa{#1}\ifx\@tempa\@empty \href
  {http://dx.doi.org/#2} {doi:#2}\else \href {http://dx.doi.org/#2} {#1}\fi
  \endgroup}
\def\mn@eprint#1#2{\mn@eprint@#1:#2::\@nil}
\def\mn@eprint@arXiv#1{\href {http://arxiv.org/abs/#1} {{\tt arXiv:#1}}}
\def\mn@eprint@dblp#1{\href {http://dblp.uni-trier.de/rec/bibtex/#1.xml}
  {dblp:#1}}
\def\mn@eprint@#1:#2:#3:#4\@nil{\def\@tempa {#1}\def\@tempb {#2}\def\@tempc
  {#3}\ifx \@tempc \@empty \let \@tempc \@tempb \let \@tempb \@tempa \fi \ifx
  \@tempb \@empty \def\@tempb {arXiv}\fi \@ifundefined
  {mn@eprint@\@tempb}{\@tempb:\@tempc}{\expandafter \expandafter \csname
  mn@eprint@\@tempb\endcsname \expandafter{\@tempc}}}

\bibitem[\protect\citeauthoryear{Abazajian \& Kaplinghat}{Abazajian \&
  Kaplinghat}{2012}]{Abazajian12}
Abazajian K.~N.,  Kaplinghat M.,  2012, \prd, 86, 083511

\bibitem[\protect\citeauthoryear{{Abbott} et~al.,}{{Abbott}
  et~al.}{2016}]{Abbott16}
{Abbott} B.~P.,  et~al., 2016, \mn@doi [Physical Review Letters]
  {10.1103/PhysRevLett.116.061102}, \href
  {https://ui.adsabs.harvard.edu/abs/2016PhRvL.116f1102A} {116, 061102}

\bibitem[\protect\citeauthoryear{{Alpar}, {Cheng}, {Ruderman}  \&
  {Shaham}}{{Alpar} et~al.}{1982}]{Alpar82}
{Alpar} M.~A.,  {Cheng} A.~F.,  {Ruderman} M.~A.,   {Shaham} J.,  1982, \mn@doi
  [\nat] {10.1038/300728a0}, \href
  {https://ui.adsabs.harvard.edu/abs/1982Natur.300..728A} {300, 728}

\bibitem[\protect\citeauthoryear{Antoniadis}{Antoniadis}{2021}]{Antoniadis21}
Antoniadis J.,  2021, Monthly Notices of the Royal Astronomical Society, 501,
  1116

\bibitem[\protect\citeauthoryear{Antoniadis, Van~Kerkwijk, Koester, Freire,
  Wex, Tauris, Kramer  \& Bassa}{Antoniadis et~al.}{2012}]{Antoniadis12}
Antoniadis J.,  Van~Kerkwijk M.,  Koester D.,  Freire P.,  Wex N.,  Tauris T.,
  Kramer M.,   Bassa C.,  2012, \mnras, 423, 3316

\bibitem[\protect\citeauthoryear{{Antoniadis} et~al.,}{{Antoniadis}
  et~al.}{2013}]{Antoniadis13}
{Antoniadis} J.,  et~al., 2013, \mn@doi [Science] {10.1126/science.1233232},
  \href {https://ui.adsabs.harvard.edu/abs/2013Sci...340..448A} {340, 448}

\bibitem[\protect\citeauthoryear{Antoniadis et~al.,}{Antoniadis
  et~al.}{2022}]{Antoniadis22}
Antoniadis J.,  et~al., 2022, \mnras, 510, 4873

\bibitem[\protect\citeauthoryear{Armstrong, Rickett  \& Spangler}{Armstrong
  et~al.}{1995}]{Armstrong95}
Armstrong J.,  Rickett B.,   Spangler S.,  1995, \apj, 443, 209

\bibitem[\protect\citeauthoryear{Arzoumanian et~al.,}{Arzoumanian
  et~al.}{2018}]{Arzoumanian18a}
Arzoumanian Z.,  et~al., 2018, \apjs, 235, 37

\bibitem[\protect\citeauthoryear{Arzoumanian et~al.,}{Arzoumanian
  et~al.}{2020}]{Arzoumanian20}
Arzoumanian Z.,  et~al., 2020, \apjl, 905, L34

\bibitem[\protect\citeauthoryear{Ashton et~al.,}{Ashton
  et~al.}{2019}]{Ashton19}
Ashton G.,  et~al., 2019, \mn@doi [The Astrophysical Journal Supplement Series]
  {10.3847/1538-4365/ab06fc}, 241, 27

\bibitem[\protect\citeauthoryear{Bailer-Jones}{Bailer-Jones}{2015}]{Bailer-Jones15}
Bailer-Jones C.~A.,  2015, \mn@doi [\pasp] {10.1086/683116}, 127, 994

\bibitem[\protect\citeauthoryear{Bailer-Jones, Rybizki, Fouesneau, Demleitner
  \& Andrae}{Bailer-Jones et~al.}{2021}]{Bailer-Jones21}
Bailer-Jones C.,  Rybizki J.,  Fouesneau M.,  Demleitner M.,   Andrae R.,
  2021, \mn@doi [\aj] {10.3847/1538-3881/abd806}, 161, 147

\bibitem[\protect\citeauthoryear{Bailes et~al.,}{Bailes
  et~al.}{2020}]{Bailes20}
Bailes M.,  et~al., 2020, \pasa, 37

\bibitem[\protect\citeauthoryear{Bailyn \& Grindlay}{Bailyn \&
  Grindlay}{1990}]{Bailyn90}
Bailyn C.~D.,  Grindlay J.~E.,  1990, \apj, 353, 159

\bibitem[\protect\citeauthoryear{{Bartel}, {Herring}, {Ratner}, {Shapiro}  \&
  {Corey}}{{Bartel} et~al.}{1986}]{Bartel86}
{Bartel} N.,  {Herring} T.~A.,  {Ratner} M.~I.,  {Shapiro} I.~I.,   {Corey}
  B.~E.,  1986, \mn@doi [\nat] {10.1038/319733a0}, \href
  {https://ui.adsabs.harvard.edu/abs/1986Natur.319..733B} {319, 733}

\bibitem[\protect\citeauthoryear{Bassa et~al.,}{Bassa et~al.}{2016}]{Bassa16}
Bassa C.,  et~al., 2016, \mnras, 460, 2207

\bibitem[\protect\citeauthoryear{Baumgardt \& Vasiliev}{Baumgardt \&
  Vasiliev}{2021}]{Baumgardt21}
Baumgardt H.,  Vasiliev E.,  2021, \mnras, 505, 5957

\bibitem[\protect\citeauthoryear{Beasley \& Conway}{Beasley \&
  Conway}{1995}]{Beasley95}
Beasley A.,  Conway J.,  1995, in Very Long Baseline Interferometry and the
  VLBA. p.~327

\bibitem[\protect\citeauthoryear{Becker, White  \& Helfand}{Becker
  et~al.}{1995}]{Becker95}
Becker R.~H.,  White R.~L.,   Helfand D.~J.,  1995, \apj, 450, 559

\bibitem[\protect\citeauthoryear{{Bell} \& {Bailes}}{{Bell} \&
  {Bailes}}{1996}]{Bell96}
{Bell} J.~F.,  {Bailes} M.,  1996, \mn@doi [\apjl] {10.1086/309862}, \href
  {https://ui.adsabs.harvard.edu/abs/1996ApJ...456L..33B} {456, L33}

\bibitem[\protect\citeauthoryear{Binney \& Tremaine}{Binney \&
  Tremaine}{2011}]{Binney11}
Binney J.,  Tremaine S.,  2011, Galactic dynamics.
Princeton university press

\bibitem[\protect\citeauthoryear{Boodram \& Heinke}{Boodram \&
  Heinke}{2022}]{Boodram22}
Boodram O.,  Heinke C.~O.,  2022, \mnras, 512, 4239

\bibitem[\protect\citeauthoryear{Bovy}{Bovy}{2020}]{Bovy20}
Bovy J.,  2020, arXiv preprint arXiv:2012.02169

\bibitem[\protect\citeauthoryear{Bower et~al.,}{Bower et~al.}{2014}]{Bower14}
Bower G.~C.,  et~al., 2014, \apjl, 780, L2

\bibitem[\protect\citeauthoryear{Brisken, Benson, Goss  \& Thorsett}{Brisken
  et~al.}{2002}]{Brisken02}
Brisken W.~F.,  Benson J.~M.,  Goss W.,   Thorsett S.,  2002, \apj, 571, 906

\bibitem[\protect\citeauthoryear{{Burgay} et~al.,}{{Burgay}
  et~al.}{2003}]{Burgay03}
{Burgay} M.,  et~al., 2003, \mn@doi [\nat] {10.1038/nature02124}, \href
  {https://ui.adsabs.harvard.edu/abs/2003Natur.426..531B} {426, 531}

\bibitem[\protect\citeauthoryear{Burgay et~al.,}{Burgay
  et~al.}{2006}]{Burgay06}
Burgay M.,  et~al., 2006, \mnras, 368, 283

\bibitem[\protect\citeauthoryear{Cameron et~al.,}{Cameron
  et~al.}{2018}]{Cameron18}
Cameron A.,  et~al., 2018, \mnras\ Letters, 475, L57

\bibitem[\protect\citeauthoryear{{Carr}}{{Carr}}{1980}]{Carr80}
{Carr} B.~J.,  1980, \aap, \href
  {https://ui.adsabs.harvard.edu/abs/1980A%26A....89....6C} {89, 6}

\bibitem[\protect\citeauthoryear{Chakrabarti, Chang, Lam, Vigeland  \&
  Quillen}{Chakrabarti et~al.}{2021}]{Chakrabarti21}
Chakrabarti S.,  Chang P.,  Lam M.~T.,  Vigeland S.~J.,   Quillen A.~C.,  2021,
  \apjl, 907, L26

\bibitem[\protect\citeauthoryear{Charlot et~al.,}{Charlot
  et~al.}{2020}]{Charlot20}
Charlot P.,  et~al., 2020, \aap, 644, A159

\bibitem[\protect\citeauthoryear{Chatterjee, Cordes, Vlemmings, Arzoumanian,
  Goss  \& Lazio}{Chatterjee et~al.}{2004}]{Chatterjee04}
Chatterjee S.,  Cordes J.,  Vlemmings W.,  Arzoumanian Z.,  Goss W.,   Lazio
  T.,  2004, \apj, 604, 339

\bibitem[\protect\citeauthoryear{{Chatterjee} et~al.,}{{Chatterjee}
  et~al.}{2009}]{Chatterjee09}
{Chatterjee} S.,  et~al., 2009, \mn@doi [\apj] {10.1088/0004-637X/698/1/250},
  \href {https://ui.adsabs.harvard.edu/abs/2009ApJ...698..250C} {698, 250}

\bibitem[\protect\citeauthoryear{Chen et~al.,}{Chen et~al.}{2021}]{Chen21}
Chen S.,  et~al., 2021, \mnras, 508, 4970

\bibitem[\protect\citeauthoryear{Condon, Cotton, Greisen, Yin, Perley, Taylor
  \& Broderick}{Condon et~al.}{1998}]{Condon98}
Condon J.~J.,  Cotton W.,  Greisen E.,  Yin Q.,  Perley R.~A.,  Taylor G.,
  Broderick J.,  1998, \aj, 115, 1693

\bibitem[\protect\citeauthoryear{Cordes \& Lazio}{Cordes \&
  Lazio}{2002}]{Cordes02}
Cordes J.~M.,  Lazio T. J.~W.,  2002, arXiv preprint astro-ph/0207156

\bibitem[\protect\citeauthoryear{Dehnen \& Binney}{Dehnen \&
  Binney}{1998}]{Dehnen98}
Dehnen W.,  Binney J.,  1998, \mnras, 294, 429

\bibitem[\protect\citeauthoryear{Deller, Verbiest, Tingay  \& Bailes}{Deller
  et~al.}{2008}]{Deller08}
Deller A.,  Verbiest J.,  Tingay S.,   Bailes M.,  2008, \apjl, 685, L67

\bibitem[\protect\citeauthoryear{{Deller}, {Bailes}  \& {Tingay}}{{Deller}
  et~al.}{2009}]{Deller09}
{Deller} A.~T.,  {Bailes} M.,   {Tingay} S.~J.,  2009, \mn@doi [Science]
  {10.1126/science.1167969}, \href
  {https://ui.adsabs.harvard.edu/abs/2009Sci...323.1327D} {323, 1327}

\bibitem[\protect\citeauthoryear{{Deller} et~al.,}{{Deller}
  et~al.}{2011}]{Deller11a}
{Deller} A.~T.,  et~al., 2011, \mn@doi [\pasp] {10.1086/658907}, \href
  {https://ui.adsabs.harvard.edu/abs/2011PASP..123..275D} {123, 275}

\bibitem[\protect\citeauthoryear{Deller, Boyles, Lorimer, Kaspi, McLaughlin,
  Ransom, Stairs  \& Stovall}{Deller et~al.}{2013}]{Deller13}
Deller A.,  Boyles J.,  Lorimer D.,  Kaspi V.,  McLaughlin M.,  Ransom S.,
  Stairs I.,   Stovall K.,  2013, \apj, 770, 145

\bibitem[\protect\citeauthoryear{{Deller} et~al.,}{{Deller}
  et~al.}{2016}]{Deller16}
{Deller} A.~T.,  et~al., 2016, \mn@doi [\apj] {10.3847/0004-637X/828/1/8},
  \href {http://adsabs.harvard.edu/abs/2016ApJ...828....8D} {828, 8}

\bibitem[\protect\citeauthoryear{{Deller}, {Weisberg}, {Nice}  \&
  {Chatterjee}}{{Deller} et~al.}{2018}]{Deller18}
{Deller} A.~T.,  {Weisberg} J.~M.,  {Nice} D.~J.,   {Chatterjee} S.,  2018,
  \mn@doi [\apj] {10.3847/1538-4357/aacf95}, \href
  {https://ui.adsabs.harvard.edu/abs/2018ApJ...862..139D} {862, 139}

\bibitem[\protect\citeauthoryear{{Deller} et~al.,}{{Deller}
  et~al.}{2019}]{Deller19}
{Deller} A.~T.,  et~al., 2019, \mn@doi [\apj] {10.3847/1538-4357/ab11c7}, \href
  {https://ui.adsabs.harvard.edu/abs/2019ApJ...875..100D} {875, 100}

\bibitem[\protect\citeauthoryear{{Desvignes} et~al.,}{{Desvignes}
  et~al.}{2016}]{Desvignes16}
{Desvignes} G.,  et~al., 2016, \mn@doi [\mnras] {10.1093/mnras/stw483}, \href
  {http://adsabs.harvard.edu/abs/2016MNRAS.458.3341D} {458, 3341}

\bibitem[\protect\citeauthoryear{{Detweiler}}{{Detweiler}}{1979}]{Detweiler79}
{Detweiler} S.,  1979, \mn@doi [\apj] {10.1086/157593}, \href
  {https://ui.adsabs.harvard.edu/abs/1979ApJ...234.1100D} {234, 1100}

\bibitem[\protect\citeauthoryear{Ding}{Ding}{2022}]{Ding22a}
Ding H.,  2022, PhD thesis, Swinburne University of Technology

\bibitem[\protect\citeauthoryear{Ding et~al.,}{Ding et~al.}{2020a}]{Ding20c}
Ding H.,  et~al., 2020a, \mn@doi [\mnras] {10.1093/mnras/staa2531}, 498, 3736

\bibitem[\protect\citeauthoryear{Ding, Deller, Freire, Kaplan, Lazio, Shannon
  \& Stappers}{Ding et~al.}{2020b}]{Ding20}
Ding H.,  Deller A.~T.,  Freire P.,  Kaplan D.~L.,  Lazio T. J.~W.,  Shannon
  R.,   Stappers B.,  2020b, \mn@doi [\apj] {10.3847/1538-4357/ab8f27}, 896, 85

\bibitem[\protect\citeauthoryear{Ding, Deller  \& Miller-Jones}{Ding
  et~al.}{2021a}]{Ding21}
Ding H.,  Deller A.~T.,   Miller-Jones J. C.~A.,  2021a, \mn@doi [\pasa]
  {10.1017/pasa.2021.37}, 38, e048

\bibitem[\protect\citeauthoryear{Ding, Deller, Fonseca, Stairs, Stappers  \&
  Lyne}{Ding et~al.}{2021b}]{Ding21a}
Ding H.,  Deller A.~T.,  Fonseca E.,  Stairs I.~H.,  Stappers B.,   Lyne A.,
  2021b, \mn@doi [\apjl] {10.3847/2041-8213/ac3091}, 921, L19

\bibitem[\protect\citeauthoryear{Ding, Deller, Lower  \& Shannon}{Ding
  et~al.}{2022}]{Ding22}
Ding H.,  Deller A.,  Lower M.,   Shannon R.,  2022, arXiv preprint
  arXiv:2201.07376

\bibitem[\protect\citeauthoryear{Eardley}{Eardley}{1975}]{Eardley75}
Eardley D.~M.,  1975, \apj, 196, L59

\bibitem[\protect\citeauthoryear{Edwards \& Bailes}{Edwards \&
  Bailes}{2001}]{Edwards01}
Edwards R.~T.,  Bailes M.,  2001, \apj, 553, 801

\bibitem[\protect\citeauthoryear{Edwards, Hobbs  \& Manchester}{Edwards
  et~al.}{2006}]{Edwards06}
Edwards R.~T.,  Hobbs G.,   Manchester R.,  2006, \mnras, 372, 1549

\bibitem[\protect\citeauthoryear{Faisal~Alam et~al.,}{Faisal~Alam
  et~al.}{2020}]{Faisal-Alam20}
Faisal~Alam M.,  et~al., 2020, arXiv, pp arXiv--2005

\bibitem[\protect\citeauthoryear{Ferdman et~al.,}{Ferdman
  et~al.}{2020}]{Ferdman20}
Ferdman R.,  et~al., 2020, Nature, 583, 211

\bibitem[\protect\citeauthoryear{{Fermi-LAT Collaboration}}{{Fermi-LAT
  Collaboration}}{2022}]{Fermi-LAT-Collaboration22}
{Fermi-LAT Collaboration} 2022, Science, 376, 521

\bibitem[\protect\citeauthoryear{Fienga, Avdellidou  \& Hanu{\v{s}}}{Fienga
  et~al.}{2020}]{Fienga20a}
Fienga A.,  Avdellidou C.,   Hanu{\v{s}} J.,  2020, \mnras, 492, 589

\bibitem[\protect\citeauthoryear{Fomalont \& Kopeikin}{Fomalont \&
  Kopeikin}{2003}]{Fomalont03}
Fomalont E.~B.,  Kopeikin S.~M.,  2003, \apj, 598, 704

\bibitem[\protect\citeauthoryear{Fonseca, Stairs  \& Thorsett}{Fonseca
  et~al.}{2014}]{Fonseca14}
Fonseca E.,  Stairs I.~H.,   Thorsett S.~E.,  2014, \apj, 787, 82

\bibitem[\protect\citeauthoryear{{Foster} \& {Backer}}{{Foster} \&
  {Backer}}{1990}]{Foster90}
{Foster} R.~S.,  {Backer} D.~C.,  1990, \mn@doi [\apj] {10.1086/169195}, \href
  {https://ui.adsabs.harvard.edu/abs/1990ApJ...361..300F} {361, 300}

\bibitem[\protect\citeauthoryear{Foster, Cadwell, Wolszczan  \&
  Anderson}{Foster et~al.}{1995}]{Foster95}
Foster R.,  Cadwell B.,  Wolszczan A.,   Anderson S.,  1995, \apj, 454, 826

\bibitem[\protect\citeauthoryear{{Freire} et~al.,}{{Freire}
  et~al.}{2012}]{Freire12}
{Freire} P.~C.~C.,  et~al., 2012, \mn@doi [\mnras]
  {10.1111/j.1365-2966.2012.21253.x}, \href
  {https://ui.adsabs.harvard.edu/abs/2012MNRAS.423.3328F} {423, 3328}

\bibitem[\protect\citeauthoryear{{Gaia Collaboration} et~al.,}{{Gaia
  Collaboration} et~al.}{2018}]{Gaia-Collaboration18a}
{Gaia Collaboration} et~al., 2018, Astronomy \& Astrophysics, 616

\bibitem[\protect\citeauthoryear{{Gaia Collaboration} et~al.,}{{Gaia
  Collaboration} et~al.}{2022}]{Gaia-Collaboration22}
{Gaia Collaboration} et~al., 2022, \aap, \href
  {https://ui.adsabs.harvard.edu/abs/2022arXiv220800211G} {p. arXiv:2208.00211}

\bibitem[\protect\citeauthoryear{Gautam, Crocker, Ferrario, Ruiter, Ploeg,
  Gordon  \& Macias}{Gautam et~al.}{2022}]{Gautam22}
Gautam A.,  Crocker R.~M.,  Ferrario L.,  Ruiter A.~J.,  Ploeg H.,  Gordon C.,
   Macias O.,  2022, \mn@doi [Nature Astronomy] {10.1038/s41550-022-01658-3},
  6, 703

\bibitem[\protect\citeauthoryear{Gold}{Gold}{1968}]{Gold68}
Gold T.,  1968, \mn@doi [Nature] {10.1038/218731a0}, 218, 731

\bibitem[\protect\citeauthoryear{Goncharov et~al.,}{Goncharov
  et~al.}{2021}]{Goncharov21}
Goncharov B.,  et~al., 2021, \apjl, 917, L19

\bibitem[\protect\citeauthoryear{Gonzalez et~al.,}{Gonzalez
  et~al.}{2011}]{Gonzalez11}
Gonzalez M.,  et~al., 2011, \apj, 743, 102

\bibitem[\protect\citeauthoryear{{Gravity Collaboration} et~al.,}{{Gravity
  Collaboration} et~al.}{2018}]{Gravity-Collaboration18}
{Gravity Collaboration} et~al., 2018, \mn@doi [\aap]
  {10.1051/0004-6361/201833718}, \href
  {https://ui.adsabs.harvard.edu/abs/2018A%26A...615L..15G} {615, L15}

\bibitem[\protect\citeauthoryear{{Greisen}}{{Greisen}}{2003}]{Greisen03}
{Greisen} E.~W.,  2003, in {Heck} A.,  ed.,  Astrophysics and Space Science
  Library Vol. 285, Information Handling in Astronomy - Historical Vistas.
  Springer, p.~109, \mn@doi{10.1007/0-306-48080-8_7}

\bibitem[\protect\citeauthoryear{Guillemot et~al.,}{Guillemot
  et~al.}{2016}]{Guillemot16}
Guillemot L.,  et~al., 2016, \aap, 587, A109

\bibitem[\protect\citeauthoryear{Guo et~al.,}{Guo et~al.}{2021}]{Guo21}
Guo Y.,  et~al., 2021, \aap, 654, A16

\bibitem[\protect\citeauthoryear{Harris et~al.,}{Harris
  et~al.}{2020}]{Harris20}
Harris C.~R.,  et~al., 2020, Nature, 585, 357

\bibitem[\protect\citeauthoryear{Helfand, Taylor, Backus  \& Cordes}{Helfand
  et~al.}{1980}]{Helfand80}
Helfand D.,  Taylor J.,  Backus P.,   Cordes J.,  1980, \apj, 237, 206

\bibitem[\protect\citeauthoryear{{Herbst}}{{Herbst}}{1975}]{Herbst75}
{Herbst} W.,  1975, \mn@doi [\aj] {10.1086/111771}, \href
  {https://ui.adsabs.harvard.edu/abs/1975AJ.....80..503H} {80, 503}

\bibitem[\protect\citeauthoryear{Hewish, Bell, Pilkington, Scott  \&
  Collins}{Hewish et~al.}{1969}]{Hewish69}
Hewish A.,  Bell S.,  Pilkington J.,  Scott P.,   Collins R.,  1969, Nature,
  224, 472

\bibitem[\protect\citeauthoryear{Hobbs, Lorimer, Lyne  \& Kramer}{Hobbs
  et~al.}{2005}]{Hobbs05}
Hobbs G.,  Lorimer D.,  Lyne A.,   Kramer M.,  2005, \mnras, 360, 974

\bibitem[\protect\citeauthoryear{Hobbs, Lyne  \& Kramer}{Hobbs
  et~al.}{2010}]{Hobbs10}
Hobbs G.,  Lyne A.,   Kramer M.,  2010, \mnras, 402, 1027

\bibitem[\protect\citeauthoryear{{Hulse} \& {Taylor}}{{Hulse} \&
  {Taylor}}{1975}]{Hulse75}
{Hulse} R.~A.,  {Taylor} J.~H.,  1975, \mn@doi [\apjl] {10.1086/181708}, \href
  {https://ui.adsabs.harvard.edu/abs/1975ApJ...195L..51H} {195, L51}

\bibitem[\protect\citeauthoryear{Hyland, Reid, Ellingsen, Rioja, Dodson, Orosz,
  Masson  \& MacCallum}{Hyland et~al.}{2022}]{Hyland22}
Hyland L.~J.,  Reid M.~J.,  Ellingsen S.~P.,  Rioja M.~J.,  Dodson R.,  Orosz
  G.,  Masson C.~R.,   MacCallum J.,  2022, arXiv preprint arXiv:2205.00092

\bibitem[\protect\citeauthoryear{Igoshev, Verbunt  \& Cator}{Igoshev
  et~al.}{2016}]{Igoshev16}
Igoshev A.,  Verbunt F.,   Cator E.,  2016, Astronomy \& Astrophysics, 591,
  A123

\bibitem[\protect\citeauthoryear{Imai, Sakai, Nakanishi, Sakanoue, Honma  \&
  Miyaji}{Imai et~al.}{2012}]{Imai12}
Imai H.,  Sakai N.,  Nakanishi H.,  Sakanoue H.,  Honma M.,   Miyaji T.,  2012,
  \pasj, 64, 142

\bibitem[\protect\citeauthoryear{Jacoby, Bailes, Ord, Edwards  \&
  Kulkarni}{Jacoby et~al.}{2009}]{Jacoby09}
Jacoby B.,  Bailes M.,  Ord S.,  Edwards R.,   Kulkarni S.,  2009, \apj, 699

\bibitem[\protect\citeauthoryear{Janssen, Stappers, Kramer, Nice, Jessner,
  Cognard  \& Purver}{Janssen et~al.}{2008}]{Janssen08}
Janssen G.,  Stappers B.,  Kramer M.,  Nice D.,  Jessner A.,  Cognard I.,
  Purver M.,  2008, \aap, 490, 753

\bibitem[\protect\citeauthoryear{{Jennings}, {Kaplan}, {Chatterjee}, {Cordes}
  \& {Deller}}{{Jennings} et~al.}{2018}]{Jennings18}
{Jennings} R.~J.,  {Kaplan} D.~L.,  {Chatterjee} S.,  {Cordes} J.~M.,
  {Deller} A.~T.,  2018, \mn@doi [\apj] {10.3847/1538-4357/aad084}, \href
  {https://ui.adsabs.harvard.edu/abs/2018ApJ...864...26J} {864, 26}

\bibitem[\protect\citeauthoryear{Kaplan et~al.,}{Kaplan
  et~al.}{2016}]{Kaplan16}
Kaplan D.~L.,  et~al., 2016, \apj, 826, 86

\bibitem[\protect\citeauthoryear{Kerr et~al.,}{Kerr et~al.}{2020}]{Kerr20}
Kerr M.,  et~al., 2020, \pasa, 37

\bibitem[\protect\citeauthoryear{{Kettenis}, {van Langevelde}, {Reynolds}  \&
  {Cotton}}{{Kettenis} et~al.}{2006}]{Kettenis06}
{Kettenis} M.,  {van Langevelde} H.~J.,  {Reynolds} C.,   {Cotton} B.,  2006,
  in {Gabriel} C.,  {Arviset} C.,  {Ponz} D.,   {Enrique} S.,  eds,
  Astronomical Society of the Pacific Conference Series Vol. 351, Astronomical
  Data Analysis Software and Systems XV. p.~497

\bibitem[\protect\citeauthoryear{Kirsten, Vlemmings, Campbell, Kramer  \&
  Chatterjee}{Kirsten et~al.}{2015}]{Kirsten15}
Kirsten F.,  Vlemmings W.,  Campbell R.~M.,  Kramer M.,   Chatterjee S.,  2015,
  \aap, 577, A111

\bibitem[\protect\citeauthoryear{Kopeikin}{Kopeikin}{1996}]{Kopeikin96}
Kopeikin S.,  1996, \apjl, 467, L93

\bibitem[\protect\citeauthoryear{Kramer et~al.,}{Kramer
  et~al.}{2021}]{Kramer21a}
Kramer M.,  et~al., 2021, Physical Review X, 11, 041050

\bibitem[\protect\citeauthoryear{{Lazaridis} et~al.,}{{Lazaridis}
  et~al.}{2009}]{Lazaridis09}
{Lazaridis} K.,  et~al., 2009, \mn@doi [\mnras]
  {10.1111/j.1365-2966.2009.15481.x}, \href
  {http://adsabs.harvard.edu/abs/2009MNRAS.400..805L} {400, 805}

\bibitem[\protect\citeauthoryear{Lazarus et~al.,}{Lazarus
  et~al.}{2016}]{Lazarus16}
Lazarus P.,  et~al., 2016, \apj, 831, 150

\bibitem[\protect\citeauthoryear{Lentati, Kerr, Dai, Shannon, Hobbs  \&
  Os{\l}owski}{Lentati et~al.}{2017}]{Lentati17}
Lentati L.,  Kerr M.,  Dai S.,  Shannon R.,  Hobbs G.,   Os{\l}owski S.,  2017,
  \mnras, 468, 1474

\bibitem[\protect\citeauthoryear{Lestrade, Rogers, Whitney, Niell, Phillips  \&
  Preston}{Lestrade et~al.}{1990}]{Lestrade90}
Lestrade J.-F.,  Rogers A.,  Whitney A.,  Niell A.,  Phillips R.,   Preston R.,
   1990, \aj, 99, 1663

\bibitem[\protect\citeauthoryear{Li et~al.,}{Li et~al.}{2018}]{Li18}
Li Z.,  et~al., 2018, \mnras, 476, 399

\bibitem[\protect\citeauthoryear{Lindegren et~al.,}{Lindegren
  et~al.}{2021}]{Lindegren21}
Lindegren L.,  et~al., 2021, \mn@doi [\aap] {10.1051/0004-6361/202039653}, 649,
  A4

\bibitem[\protect\citeauthoryear{{Lobanov}}{{Lobanov}}{1998}]{Lobanov98}
{Lobanov} A.~P.,  1998, \aap, \href
  {https://ui.adsabs.harvard.edu/abs/1998A%26A...330...79L} {330, 79}

\bibitem[\protect\citeauthoryear{Lorimer et~al.,}{Lorimer
  et~al.}{1995}]{Lorimer95a}
Lorimer D.,  et~al., 1995, \apj, 439, 933

\bibitem[\protect\citeauthoryear{Lorimer et~al.,}{Lorimer
  et~al.}{2006}]{Lorimer06a}
Lorimer D.,  et~al., 2006, \mnras, 372, 777

\bibitem[\protect\citeauthoryear{Lundgren, Cordes, Foster, Wolszczan  \&
  Camilo}{Lundgren et~al.}{1996}]{Lundgren96}
Lundgren S.,  Cordes J.,  Foster R.,  Wolszczan A.,   Camilo F.,  1996, \apj,
  458, L33

\bibitem[\protect\citeauthoryear{{Lutz} \& {Kelker}}{{Lutz} \&
  {Kelker}}{1973}]{Lutz73}
{Lutz} T.~E.,  {Kelker} D.~H.,  1973, \mn@doi [\pasp] {10.1086/129506}, \href
  {https://ui.adsabs.harvard.edu/abs/1973PASP...85..573L} {85, 573}

\bibitem[\protect\citeauthoryear{Lyne, Brinklow, Middleditch, Kulkarni, Backer
  \& Clifton}{Lyne et~al.}{1987}]{Lyne87a}
Lyne A.,  Brinklow A.,  Middleditch J.,  Kulkarni S.,  Backer D.~a.,   Clifton
  T.,  1987, Nature, 328, 399

\bibitem[\protect\citeauthoryear{Macquart et~al.,}{Macquart
  et~al.}{2020}]{Macquart20}
Macquart J.-P.,  et~al., 2020, Nature, 581, 391

\bibitem[\protect\citeauthoryear{{Madison}, {Chatterjee}  \&
  {Cordes}}{{Madison} et~al.}{2013}]{Madison13}
{Madison} D.~R.,  {Chatterjee} S.,   {Cordes} J.~M.,  2013, \mn@doi [\apj]
  {10.1088/0004-637X/777/2/104}, \href
  {http://adsabs.harvard.edu/abs/2013ApJ...777..104M} {777, 104}

\bibitem[\protect\citeauthoryear{Mall et~al.,}{Mall et~al.}{2022}]{Mall22}
Mall G.,  et~al., 2022, \mn@doi [\mnras] {10.1093/mnras/stac096}, 511, 1104

\bibitem[\protect\citeauthoryear{{Manchester}, {Hobbs}, {Teoh}  \&
  {Hobbs}}{{Manchester} et~al.}{2005}]{Manchester05}
{Manchester} R.~N.,  {Hobbs} G.~B.,  {Teoh} A.,   {Hobbs} M.,  2005, \mn@doi
  [\aj] {10.1086/428488}, \href
  {https://ui.adsabs.harvard.edu/abs/2005AJ....129.1993M} {129, 1993}

\bibitem[\protect\citeauthoryear{Mannings et~al.,}{Mannings
  et~al.}{2021}]{Mannings21}
Mannings A.~G.,  et~al., 2021, \apj, 917, 75

\bibitem[\protect\citeauthoryear{McMillan}{McMillan}{2017}]{McMillan17}
McMillan P.~J.,  2017, \mnras, 465, 76

\bibitem[\protect\citeauthoryear{{Mingarelli}, {Anderson}, {Bedell}  \&
  {Spergel}}{{Mingarelli} et~al.}{2018}]{Mingarelli18}
{Mingarelli} C.~M.~F.,  {Anderson} L.,  {Bedell} M.,   {Spergel} D.~N.,  2018,
  arXiv e-prints, \href {https://ui.adsabs.harvard.edu/abs/2018arXiv181206262M}
  {}

\bibitem[\protect\citeauthoryear{{Nice} \& {Taylor}}{{Nice} \&
  {Taylor}}{1995}]{Nice95}
{Nice} D.~J.,  {Taylor} J.~H.,  1995, \mn@doi [\apj] {10.1086/175367}, \href
  {https://ui.adsabs.harvard.edu/abs/1995ApJ...441..429N} {441, 429}

\bibitem[\protect\citeauthoryear{Nice, Splaver  \& Stairs}{Nice
  et~al.}{2001}]{Nice01}
Nice D.~J.,  Splaver E.~M.,   Stairs I.~H.,  2001, \apj, 549, 516

\bibitem[\protect\citeauthoryear{Ocker, Cordes  \& Chatterjee}{Ocker
  et~al.}{2020}]{Ocker20}
Ocker S.~K.,  Cordes J.~M.,   Chatterjee S.,  2020, \apj, 897, 124

\bibitem[\protect\citeauthoryear{Pacini}{Pacini}{1968}]{Pacini68}
Pacini F.,  1968, Nature, 219, 145

\bibitem[\protect\citeauthoryear{Pallanca, Mignani, Dalessandro, Ferraro,
  Lanzoni, Possenti, Burgay  \& Sabbi}{Pallanca et~al.}{2012}]{Pallanca12}
Pallanca C.,  Mignani R.,  Dalessandro E.,  Ferraro F.,  Lanzoni B.,  Possenti
  A.,  Burgay M.,   Sabbi E.,  2012, \apj, 755, 180

\bibitem[\protect\citeauthoryear{Park, Folkner, Williams  \& Boggs}{Park
  et~al.}{2021}]{Park21}
Park R.~S.,  Folkner W.~M.,  Williams J.~G.,   Boggs D.~H.,  2021, The
  Astronomical Journal, 161, 105

\bibitem[\protect\citeauthoryear{Pearson}{Pearson}{1895}]{Pearson95}
Pearson K.,  1895, in Royal Society Proceedings. p.~214

\bibitem[\protect\citeauthoryear{Perera et~al.,}{Perera
  et~al.}{2019}]{Perera19}
Perera B.,  et~al., 2019, \mnras, 490, 4666

\bibitem[\protect\citeauthoryear{Perger et~al.,}{Perger
  et~al.}{2018}]{Perger18}
Perger K.,  et~al., 2018, \mnras, 477, 1065

\bibitem[\protect\citeauthoryear{Peters}{Peters}{1964}]{Peters64}
Peters P.~C.,  1964, Physical Review, 136, B1224

\bibitem[\protect\citeauthoryear{Phillips, Ravi, Ebadi  \& Walsworth}{Phillips
  et~al.}{2021}]{Phillips21}
Phillips D.~F.,  Ravi A.,  Ebadi R.,   Walsworth R.~L.,  2021, \prl, 126,
  141103

\bibitem[\protect\citeauthoryear{Piffl et~al.,}{Piffl et~al.}{2014}]{Piffl14}
Piffl T.,  et~al., 2014, \mnras, 445, 3133

\bibitem[\protect\citeauthoryear{Pradel, Charlot  \& Lestrade}{Pradel
  et~al.}{2006}]{Pradel06}
Pradel N.,  Charlot P.,   Lestrade J.-F.,  2006, \aap, 452, 1099

\bibitem[\protect\citeauthoryear{Reardon, Coles, Hobbs, Ord, Kerr, Bailes, Bhat
   \& Venkatraman~Krishnan}{Reardon et~al.}{2019}]{Reardon19}
Reardon D.,  Coles W.,  Hobbs G.,  Ord S.,  Kerr M.,  Bailes M.,  Bhat N.,
  Venkatraman~Krishnan V.,  2019, \mnras, 485, 4389

\bibitem[\protect\citeauthoryear{Reardon et~al.,}{Reardon
  et~al.}{2020}]{Reardon20}
Reardon D.~J.,  et~al., 2020, \apj, 904, 104

\bibitem[\protect\citeauthoryear{Reardon et~al.,}{Reardon
  et~al.}{2021}]{Reardon21}
Reardon D.,  et~al., 2021, \mnras, 507, 2137

\bibitem[\protect\citeauthoryear{Reid, Menten, Brunthaler, Zheng, Moscadelli
  \& Xu}{Reid et~al.}{2009}]{Reid09a}
Reid M.,  Menten K.,  Brunthaler A.,  Zheng X.,  Moscadelli L.,   Xu Y.,  2009,
  \apj, 693, 397

\bibitem[\protect\citeauthoryear{Reid et~al.,}{Reid et~al.}{2019}]{Reid19}
Reid M.,  et~al., 2019, \apj, 885, 131

\bibitem[\protect\citeauthoryear{Rioja, Dodson, Orosz, Imai  \& Frey}{Rioja
  et~al.}{2017}]{Rioja17}
Rioja M.~J.,  Dodson R.,  Orosz G.,  Imai H.,   Frey S.,  2017, \aj, 153, 105

\bibitem[\protect\citeauthoryear{{Roebber}}{{Roebber}}{2019}]{Roebber19}
{Roebber} E.,  2019, \mn@doi [\apj] {10.3847/1538-4357/ab100e}, \href
  {https://ui.adsabs.harvard.edu/abs/2019ApJ...876...55R} {876, 55}

\bibitem[\protect\citeauthoryear{Sayer, Nice  \& Taylor}{Sayer
  et~al.}{1997}]{Sayer97}
Sayer R.,  Nice D.,   Taylor J.,  1997, \apj, 474, 426

\bibitem[\protect\citeauthoryear{{Sesana}, {Vecchio}  \& {Colacino}}{{Sesana}
  et~al.}{2008}]{Sesana08}
{Sesana} A.,  {Vecchio} A.,   {Colacino} C.~N.,  2008, \mn@doi [\mnras]
  {10.1111/j.1365-2966.2008.13682.x}, \href
  {https://ui.adsabs.harvard.edu/abs/2008MNRAS.390..192S} {390, 192}

\bibitem[\protect\citeauthoryear{{Shepherd}, {Pearson}  \& {Taylor}}{{Shepherd}
  et~al.}{1994}]{Shepherd94}
{Shepherd} M.~C.,  {Pearson} T.~J.,   {Taylor} G.~B.,  1994, in Bulletin of the
  American Astronomical Society. pp 987--989

\bibitem[\protect\citeauthoryear{{Shklovskii}}{{Shklovskii}}{1970}]{Shklovskii70}
{Shklovskii} I.~S.,  1970, \sovast, \href
  {https://ui.adsabs.harvard.edu/abs/1970SvA....13..562S} {13, 562}

\bibitem[\protect\citeauthoryear{Siemens, Ellis, Jenet  \& Romano}{Siemens
  et~al.}{2013}]{Siemens13}
Siemens X.,  Ellis J.,  Jenet F.,   Romano J.~D.,  2013, Classical and Quantum
  Gravity, 30, 224015

\bibitem[\protect\citeauthoryear{{Sokolovsky}, {Kovalev}, {Pushkarev}  \&
  {Lobanov}}{{Sokolovsky} et~al.}{2011}]{Sokolovsky11}
{Sokolovsky} K.~V.,  {Kovalev} Y.~Y.,  {Pushkarev} A.~B.,   {Lobanov} A.~P.,
  2011, \mn@doi [\aap] {10.1051/0004-6361/201016072}, \href
  {https://ui.adsabs.harvard.edu/abs/2011A%26A...532A..38S} {532, A38}

\bibitem[\protect\citeauthoryear{Stovall et~al.,}{Stovall
  et~al.}{2018}]{Stovall18}
Stovall K.,  et~al., 2018, \apjl, 854, L22

\bibitem[\protect\citeauthoryear{Tauris et~al.,}{Tauris
  et~al.}{2017}]{Tauris17}
Tauris T.,  et~al., 2017, \mn@doi [\apj] {10.3847/1538-4357/aa7e89}, 846, 170

\bibitem[\protect\citeauthoryear{Taylor \& Weisberg}{Taylor \&
  Weisberg}{1982}]{Taylor82}
Taylor J.~H.,  Weisberg J.~M.,  1982, \apj, 253, 908

\bibitem[\protect\citeauthoryear{{The Astropy Collaboration} et~al.,}{{The
  Astropy Collaboration} et~al.}{2018}]{The-Astropy-Collaboration18}
{The Astropy Collaboration} et~al., 2018, \mn@doi [\aj]
  {10.3847/1538-3881/aabc4f}, \href
  {http://adsabs.harvard.edu/abs/2018AJ....156..123T} {156, 123}

\bibitem[\protect\citeauthoryear{Tiburzi et~al.,}{Tiburzi
  et~al.}{2016}]{Tiburzi16}
Tiburzi C.,  et~al., 2016, \mnras, 455, 4339

\bibitem[\protect\citeauthoryear{Vallisneri et~al.,}{Vallisneri
  et~al.}{2020}]{Vallisneri20}
Vallisneri M.,  et~al., 2020, \apj, 893, 112

\bibitem[\protect\citeauthoryear{Vasiliev \& Baumgardt}{Vasiliev \&
  Baumgardt}{2021}]{Vasiliev21}
Vasiliev E.,  Baumgardt H.,  2021, \mnras, 505, 5978

\bibitem[\protect\citeauthoryear{Verbiest, Weisberg, Chael, Lee  \&
  Lorimer}{Verbiest et~al.}{2012}]{Verbiest12}
Verbiest J.,  Weisberg J.,  Chael A.,  Lee K.,   Lorimer D.,  2012, \apj, 755,
  39

\bibitem[\protect\citeauthoryear{{Vigeland}, {Deller}, {Kaplan}, {Istrate},
  {Stappers}  \& {Tauris}}{{Vigeland} et~al.}{2018}]{Vigeland18}
{Vigeland} S.~J.,  {Deller} A.~T.,  {Kaplan} D.~L.,  {Istrate} A.~G.,
  {Stappers} B.~W.,   {Tauris} T.~M.,  2018, \mn@doi [\apj]
  {10.3847/1538-4357/aaaa73}, \href
  {https://ui.adsabs.harvard.edu/abs/2018ApJ...855..122V} {855, 122}

\bibitem[\protect\citeauthoryear{Virtanen et~al.,}{Virtanen
  et~al.}{2020}]{Virtanen20}
Virtanen P.,  et~al., 2020, \mn@doi [Nature Methods]
  {10.1038/s41592-019-0686-2}, \href {https://rdcu.be/b08Wh} {17, 261}

\bibitem[\protect\citeauthoryear{{Wang} et~al.,}{{Wang} et~al.}{2017}]{Wang17}
{Wang} J.~B.,  et~al., 2017, \mn@doi [\mnras] {10.1093/mnras/stx837}, \href
  {https://ui.adsabs.harvard.edu/abs/2017MNRAS.469..425W} {469, 425}

\bibitem[\protect\citeauthoryear{{Weisberg} \& {Huang}}{{Weisberg} \&
  {Huang}}{2016}]{Weisberg16}
{Weisberg} J.~M.,  {Huang} Y.,  2016, \mn@doi [\apj]
  {10.3847/0004-637X/829/1/55}, \href
  {https://ui.adsabs.harvard.edu/abs/2016ApJ...829...55W} {829, 55}

\bibitem[\protect\citeauthoryear{{Wolszczan}}{{Wolszczan}}{1991}]{Wolszczan91}
{Wolszczan} A.,  1991, \mn@doi [\nat] {10.1038/350688a0}, \href
  {https://ui.adsabs.harvard.edu/abs/1991Natur.350..688W} {350, 688}

\bibitem[\protect\citeauthoryear{Yang, Paragi, van~der Horst, Gurvits,
  Campbell, Giannios, An  \& Komossa}{Yang et~al.}{2016}]{Yang16}
Yang J.,  Paragi Z.,  van~der Horst A.,  Gurvits L.,  Campbell R.,  Giannios
  D.,  An T.,   Komossa S.,  2016, \mnras, 462, L66

\bibitem[\protect\citeauthoryear{{Yao}, {Manchester}  \& {Wang}}{{Yao}
  et~al.}{2017}]{Yao17}
{Yao} J.~M.,  {Manchester} R.~N.,   {Wang} N.,  2017, \mn@doi [\apj]
  {10.3847/1538-4357/835/1/29}, \href
  {https://ui.adsabs.harvard.edu/abs/2017ApJ...835...29Y} {835, 29}

\bibitem[\protect\citeauthoryear{Zhang, An  \& Frey}{Zhang
  et~al.}{2020}]{Zhang20c}
Zhang Y.,  An T.,   Frey S.,  2020, Science Bulletin, 65, 525

\bibitem[\protect\citeauthoryear{Zhao, Freire, Kramer, Shao  \& Wex}{Zhao
  et~al.}{2022}]{Zhao22}
Zhao J.,  Freire P.~C.,  Kramer M.,  Shao L.,   Wex N.,  2022, Classical and
  Quantum Gravity, 39, 11LT01

\bibitem[\protect\citeauthoryear{{Zhu} et~al.,}{{Zhu} et~al.}{2015}]{Zhu15}
{Zhu} W.~W.,  et~al., 2015, \mn@doi [\apj] {10.1088/0004-637X/809/1/41}, \href
  {https://ui.adsabs.harvard.edu/abs/2015ApJ...809...41Z} {809, 41}

\bibitem[\protect\citeauthoryear{Zhu et~al.,}{Zhu et~al.}{2019}]{Zhu19}
Zhu W.,  et~al., 2019, \mnras, 482, 3249

\bibitem[\protect\citeauthoryear{{van der Wateren} et~al.,}{{van der Wateren}
  et~al.}{2022}]{van-der-Wateren22}
{van der Wateren} E.,  et~al., 2022, \aap, 661, A57

\makeatother
\end{thebibliography}




\appendix




\bsp	
\label{lastpage}
\end{document}